\documentclass[10pt]{article}
\usepackage{subfig,amsmath,amssymb,amsthm,graphicx,tikz,bbm,bm,geometry,setspace}
\usepackage[authoryear,longnamesfirst]{natbib}
\newtheorem{lemma}{Lemma}
\newtheorem{definition}{Definition}
\newtheorem{theorem}{Theorem}
\newtheorem{corollary}{Corollary}
\newtheorem{assumption}{Assumption}
\DeclareMathOperator*{\argmin}{arg\,min}
\begin{document}
\title{Instrumental Variable Quantile Regression with Misclassification\thanks{First version: October, 2016. I am grateful to  Peter C. B. Phillips, Arthur Lewbel and four anonymous referees for comments and suggestions that have significantly improved this paper. I would like to thank Alexandre Belloni, St\'ephane Bonhomme, Federico A. Bugni, A. Colin Cameron, V. Joseph Hotz, Shakeeb Khan, Jia Li, Matthew A. Masten, Arnaud Maurel, Adam M. Rosen, and seminar participants at Duke, UC Berkeley, Shanghai University of Finance and Economics, Hakodate Conference in Econometrics, IAAE Annual Conference, Econometric Society North American Summer Meeting, and Econometric Society Asian Meeting for very helpful comments. The usual disclaimer applies.}}
\author{Takuya Ura\thanks{Department of Economics, University of California, Davis, One Shields Avenue, Davis, CA 95616-5270; Email: takura@ucdavis.edu}}
\date{\today}
\maketitle
\begin{abstract}
This paper investigates the instrumental  variable quantile regression model \citep{chernozhukov/hansen:2005,chernozhukov/hansen:2013} with a binary endogenous treatment. 
It offers two identification results when the treatment status is not directly observed. 
The first result is that, remarkably, the reduced-form quantile regression of the outcome variable on the instrumental variable provides a lower bound on the structural quantile treatment effect under the stochastic monotonicity condition \citep{small/tan:2007,dinardo/lee:2011}.
This result is relevant, not only when the treatment variable is subject to misclassification, but also when any measurement of the treatment variable is not available. 
The second result is for the structural quantile function when the treatment status  is measured with error; the sharp identified set is characterized by a set of moment conditions under widely-used assumptions on the measurement error. 
Furthermore, an inference method is provided in the presence of other covariates. 
\begin{description}\item Keywords: Misclassification; Instrumental variable quantile regression; Partial identification; Binary endogenous treatment \end{description}
\end{abstract}

\newpage 
\section{Introduction}\label{sec1}
The instrumental  variable quantile regression model \citep{chernozhukov/hansen:2005,chernozhukov/hansen:2013} aims to investigate heterogeneous treatment effects in the presence of an endogenous binary treatment variable.  
In many empirical applications, the treatment variable is potentially mismeasured, so it is empirically relevant how researchers can use the instrumental  variable quantile regression model with a mismeasured treatment variable.  
For example, \cite{chernozhukov/hansen:2004} use the instrumental  variable quantile regression model to investigate the quantile treatment effect of 401(k) participation on saving behaviors, but the pension plan type is subject to measurement error in survey datasets.
Using the Health and Retirement Study, \cite{gustman/steinmeier/tabatabai:2007} estimate that around one-fourth of the survey respondents misclassified their pension plan type. 
To the best of my knowledge, however, no paper has investigated the instrumental  variable quantile regression model when a binary regressor is potentially misclassified and endogenous. 

This paper has two identification results on the structural quantile function under the rank similarity condition.\footnote{\cite{wuthrich:2016} investigates the instrumental  variable quantile regression model without the rank similarity condition and characterizes the estimand of \cite{chernozhukov/hansen:2005} when the rank similarity condition fails.
\cite{dong/shen:2019}, \cite{frandsen/lefgren:2018}, \cite{kim/park:2016}, and \cite{yu:2017} propose a test for the rank invariance or similarity condition.}  
The first identification result considers a reduced-form parameter, that is, the coefficient of the instrumental  variable when running the quantile regression of  the outcome variable on the instrumental  variable.
Under the rank similarity condition and the stochastic monotonicity condition \citep{small/tan:2007,dinardo/lee:2011}, this reduced-form parameter can be used as a lower bound for the structural quantile treatment effect. 
Although it has been used by empirical studies \citep[e.g.,][]{bitler/hoynes/domina:2016}, the reduced-form quantile regression on the instrumental variable has not been formally related to the structural quantile treatment effect.
Moreover, this result does not depend on the treatment variable or its measurement, and therefore it is relevant even when a measurement does not exist. 

The second identification result is to derive moment conditions for the structural quantile function when the treatment variable is measured with exogenous errors. 
The exogeneity of the measurement error is widely assumed in the measurement error literature \citep[e.g.,][]{bound/brown/mathiowetz:2001} and yields exclusion restrictions similar to \cite{henry/kitamura/salanie:2014}. 
Given the structure of the moment conditions, the structural quantile function can be under-identified even if the order condition for point identification holds. 
In other words, additional assumptions or variables are necessary to achieve point identification for the structural quantile function. 
As an example of additional restrictions, this paper considers two observed measurements for one latent treatment variable, and point-identify the structural quantile function.  
Point identification results from combining two existing methods:  misclassification correction techniques \citep{mahajan:2006,lewbel:2007,hu:2008}, and the identification results in \cite{chernozhukov/hansen:2005,chernozhukov/hansen:2013}.

Based on the partial identification result, an inference method for the structural quantile function is provided by incorporating the misclassification probabilities in the inference method of \cite{chernozhukov/hansen:2008}. 
The proposed inference method can include covariates other than the treatment variable, and it is computationally feasible by imposing a linear-in-parameters structure on the structural quantile function. 
Simulation studies and an empirical illustration demonstrate the finite sample performance for the proposed inference method.

Related to this paper, several papers have considered the problem of mis-measured regressors in the quantile regression framework, e.g., \cite{chesher:1991,chesher:2017}, \cite{schennach:2008}, \cite{galvao/montes-rojas:2009}, \cite{wei/carroll:2009}, \cite{firpo/galvao/song:2015}, and \cite{song:2016}.
They focus on the case in which the mismeasured regressor is continuously distributed, whereas this paper focuses on a binary treatment variable in which the measurement error has to be nonclassical.

\cite{mahajan:2006}, \cite{lewbel:2007}, and \cite{hu:2008} consider identification of the average treatment effect (or, more generally, the conditional density function of the outcome variable given the true treatment variable) when a discrete treatment variable is mismeasured. 
Their identification strategy is based on the assumption that the true treatment variable (or the individual treatment effect in Lewbel, 2007) is exogenous, and there is no straightforward way to modify their results to the endogenous treatment.

\cite{calvi/lewbel/tommasi:2017,yanagi:2017,ura:2015} investigate the local average treatment effect model with a mismeasured binary treatment. The local average treatment effect model is also a model for heterogeneous treatment effects in the presence of endogeneity but has a different structure than the instrumental  variable quantile regression model.

\cite{frazis/loewenstein:2003}, \cite{ditragliaand/garica-jimeno:2015} and \cite{Nguimkeu/Denteh/Tchernis:2017} study a linear regression model in which a binary regressor is potentially misclassified and endogenous. 
Their approach is based on a homogenous treatment effect, which does not hold in the quantile treatment effect framework.

The remainder of this paper is organized as follows. 
Section 2 introduces the instrumental  variable quantile regression model \citep{chernozhukov/hansen:2005,chernozhukov/hansen:2013} with a misclassified treatment variable.  
Section 3 studies the reduced-form quantile regression of the outcome variable on the instrumental variable. 
Section 4 presents the identified set for the structural quantile function. 
Section 5 proposes an inference method based on the identification analysis in Section 4. 
Section 6 provides an empirical illustration and simulation studies. 
Section 7 concludes. 
The appendix includes the proofs and additional results.

The rest of this paper uses the following notations. 
$Pr$ denotes the true probability measure for the observed and unobserved random variables.   
$Q_{RV_1\mid RV_2}(\tau)$ is the $\tau$-th conditional quantile of a continuous random variable $RV_1$ given a random variable $RV_2$. 
$F_{RV_1\mid RV_2}$ is the conditional cumulative distribution function of a random variable $RV_1$ given a random variable $RV_2$. 
$f_{RV_1\mid RV_2}$ is the conditional  probability density function of a random variable $RV_1$ given a random variable $RV_2$. 

\section{Instrumental variable quantile regression model with misclassification}
This section presents notations and assumptions that are essentially those of the instrumental variable quantile regression model \citep{chernozhukov/hansen:2005,chernozhukov/hansen:2013}, though the treatment variable $D^\ast$ is not observed in this paper. 
For the sake of simplicity, covariates are omitted other than the treatment variable when analyzing the identification problem. 
In the instrumental variable quantile regression model, $Y$ is an outcome variable, $D^\ast$ is a binary (true but latent) treatment variable taking values in $\{0,1\}$, and $Z$ is an instrumental  variable. $D^\ast=1$ means that the individual is treated; otherwise, $D^\ast=0$.  

In \cite{chernozhukov/hansen:2004} and Section \ref{sec:empirical_section} of this paper, the outcome variable $Y$ is the net amount of financial assets in dollars, the treatment variable $D^\ast$ is the participation status in a 401(k) program, and the instrumental  variable $Z$ is the 401(k) eligibility indicator of whether an employer offers a 401(k) program to employees. 

The goal of this paper is to investigate treatment effects of $D^\ast$ on $Y$. 
The error term $(U_0,U_1)$ and the (unknown) structural quantile function $q(d^\ast,u)$ are used to model the relationship between the outcome variable $Y$ and  the binary treatment variable $D^\ast$: 
$$
Y=q(D^\ast,U)\mbox{ where }U=(1-D^\ast)U_0+D^\ast U_1.
$$
The random variable $q(d^\ast,U_{d^\ast})$ is the potential outcome variable when $D^\ast=d^\ast$.
The parameter of interest is the $\tau$-th quantile of the counterfactual outcome variable, $q(d^\ast,\tau)$, for a given value of $\tau\in(0,1)$.

To identify $q(d^\ast,u)$ even partially, it is necessary to impose some structure on the unknown function, $q(d^\ast,u)$, and on the unobserved variable, $U_{d^\ast}$.
The following assumptions are based on \cite{chernozhukov/hansen:2005,chernozhukov/hansen:2013}. 
Unlike their papers, the condition is local at the given value of $\tau$, which is sufficient for deriving the testable implication in \citet[][Theorem 1]{chernozhukov/hansen:2005} for the structural quantile function at $\tau$.\footnote{The local restriction is a weaker condition than full independence between $Z$ and $U_{d^\ast}$ \citep{chesher:2003}. Technically speaking, $U_{d^\ast}$'s are not necessarily uniform random variables in this paper since the assumptions are only about the given value of $\tau$, but the subsequent discussions in terms of the quantile treatment effect hold only when $U_{d^\ast}$'s are uniform random variables.}

\begin{assumption}\label{CHassumption1}
The mapping $u\mapsto q(d^\ast,u)$ is strictly increasing and left-continuous for every $u\in[0,1]$, and has the inverse $y\mapsto q^{-1}(y,d^\ast)$.
\end{assumption}
\begin{assumption}\label{CHassumption2}
(i) $Pr(U_0\leq \tau\mid Z)=Pr(U_1\leq \tau\mid Z)=\tau$.
(ii) $Pr(U_0\leq\tau\mid D^\ast,Z)=Pr(U_1\leq\tau\mid D^\ast,Z)$.
\end{assumption}

Assumption \ref{CHassumption1} requires that the outcome variable $Y$ is continuously distributed. 
Assumption \ref{CHassumption2} (i) is the exogeneity of the instrumental  variable $Z$.
This assumption allows for the endogeneity of the treatment variable $D^\ast$.
Assumption \ref{CHassumption2} (ii) is the rank similarity condition on $U_{d^\ast}\leq\tau$. 
It is a relaxation of the rank invariance condition, $U_0=U_1$, that an individual's rank, $U_{d^\ast}$, is the same regardless of whether the individual is treated or controlled.  
The rank similarity condition allows the two unobserved heterogeneity terms, $U_1$ and $U_0$, to be different, though it still requires them to have the same distribution given the endogenous treatment assignment and the instrumental  variable. 
It enables the $\tau$-th quantile of the counterfactual outcome variable, $q(d^\ast,\tau)$, to be compared between the control group ($d^\ast=0$) and the treatment group ($d^\ast=1$).
The rank similarity condition is a restriction on the unobserved heterogeneity in the outcome variable equation and has been widely used for investigating heterogenous treatment effects \citep[e.g.,][]{doksum:1974,heckman/smith/clements:1997,chernozhukov/hansen:2004}.

Under the rank similarity condition, \cite{chernozhukov/hansen:2005} obtain the following relationship between the distribution of $(Y,D^\ast,Z)$ and the structural quantile function $q(d^\ast,\tau)$.
\begin{lemma}[{\citealp[][Theorem 1]{chernozhukov/hansen:2005}}]\label{Thm1CH}
Under Assumptions \ref{CHassumption1} and \ref{CHassumption2}, 
\begin{equation}
Pr(U\leq \tau\mid Z)=\tau
\mbox{ and  }
Pr(Y\leq q(D^\ast,\tau)\mid Z)=\tau.\label{originalTestableImpe}
\end{equation}
\end{lemma}

The rest of this paper adds the complication that the binary treatment variable $D^\ast$ may not be observed. Then the equality (\ref{originalTestableImpe}) cannot be directly used for identifying the structural quantile function.

\section{Quantile regression of $Y$ on $Z$}\label{sec:reducdee_fourm}
This section does not use any measurement of $D^\ast$; instead the relationship between the outcome variable $Y$ and the instrumental variable $Z$ is used to provide a lower bound on the structural quantile treatment effect $q(1,\tau)-q(0,\tau)$. 
Namely, $Q_{Y\mid Z=z_1}(\tau)-Q_{Y\mid Z=z_0}(\tau)$ can be a lower bound on $q(1,\tau)-q(0,\tau)$, when $Z$ is a binary variable taking  $z_0$ and $z_1$. 
This analysis provides a new structural interpretation to $Q_{Y\mid Z=z_1}(\tau)-Q_{Y\mid Z=z_0}(\tau)$, which is computed as the regression coefficient from the quantile regression of $Y$ on $Z$.  
When $Z$ takes more than two values, the discussion in this section can be applied by selecting any two values in the support of $Z$ or partitioning the support into two parts. It is worth clarifying that the results in this section and the following sections are valid regardless of whether $Z$ is binary, discrete, continuous, or mixed. 

The result in this section uses the stochastic monotonicity condition (\citealp{small/tan:2007} and \citealp{dinardo/lee:2011}). 
It assumes a positive relationship between the treatment variable $D^\ast$ and the instrumental  variable $Z$ in which, for every possible realization $u$ of $U$,  the probability of being treated $f_{U_1,D^\ast\mid Z=z}(u,1)$ is weakly increasing in $z$, and the probability of being untreated $f_{U_0,D^\ast\mid Z=z}(u,0)$ is weakly decreasing in $z$. 
The condition is weaker than the deterministic monotonicity condition \citep{imbens/angrist:1994,angrist/imbens/rubin:1996} because it allows for defiers, i.e., some individuals who change $D^\ast$ from $1$ to $0$ when $Z$ increases.
\begin{definition}\label{def:stochastic_monotonicty}
The stochastic monotonicity condition is that 
\begin{equation}
f_{U_0,D^\ast\mid Z=z_1}(u,0)\leq f_{U_0,D^\ast\mid Z=z_0}(u,0)\mbox{ and }f_{U_1,D^\ast\mid Z=z_1}(u,1)\geq f_{U_1,D^\ast\mid Z=z_0}(u,1)\label{stochMono2}
\end{equation}
for every $u\in[0,1]$.
\end{definition}

Theorem \ref{convcomb} shows that, under the stochastic monotonicity condition in (\ref{stochMono2}),  $Q_{Y\mid Z=z_1}(\tau)-Q_{Y\mid Z=z_0}(\tau)$ is biased towards zero compared to the structural quantile treatment effect $q(1,\tau)-q(0,\tau)$.  
\begin{theorem}\label{convcomb}
Suppose that Assumptions \ref{CHassumption1} and \ref{CHassumption2} hold and that the stochastic monotonicity condition holds.  
\begin{enumerate}
\item[(a)] There is some unknown constant $\kappa\in[0,1]$ such that 
\begin{equation}\label{attenuationEq}
Q_{Y\mid Z=z_1}(\tau)-Q_{Y\mid Z=z_0}(\tau)=\kappa\times (q(1,\tau)-q(0,\tau)).
\end{equation}
\item[(b)] If $f_{U_0,D^\ast\mid Z=z_1}(u,0)<f_{U_0,D^\ast\mid Z=z_0}(u,0)$ and $f_{U_1,D^\ast\mid Z=z_1}(u,1)>f_{U_1,D^\ast\mid Z=z_0}(u,1)$ in a neighborhood of $\tau$,
then $\kappa\ne 0$.
\end{enumerate}
\end{theorem}

This theorem provides a one-sided bound on $q(1,\tau)-q(0,\tau)$: 
$q(1,\tau)-q(0,\tau)\geq Q_{Y\mid Z=z_1}(\tau)-Q_{Y\mid Z=z_0}(\tau)$ if $Q_{Y\mid Z=z_1}(\tau)-Q_{Y\mid Z=z_0}(\tau)\geq 0$; and 
$q(1,\tau)-q(0,\tau)\leq Q_{Y\mid Z=z_1}(\tau)-Q_{Y\mid Z=z_0}(\tau)$ if $Q_{Y\mid Z=z_1}(\tau)-Q_{Y\mid Z=z_0}(\tau)\leq 0$. 
This bound gives researchers a justification for using $Q_{Y\mid Z=z_1}(\tau)-Q_{Y\mid Z=z_0}(\tau)$, which is a lower bound for the structural quantile treatment effect. 
Note that $Q_{Y\mid Z=z_1}(\tau)-Q_{Y\mid Z=z_0}(\tau)$ is a simple object to compute; it is obtained as the quantile regression coefficient on $Z$ and various statistical software packages include linear and nonlinear quantile regressions (e.g., the qreg command in Stata). 

The stochastic monotonicity condition cannot be removed from Theorem \ref{convcomb}, but Lemma \ref{convcomb_lemma} in the appendix shows that Eq. (\ref{attenuationEq}) holds with $\kappa\in[-1,1]$ even if the stochastic monotonicity condition does not hold. 
In other words, the inequality $|Q_{Y\mid Z=z_1}(\tau)-Q_{Y\mid Z=z_0}(\tau)|\leq |q(1,\tau)-q(0,\tau))|$ still holds without the stochastic monotonicity condition. 
It is possible to use this inequality to test the significance of $D^\ast$ by testing $Q_{Y\mid Z=z_1}(\tau)-Q_{Y\mid Z=z_0}(\tau)=0$.

\section{Identified set for the structural quantile function}
This section considers use of a potentially misclassified treatment variable $D$ and provides the sharp identified set for the structural quantile function $q(\cdot,\tau)$. 
To extract some information about the true treatment $D^\ast$ from its measurement $D$, the following restrictions on the misclassification probabilities are imposed.
\begin{assumption}\label{missclass}
(i) $(Pr(D=1\mid D^\ast=0,Y,Z),Pr(D=0\mid D^\ast=1,Y,Z))=(\pi_0,\pi_1)$ for some constants $(\pi_0,\pi_1)$. 
(ii) $Pr(D\ne D^\ast\mid D^\ast=0,Y,Z)+Pr(D\ne D^\ast\mid D^\ast=1,Y,Z)<1$.
\end{assumption}
Assumption \ref{missclass} (i) is that the measurement error does not depend on $(Y,Z)$. 
Assumption \ref{missclass} (ii) is that the measurement $D$ is positively correlated with the true treatment variable $D^\ast$.
These assumptions are widely used in the literature on misclassification (e.g., \citealp{hausman/abrevaya/schott-morton:1998}, \citealp{mahajan:2006}; \citealp{lewbel:2007}; and \citealp{hu:2008}).

The sharp identified set for $q(\cdot,\tau)$ is the set of values for $q(\cdot,\tau)$ that exhausts all the information from the model and the data distribution. 
Let $\mathcal{Q}$ be a subset of the set of functions of $\{0,1\}\times[0,1]$ to $\mathbb{R}$, and $\mathcal{P}^\ast$ be a subset of the set of probability distributions for $(D,Z,U_0,U_1,D^\ast)$.\footnote{
$\mathcal{Q}$ and $\mathcal{P}^\ast$ are subsets because there can be restrictions on $q$ and the distribution for $(D,Z,U_0,U_1,D^\ast)$. 
Also note that the distribution for $(D,Z,U_0,U_1,D^\ast)$ can be characterized by 
$$
Pr(D=d,Z\leq z,U_0\leq u_0,U_1\leq u_0,D^\ast=d^\ast)
$$
and the distribution for $(Y,D,Z)$ can be characterized by 
$$
Pr(Y\leq y,D=d,Z\leq z).
$$
The distribution $P$ for $(Y,D,Z)$ is induced by $(q,P^\ast)$ via  
\begin{eqnarray*}
P(Y\leq y,D=0,Z\leq z)&=&P^\ast(D=0,q(0,U_0)\leq y,D^\ast=0,Z\leq z)+P^\ast(D=0,q(1,U_1)\leq y,D^\ast=1,Z\leq z)\\
P(Y\leq y,D=1,Z\leq z)&=&P^\ast(D=1,q(0,U_0)\leq y,D^\ast=0,Z=z)+P^\ast(D=1,q(1,U_1)\leq y,D^\ast=1,Z\leq z).
\end{eqnarray*}
}
Given a distribution $P$ for $(Y,D,Z)$, the sharp identified set for $(q,P^\ast)$ is the set of elements $(\tilde{q},\tilde{P}^\ast)$ of $\mathcal{Q}\times\mathcal{P}^\ast$ such that $P$ is the distribution for $(\tilde{q}(D^\ast,U),D,Z)$ under $\tilde{P}^\ast$.\footnote{In this paper, $\tilde{q}$ is a generic element of $\mathcal{Q}$ and $\tilde{P}^\ast$ is a generic element of $\mathcal{P}^\ast$, whereas $q$ is the true structural quantile function and $P^\ast$ is the true distribution for $(D,Z,U_0,U_1,D^\ast)$.}
The sharp identified set for $q(\cdot,\tau)$ is defined as the projection of the sharp identified set for $(q,P^\ast)$ on the component $q(\cdot,\tau)$. 

The following theorem characterizes the sharp identified set for $q(\cdot,\tau)$ under Assumptions \ref{CHassumption1}, \ref{CHassumption2}, and \ref{missclass}, by using moment equalities and inequalities.  
\begin{theorem}\label{IS1}
Assume that all the elements in $\mathcal{Q}\times\mathcal{P}^\ast$ satisfy Assumptions \ref{CHassumption1}, \ref{CHassumption2}, and \ref{missclass}.
(a) Given a distribution $P$ for the observed variables, if $(y_0,y_1)$ belongs to the sharp identified set for $q(\cdot,\tau)$, then 
\begin{equation}
P(Y\leq y_D\mid Z)-\tau=p_1(P(Y\leq y_0\mid Z)-\tau)+p_0(P(Y\leq y_1\mid Z)-\tau)\label{ModifTestImp}
\end{equation}
for some $(p_0,p_1)$ with $p_0+p_1<1$
such that 
$$
0\leq p_0\leq P(D=1\mid Y,Z)\mbox{ a.s.}
\mbox{ and }
0\leq p_1\leq P(D=0\mid Y,Z)\mbox{ a.s.}
$$
(b) The converse is also true if $\mathcal{Q}\times\mathcal{P}^\ast$ includes all $(q,P^\ast)$'s satisfying Assumptions \ref{CHassumption1}, \ref{CHassumption2}, and \ref{missclass}.
\end{theorem}

The moment equality condition in Eq. (\ref{ModifTestImp}) is equivalent to the main testable implication in Chernozhukov and Hansen (2005) when $(p_0,p_1)=(\pi_0,\pi_1)$ where $(\pi_0,\pi_1)$ are the true unknown misclassification probabilities defined in Assumption \ref{missclass}.
The moment inequality conditions about $(\pi_0,\pi_1)$ are derived from the following calculations: 
\begin{eqnarray*}
&&
Pr(D=1\mid Y,Z)
=
\pi_0+(1-\pi_0-\pi_1)Pr(D^\ast=1\mid Y,Z)
\geq
\pi_0
\\
&&
Pr(D=0\mid Y,Z)
=
\pi_1+(1-\pi_0-\pi_1)Pr(D^\ast=0\mid Y,Z)
\geq
\pi_1.
\end{eqnarray*}

As a corollary to Theorem \ref{IS1}, it is possible to compare the identified set for $q(\cdot,\tau)$ and the estimand in \cite{chernozhukov/hansen:2005}, which does not consider measurement error in $D$.
\begin{corollary}\label{coro1}
Suppose all the assumptions in Theorem \ref{IS1} (b) hold. 
Every solution $(y_0,y_1)$ to $P(Y\leq y_D\mid Z)-\tau=0$, belongs to the sharp identified set for $q(\cdot,\tau)$. 
\end{corollary}

As another corollary to Theorem \ref{IS1}, it is possible to relate $Q_{Y\mid Z}(\tau)$ to the identified set for $q(\cdot,\tau)$. 
Although $Q_{Y\mid Z=z_1}(\tau)-Q_{Y\mid Z=z_0}(\tau)$ can be used as a lower bound for $q(1,\tau)-q(0,\tau)$, it does not always belong to the identified set. 
\begin{corollary}\label{coro2}
Consider two points, $z_0$ and $z_1$, in the support of $Z$, and suppose all the assumptions in Theorem \ref{IS1} (b) hold. 
Then $(y_0,y_1)=(Q_{Y\mid Z=z_0}(\tau),Q_{Y\mid Z=z_1}(\tau))$ belongs to the sharp identified set for $q(\cdot,\tau)$  if and only if 
$P(D=1\mid \{y_0<Y\leq y_1\mbox{ or }y_1<Y\leq y_0\},Z=z_0)\leq P(D=1\mid Y,Z)$ a.s. and $P(D=0\mid \{y_0<Y\leq y_1\mbox{ or }y_1<Y\leq y_0\},Z=z_1)\leq P(D=0\mid Y,Z)$ a.s. 
\end{corollary}

Note that, by the exogeneity of $Z$, $(q(0,\tau),q(1,\tau))=(Q_{Y\mid Z=z_0}(\tau),Q_{Y\mid Z=z_1}(\tau))$ if $D^\ast=1\{Z=z_1\}$. 
Corollary \ref{coro2} is roughly related to the observation that, under Assumption \ref{missclass}, $D^\ast=1\{Z=z_1\}$ implies 
\begin{eqnarray*}
P(D=1\mid \{y_0<Y\leq y_1\mbox{ or }y_1<Y\leq y_0\},Z=z_0) 
&=&
P(D=1\mid \{y_0<Y\leq y_1\mbox{ or }y_1<Y\leq y_0\},D^\ast=0) \\
&=&
\pi_0\\
&\leq&
\pi_0+(1-\pi_0-\pi_1)P(D^\ast=1\mid Y,Z)\\
&=&
P(D=1\mid Y,Z)
\\
P(D=0\mid \{y_0<Y\leq y_1\mbox{ or }y_1<Y\leq y_0\},Z=z_1)
&=&
P(D=0\mid \{y_0<Y\leq y_1\mbox{ or }y_1<Y\leq y_0\},D^\ast=1) \\
&=&
\pi_1\\
&\leq&
\pi_1+(1-\pi_0-\pi_1)P(D^\ast=0\mid Y,Z)\\
&=&
P(D=0\mid Y,Z).
\end{eqnarray*}
The precise derivations are found in the proof in the Appendix.

\subsection{Under-identification even with large variation in $Z$}\label{impossible}
This section shows that the structural quantile function is not point identified in general unless there is additional information on the model primitives $(q,P^\ast)$. 
The failure of point identification is due to the rank condition and happens regardless of the order condition based on Eq. (\ref{ModifTestImp}), where the number of the parameters is $4$, and the number of the equations is the number of the support points of $Z$. 
Theorem \ref{underident} presents this failure for a class of data generating processes. 
In particular, the theorem states that the quantile treatment effect can be under-identified if one cannot exclude that the treatment is exogenous. 
Based on the result, it is necessary to impose additional assumptions (other than the ones maintained in this paper) to achieve point identification when the treatment variable is potentially misclassified.

\begin{theorem}\label{underident}
Consider $\bar{d}^\ast=0$ or $\bar{d}^\ast=1$. 
Assume (i) the mapping $u\mapsto q^{-1}(q(1,u),0)$ is Lipschitz continuous, (ii) $q(d^\ast,\tau)\ne q(1-d^\ast,\tau)$, (iii) $(U_0,U_1)$ is independent of $(D^\ast,Z)$, (iv) $\pi_{\bar{d}^\ast}>0$, and (v) for sufficiently small $\varepsilon>0$, the following three statements holds:
\begin{enumerate} 
\item $\tilde{q}\in\mathcal{Q}$ for every strictly increasing bijection $t$ of $[0,1]$ to $[0,1]$ such that $|t(u)-u|\leq\varepsilon$ for every $u\in(0,1)$, where $\tilde{q}(d^\ast,\cdot)=q(d^\ast,t(\cdot))$ and  $\tilde{q}(1-d^\ast,\cdot)=q(1-d^\ast,\cdot)$.
\item $\tilde{P}^\ast\in\mathcal{P}^\ast$ for every distribution $\tilde{P}^\ast$ for $(D,Z,U_0,U_1,D^\ast)$ such that $\tilde{P}^\ast$ satisfies Assumption \ref{CHassumption2} and that 
$$
\tilde{P}^\ast(D=1-\bar{d}^\ast\mid U_0,U_1,D^\ast=\bar{d}^\ast,Z)=\pi_{\bar{d}^\ast}-\varepsilon
$$
$$
\tilde{P}^\ast(D=\bar{d}^\ast\mid U_0,U_1,D^\ast=1-\bar{d}^\ast,Z)=\pi_{1-\bar{d}^\ast}
$$
$$
|\tilde{P}^\ast(U_0\leq u_0,U_1\leq u_1,D^\ast=d^\ast,Z\leq z)
-
{P}^\ast(U_0\leq u_0,U_1\leq u_1,D^\ast=d^\ast,Z\leq z)|\leq\varepsilon.
$$
\end{enumerate}
Then the sharp identified set for  $q(\cdot,\tau)$ has more than one element. 
\end{theorem}

Condition (i) is a regularity condition. 
Condition (ii) is that the treatment variable can have a non-zero effect on the outcome variable at quantile index $\tau$.   
Condition (iii) is that the treatment variable can be exogenous.  
Condition (iv) is that there is a non-zero measurement error. 
Condition (v) is a condition about the size of the parameter space $\mathcal{Q}\times\mathcal{P}^\ast$. 
A sufficient condition for (v) is that $\mathcal{Q}\times\mathcal{P}^\ast$ includes all $(q,P^\ast)$'s satisfying Assumptions \ref{CHassumption1}, \ref{CHassumption2}, and \ref{missclass}.

Condition (iii) needs a careful discussion. 
Theorem \ref{underident} states that the quantile treatment effect is not always point identified unless the treatment is assumed to be endogenous.
This theorem is more relevant when one cannot exclude that $D^\ast$ is exogenous than when $D^\ast$ is known to be exogenous. 
When one cannot exclude that $D^\ast$ is exogenous, there is a possibility for the lack of point identification. 
It can be possible to point-identify the quantile treatment function if one can assume that $D^\ast$ is not exogenous.

\subsection{Point identification with second measurement}
Given the under-identification result in Theorem \ref{underident}, this section considers the case of two measurements for $D^\ast$ to achieve point identification of the structural quantile function.  
The identification strategy is based on existing results in the econometric literature. 
First, the results in \cite{mahajan:2006}, \cite{lewbel:2007}, and \cite{hu:2008} are applied to identify  $f_{Y,D^\ast\mid Z}$. 
Given identification of $f_{Y,D^\ast\mid Z}$, the identification result in \cite{chernozhukov/hansen:2005,chernozhukov/hansen:2013} recovers the structural quantile function.  

The following assumption and Lemma \ref{lemma_hu} are based on Theorem 1 in \cite{hu:2008}.
\begin{assumption}\label{point_identi_assn}
(i) The two measurements, $D$ and $V$, are conditionally independent given $D^\ast$. 
(ii) $0<f_{D^\ast\mid Z=z_1}(0)<1$. 
(iii) There are two points, $v_0$ and $v_1$, in the support of $V$ such that 
$$
\left(
\begin{array}{cc}
f_{V\mid D^\ast=0}(v_0)&f_{V\mid D^\ast=0}(v_1)\\
f_{V\mid D^\ast=1}(v_0)&f_{V\mid D^\ast=1}(v_1)\\
\end{array}
\right)
$$
is invertible. 
(iv) $P(D= D^\ast\mid D^\ast)>1/2$. 
(v) There are two points, $z_0$ and $z_1$, in the support of $Z$.
\end{assumption}

\begin{lemma}\label{lemma_hu}
Under Assumptions \ref{missclass} and \ref{point_identi_assn},
$f_{(Y,D^\ast)\mid Z}$ is point identified. 
\end{lemma}

\cite{chernozhukov/hansen:2013} provide a simple sufficient condition for the the global identification of the structural quantile function given $f_{(Y,D^\ast)\mid Z}$. The following assumption and identification result are borrowed from \citet[][Section 3.1]{chernozhukov/hansen:2013}.  
\begin{assumption}\label{CH_point_ident}
There is a cube  $\mathcal{L}$ with $(q(0,\tau),q(1,\tau))\in\mathcal{L}$ such that 
$$
\frac{f_{(Y,D^\ast)\mid Z=z_1}(y_1,1)}{f_{(Y,D^\ast)\mid Z=z_1}(y_0,0)}
>
\frac{f_{(Y,D^\ast)\mid Z=z_0}(y_1,1)}{f_{(Y,D^\ast)\mid Z=z_0}(y_0,0)}\mbox{, }
f_{(Y,D^\ast)\mid Z=z_1}(y_1,1)>0\mbox{ and }f_{(Y,D^\ast)\mid Z=z_0}(y_0,0)>0
$$
for all $(y_0,y_1)\in\mathcal{L}$.
\end{assumption}
\begin{lemma}\label{lemma_CH_31}
Under Assumptions  \ref{CHassumption1}, \ref{CHassumption2} and \ref{CH_point_ident},  
$(q(0,\tau),q(1,\tau))$ is uniquely determined from $f_{(Y,D^\ast)\mid Z}$.
\end{lemma}

By Lemmas \ref{lemma_hu} and \ref{lemma_CH_31},  the structural quantile function can be identified with two measurements for $D^\ast$. 
\begin{theorem}\label{theorem_hu_CH_31}
Under Assumptions  \ref{CHassumption1}, \ref{CHassumption2}, \ref{missclass}, \ref{point_identi_assn}, and \ref{CH_point_ident},  
$(q(0,\tau),q(1,\tau))$ is identified. 
\end{theorem}

\section{Inference procedure with covariates}\label{inference_section}
This section proposes an inference method for the structural quantile function. 
The method  extends the inference method in \cite{chernozhukov/hansen:2008} to incorporate misclassification probabilities. 
To include control variables $X$, a linear-in-parameters structure is imposed on the structural quantile function: 
\begin{equation}\label{linear_in_para_specification}
Y=q(D^\ast,X,U)\mbox{ with }
q(d^\ast,x,\tau)=\alpha_0d^\ast+x'\beta_0.
\end{equation}
This section focuses on constructing a confidence interval for $\alpha_0$. 

With control variables $X$, Assumptions \ref{CHassumption1}-\ref{missclass} are modified into the following assumptions: 
\begin{assumption}\label{CHassumption_covariate}
With probability one, the mapping $u\mapsto q(d^\ast,X,u)$ is strictly increasing and left-continuous for every $u\in[0,1]$.
\end{assumption}
\begin{assumption}
(i) $Pr(U_0\leq \tau\mid Z,X)=Pr(U_1\leq \tau\mid Z,X)=\tau$.
(ii) $Pr(U_0\leq\tau\mid D^\ast,Z,X)=Pr(U_1\leq\tau\mid D^\ast,Z,X)$.
\end{assumption}
\begin{assumption}\label{missclass_covariate}
(i) For each $d^\ast=0,1$, $Pr(D\ne D^\ast\mid D^\ast=d^\ast,Y,Z,X)$ is a constant, denoted by $\pi_{d^\ast}$. 
(ii) $Pr(D\ne D^\ast\mid D^\ast=0,Y,Z,X)+Pr(D\ne D^\ast\mid D^\ast=1,Y,Z,X)<1$.
\end{assumption}

Given $n$ i.i.d. copies $\{(Y_i,D_i,X_i,Z_i): i=1,\ldots,n\}$ of $(Y,D,X,Z)$, a confidence interval for $\alpha_0$ is constructed via the following two steps. The first step constructs a confidence interval for $(\pi_0,\pi_1)$. Given each point in the confidence interval for $(\pi_0,\pi_1)$, the second step constructs a confidence interval for $\alpha_0$. 
The size control comes from the Bonferroni correction for the first and second steps. 

The following condition is imposed on a $(1-\mathrm{size}_1)$ confidence region, $CI_1$, for $(\pi_0,\pi_1)$.  
\begin{assumption}
$\liminf_{n\rightarrow\infty}Pr((\pi_0,\pi_1)\in CI_1)\geq 1-\mathrm{size}_1$. 
\end{assumption}
In the empirical illustration, $CI_1$ is constructed by inverting the one-tailed $t$-tests based on $\pi_0\leq E[D\mid Z=0]$ and $\pi_1\leq E[1-D\mid Z=1]$, where $\mathrm{size}_1/2$ for each $t$-test. In the empirical illustration, the confidence interval for $\alpha_0$ is bounded by using $E_n[D\mid Z=0]=0$ and $E_n[1-D\mid Z=1]\approx 0.3$.

At the true value $(\alpha_0,\pi_0,\pi_1)$ of $(\alpha,p_0,p_1)$, the following testable implications \citep[cf.][]{chernozhukov/hansen:2008} hold. As in \cite{chernozhukov/hansen:2008}, it is possible to replace $Z$ with a function $g(X,Z)$ of $(X,Z)$.
\begin{lemma}\label{lemma:gamma_0_minimize}
Under Assumptions \ref{CHassumption_covariate}-\ref{missclass_covariate}, 
$$
0\in\argmin_{\gamma}\min_{\beta}Q_0(\theta;\alpha_0,\pi_0,\pi_1).
$$
where $\nu_0(y,x,z)=E[D\mid Y=y,X=x,Z=z]$, 
$t_+=1\{t\geq 0\}\cdot t$, 
$\rho_\tau(t)=(\tau-1\{t\leq 0\})t$,
$\theta=(\beta',\gamma')'$, 
$W=[X',Z']'$ and 
\begin{eqnarray*}
Q_0(\theta;\alpha,p_0,p_1)
&=&
E[\rho_\tau(Y-W'\theta)(1-p_1-D)]+E[\rho_\tau(Y-\alpha-W'\theta)(D-p_0)].
\\
&=&
E[\rho_\tau(Y-W'\theta)(1-p_1-\nu_0(Y,X,Z))_+]+E[\rho_\tau(Y-\alpha-W'\theta)(\nu_0(Y,X,Z)-p_0)_+].
\end{eqnarray*}
\end{lemma}

This paper assumes that $Q_0(\theta;\alpha_0,\pi_0,\pi_1)$ has a unique minimizer over $\theta$ for the true parameter value $(\alpha_0,\pi_0,\pi_1)$, which is implied by Assumption \ref{assn_inverty} (1). 
Since $(\alpha_0,\pi_0,\pi_1)$ is unknown, 
an estimator for $\theta$ is computed as a function of $(\alpha,p_0,p_1)$: 
$$
\hat\theta(\alpha;p_0,p_1)=\argmin_{\theta\in\Theta}Q_n(\theta;\alpha,p_0,p_1),
$$
where $\hat{\nu}(y,x,z)$ is a estimator for $\nu_0(y,x,z)$, and 
$$
Q_n(\theta;\alpha,p_0,p_1)=E_n[\rho_\tau(Y-W'\theta)(1-p_1-\hat{\nu}(Y,X,Z))_+]+E_n[\rho_\tau(Y-\alpha-W'\theta)(\hat{\nu}(Y,X,Z)-p_0)_+].
$$
The objective function $Q_n(\theta;\alpha,p_0,p_1)$ is convex in $\theta$, which is the result of using $\hat{\nu}(Y,X,Z)$ instead of $D$. This transformation comes from \cite{buchinsky/hahn:1998} and \cite{abadie/angrist/imbens:2002}.\footnote{I am thankful to a referee for proposing this transformation.} 
To simplify the arguments, a parametric model $\nu_0(y,x,z)=\nu_{\delta_0}(y,x,z)$ is imposed with a parametric estimator $\hat\delta$ for $\delta_0$ and the following assumptions.
\begin{assumption}\label{assn:condition_nu}
(i) $E[D^\ast\mid Y,X,Z]$ is bounded away from zero and one. 
(ii) $\sup_{(y,x,z)}|\hat{\nu}(y,x,z)-\nu_0(y,x,z)|=o_p(1)$ where $\hat{\nu}(y,x,z)=\nu_{\hat\delta}(y,x,z)$. 
(iii) $E_n\left[W(\xi_0(\hat{\nu})-\xi_0(\nu_0)-\Xi_0(\delta_0)(\hat\delta-\delta_0)\right]=o_p(n^{-1/2})$, where 
\begin{eqnarray*}
\xi_0(\nu)
&=&
(\tau-1\{Y-W'\theta_0\leq 0\})(1-\pi_1-\nu(Y,X,Z))+(\tau-1\{Y-\alpha_0-W'\theta_0\leq 0\})(\nu(Y,X,Z)-\pi_0)
\\
\Xi_0(\delta)
&=&
(1\{Y-W'\theta_0\leq 0\})-1\{Y-\alpha_0-W'\theta_0\leq 0\})
\frac{\partial}{\delta'}\nu_{\delta}(Y,X,Z).
\end{eqnarray*}
(iv) There are $n$ random variables, $\psi_{\delta,1},\ldots,\psi_{\delta,n}$, such that $\hat\delta-\delta_0=E_n[\psi_{\delta}]+o_p(n^{-1/2})$ with $E[\|\psi_{\delta}\|^2]<\infty$.
(v) There is an estimator, $(\widehat{\psi_{\delta,1}},\ldots,\widehat{\psi_{\delta,n}})$, for $(\psi_{\delta,1},\ldots,\psi_{\delta,n})$, that satisfies $E[\|\widehat{\psi_{\delta}}-\psi_{\delta}\|^2]=o(1)$.
(vi) There is an estimator $\widehat{\frac{\partial}{\delta'}\nu_{\delta}}$ for ${\frac{\partial}{\delta'}\nu_{\delta}}$ that satisfies $E\left[\left\|W\left(\widehat{\frac{\partial}{\delta'}\nu_{\delta}}(Y,X,Z)-{\frac{\partial}{\delta'}\nu_{\delta}}(Y,X,Z)\right)\right\|\right]=o(1)$.
(vii) $E\left[\left\|W\frac{\partial}{\delta'}\nu_{\delta}(Y,X,Z)\right\|^2\right]<\infty$.
\end{assumption}  

The optimization of $Q_n(\theta;\alpha,p_0,p_1)$ is implemented in the same way as the linear quantile regression. 
Namely, the objective function can be written as 
$Q_n(\theta;\alpha,p_0,p_1)
=n^{-1}\sum_{i=1}^{2n}\rho_\tau\left(\check{Y}_i-\check{W}_i'\theta\right)$, where 
$$
\check{Y}_i
=
\begin{cases}
Y_i(1-p_1-\hat{\nu}(Y_i,X_i,Z_i))_+&\mbox{ if }i\leq n\\
Y_{i-n}(\hat{\nu}(Y_{i-n},X_{i-n},Z_{i-n})-p_0)_+&\mbox{ if }i\geq n+1
\end{cases}
$$
$$
\check{W}_i
=
\begin{cases}
W_i(1-p_1-\hat{\nu}(Y_i,X_i,Z_i))_+&\mbox{ if }i\leq n\\
W_{i-n}(\hat{\nu}(Y_{i-n},X_{i-n},Z_{i-n})-p_0)_+&\mbox{ if }i\geq n+1.
\end{cases}
$$
for $i=1,\ldots,n,n+1,\ldots,2n$.\footnote{Note that 
$\rho_\tau(t)c=(\tau-1\{t\leq 0\})tc=(\tau-1\{ct\leq 0\})ct=\rho_\tau(ct)$ for every $t\in\mathbb{R}$ and $c\geq 0$. 
Since the weights, $(1-p_1-\hat{\nu}(Y,X,Z))_+$ and $(\hat{\nu}(Y,X,Z)-p_0)_+$, are non-negative, 
\begin{eqnarray*}
\rho_\tau(Y-W'\theta)(1-p_1-\hat{\nu}(Y,X,Z))_+
&=&
\rho_\tau(Y(1-p_1-\hat{\nu}(Y,X,Z))_+-(W(1-p_1-\hat{\nu}(Y,X,Z))_+)'\theta)
\\
\rho_\tau(Y-\alpha-W'\theta)(\hat{\nu}(Y,X,Z)-p_0)_+
&=&
\rho_\tau((Y-\alpha)(\hat{\nu}(Y,X,Z)-p_0)_+-(W(\hat{\nu}(Y,X,Z)-p_0)_+)'\theta).
\end{eqnarray*}
Therefore, 
$$
Q_n(\theta;\alpha,p_0,p_1)
=
E_n[\rho_\tau(Y-W'\theta)(1-p_1-\hat{\nu}(Y,X,Z))_+]+E_n[\rho_\tau(Y-\alpha-W'\theta)(\hat{\nu}(Y,X,Z)-p_0)_+]
=
\frac{1}{n}\sum_{i=1}^{2n}\rho_\tau\left(\check{Y}_i-\check{W}_i'\theta\right).
$$
}

The asymptotic variance for $\hat\theta(\alpha;p_0,p_1)$ is estimated with a kernel function $K(\cdot)$ and a bandwidth $h$.
For every value of $(\alpha,p_0,p_1)$, the asymptotic variance for $\hat\theta(\alpha;p_0,p_1)$ estimated by 
$$
\hat\Omega(\alpha;p_0,p_1)=
E_n[\hat\lambda(\alpha;p_0,p_1)WW']^{-1}
E_n[\hat{s}(\alpha;p_0,p_1)\hat{s}(\alpha;p_0,p_1)']
E_n[\hat\lambda(\alpha;p_0,p_1)WW']^{-1},
$$
where 
\begin{eqnarray*}
\widehat{\xi}(\alpha;p_0,p_1)
&=&
(\tau-1\{Y-W'\hat\theta(\alpha;p_0,p_1)\leq 0\})(1-p_1-\hat\nu(Y,X,Z))\\&&+(\tau-1\{Y-\alpha-W'\hat\theta(\alpha;p_0,p_1)\leq 0\})(\hat\nu(Y,X,Z)-p_0)
\\
\widehat{\Xi}(\alpha;p_0,p_1)
&=&
1\{Y-W'\hat\theta(\alpha;p_0,p_1)\leq 0\}\widehat{\frac{\partial}{\delta'}\nu_{\delta}}(Y,X,Z)
-1\{Y-\alpha-W'\hat\theta(\alpha;p_0,p_1)\leq 0\}
\widehat{\frac{\partial}{\delta'}\nu_{\delta}}(Y,X,Z)
\\
\hat{s}(\alpha;p_0,p_1)
&=&
\widehat{\xi}(\alpha;p_0,p_1)
W
+E_n\left[W\widehat{\Xi}(\alpha;p_0,p_1)\right]\widehat{\psi_{\delta}}
\\
\hat{\lambda}(\alpha;p_0,p_1)
&=&
K_h(Y-W'\hat\theta(\alpha;p_0,p_1))(1-p_1-D)
+
K_h(Y-\alpha-W'\hat\theta(\alpha;p_0,p_1))(D-p_0).
\end{eqnarray*}
Denote by $\hat\Omega_\gamma(\alpha;p_0,p_1)$ the asymptotic variance for $\hat\gamma(\alpha;p_0,p_1)$.

The proposed confidence interval for $\alpha_0$ is 
$$
CI_{\alpha}(\mathrm{size}_1+\mathrm{size}_2)=\bigcup_{(p_0,p_1)\in CI_1}\left\{\alpha\in\mathcal{A}: T(\alpha;p_0,p_1)\leq cv\right\},
$$
where $\mathcal{A}$ is the parameter space for $\alpha$, the test statistic is 
$$
T(\alpha;p_0,p_1)=n \hat\gamma(\alpha;p_0,p_1)'\hat\Omega_\gamma(\alpha;p_0,p_1)^{-1}\hat\gamma(\alpha;p_0,p_1),
$$
and the critical value $cv$ is the $(1-\mathrm{size}_2)$ quantile of the $\chi^2$ distribution with $\dim(\gamma)$ degrees of freedom. 
The proposed confidence interval satisfies the asymptotic size control under the following assumptions.
\begin{assumption}\label{para_assn_inference}
$\theta_0\equiv(\beta_0',0')'$ is in the interior of a compact parameter space $\Theta$.
\end{assumption}
\begin{assumption}\label{assn_inverty}
(i) $E\left[f_{Y-\alpha_0D^\ast\mid Z,X}(W'\theta_0)WW'\right]$ is invertible. 
(ii) $E[\|W\|^4]$ is finite.
(iii) 
$\lim_{\epsilon\rightarrow 0}Pr(|Y-W'\theta_0|\leq \epsilon\cdot\|W\|)=0$ and 
$\lim_{\epsilon\rightarrow 0}Pr(|Y-\alpha_0-W'\theta_0|\leq \epsilon\cdot\|W\|)=0$. 
(iv) There is a constant $C$ such that $\max\{f_{Y\mid X,Z,D},f_{Y\mid X,Z,D}^{(1)}\}<C$ a.s.
\end{assumption}
\begin{assumption}\label{kernel_assn}
(i) $h\rightarrow 0$ and $\sqrt{n}h\rightarrow\infty$ as $n\rightarrow\infty$. 
(ii) $K$ is differentiable with $\sup_{v}|K^{(1)}(v)|<\infty$, $\int K(v)dv=1$, $\int |K(v)v|dv<\infty$, and $\int K(v)^2dv<\infty$.
\end{assumption}
\begin{theorem}\label{cov_theorem}
Under Assumptions \ref{CHassumption_covariate}-\ref{kernel_assn}, 
$\liminf_{n\rightarrow\infty}Pr(\alpha_0\in CI_{\alpha}(\mathrm{size}_1+\mathrm{size}_2))\geq 1-(\mathrm{size}_1+\mathrm{size}_2)$.
\end{theorem}
Assumption \ref{para_assn_inference} is a regularity condition on the parameter. 
Assumption \ref{assn_inverty} (i) is that the Hessian matrix is non-singular, and it implies point identification of $\theta_0$ given the true value $(\alpha_0,\pi_0,\pi_1)$.
Assumption \ref{assn_inverty} (ii)-(iv) is a regularity condition on the distribution of the observables.  
Assumption \ref{kernel_assn} is a restriction on the bandwidth and the kernel function, which is used to estimate the asymptotic variance $\hat\Omega_\gamma(\alpha;p_0,p_1)$.

\section{Empirical illustration and Monte Carlo simulations}
This section investigates the finite sample performance of the proposed method using an existing empirical application and simulated datasets.  
As emphasized in Section \ref{impossible}, the inference results presented in this section are valid regardless of whether the structural quantile function is point or partially identified. 

\subsection{Empirical illustration}\label{sec:empirical_section}
This empirical illustration studies the quantile treatment effects of the 401(k) participation on financial savings (\citealp{chernozhukov/hansen:2004}) and consider the problem of mis-measured 401(k) participation.\footnote{\cite{ura:2015} uses the same empirical setting to investigate the local average treatment effect under treatment misclassification.} 
It uses the same dataset as \cite{chernozhukov/hansen:2004}, which is an extract from the Survey of Income and Program Participation (SIPP) of 1991. 
The sample consists of households in which at least one person is in employment, and which has no income from self-employment.
The resulting sample size is 9,915.

The model is based on \cite{chernozhukov/hansen:2004}.  
The outcome variable $Y$ is the net amount of financial assets in dollars, and the measured treatment variable $D$ is self-reported participation in a 401(k) program.
Participation in a 401(k) program may be endogenous because participants may be concerned more about retirement plans than non-participants. 
To control for endogeneity, an instrumental  variable $Z$ is 401(k) eligibility, an indicator variable of whether an employer offers a 401(k) program, where $z_1$ means eligibility and $z_0$ means ineligibility. 
The control variables include a constant, family income, age, age squared, marital status, and family size. 
The summary statistics for $(Y,D,Z)$ are in Table \ref{table_summary}.

\begin{table}[t]
\centering
\begin{tabular}{l|crr}
\hline 
\hline 
&sample size&mean&std. dev.\\
\hline
$Y$: family net financial assets (in \$1000) & 9,915 & 18.05 &  63.52\\
$D$: 401(k) participation &  9,915 &   0.26  &  0.44\\
$Z$: 401(k) eligibility & 9,915 &    0.37  &  0.48\\
    \hline
\end{tabular}
\caption{Summary statistics for $Y$, $D$ and $Z$}
\label{table_summary}
\end{table}

\begin{figure}[t]
\parbox{.4\textwidth}{
\centering
\includegraphics[height=.4\textwidth,keepaspectratio]{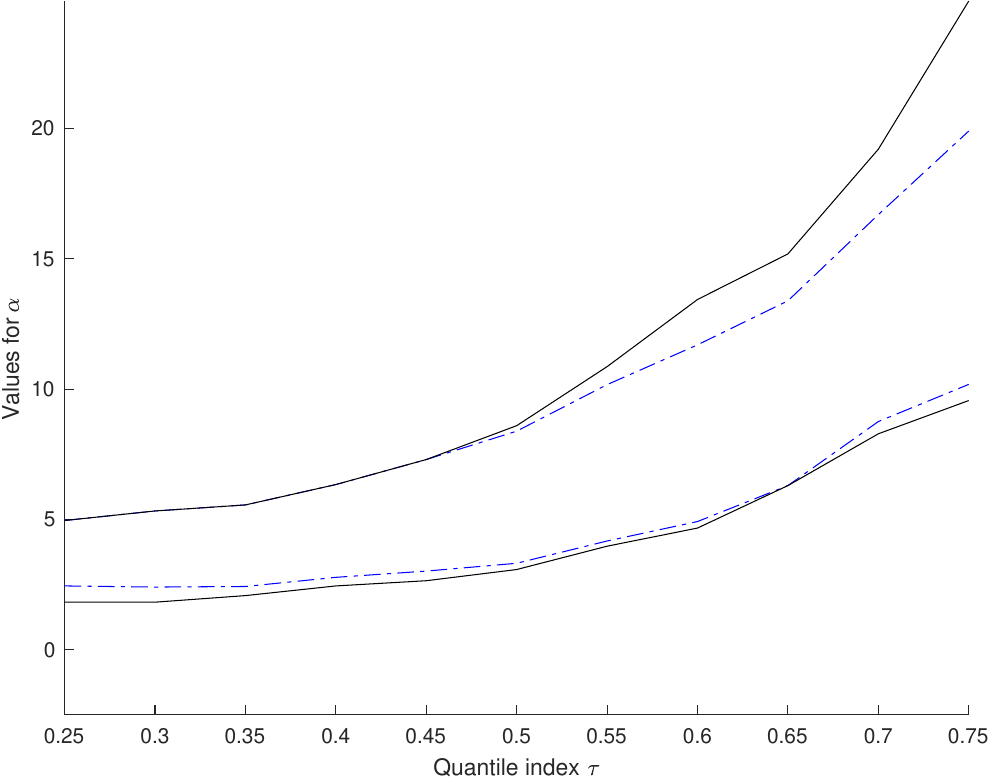}
}
\hspace{.1\textwidth}
\parbox{.4\textwidth}{
\centering
\includegraphics[height=.4\textwidth,keepaspectratio]{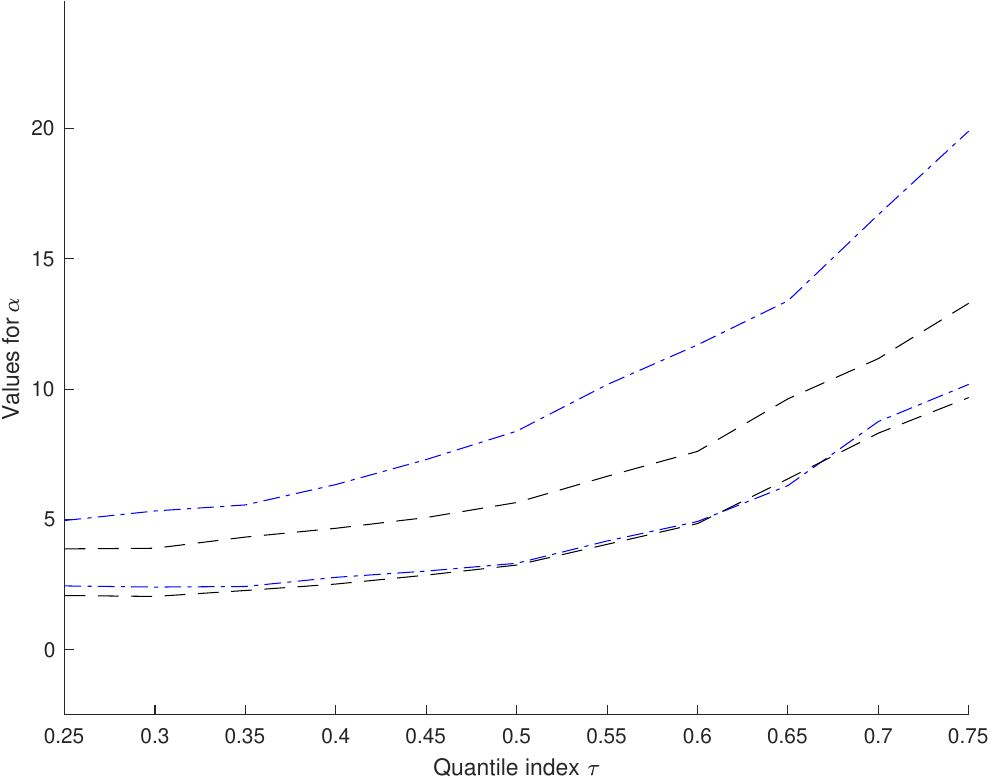}
}
\caption{Three point-wise $95\%$ confidence intervals for $\alpha_0$. The solid curves represent the proposed inference method,  the dash-dotted line ($-.$) curves represent the inference method that assumes $\pi_0=\pi_1=0$, and the dashed ($--$) curves represent the inference method that assumes $(\pi_0,\pi_1)=(0,0.31)$.}
\label{figEMP}
\end{figure}

The details for the the confidence intervals are as follows. The pre-specified sizes are $(\mathrm{size}_1,\mathrm{size}_2)=(1\%,4\%)$, where $\mathrm{size}_1$ is used to construct a confidence interval for unknown misclassification probabilities $(\pi_0,\pi_1)$ and $\mathrm{size}_2$ is used for the critical value $cv$. The confidence interval for $(\pi_0,\pi_1)$ is $CI_1=\{0\}\times [0, 0.31]$ for $(\pi_0,\pi_1)$, where $\{0\}$ follows from $E_n[D\mid Z=0]=0$ and $[0, 0.31]$ comes from the one-tailed $t$-test for $\pi_1\leq E[1-D\mid Z=1]$ with size $0.5\%$. 
The conditional probability of $D$ given $(Y,X,Z)$ is estimated by probit regression of $D$ on all the interactions of $(1,X,Z)$ and the cubic polynomials of $Y$.\footnote{The interaction terms can be written as $(1,X',Z',Y,Y\cdot X',Y\cdot Z',Y^2,Y^2\cdot X',Y^2\cdot Z',Y^3,Y^3\cdot X',Y^3\cdot Z')'$.}
Using $p_0=0$ and grid points $p_1=0,0.01,0.02,\ldots,0.31$ for $CI_1$, $CI_{\alpha}(\mathrm{size}_1+\mathrm{size}_2)$ is constructed by $\bigcup_{p_1=0,0.01,0.02,\ldots,0.31}\left\{\alpha\in\mathcal{A}: T(\alpha;p_0,p_1)\leq cv\right\}$.

Figure \ref{figEMP} (left) shows the $95\%$ confidence intervals based on Section \ref{inference_section}, with the specification in Eq. (\ref{linear_in_para_specification}) and the same list of covariates as \cite{Benjamin:2003} and \cite{chernozhukov/hansen:2004}.\footnote{$X$ includes a constant, age, age$^2$, income categories, family size, education dummies, marital status dummy, two-earner status dummy, defined benefit pension status  dummy, IRA participation status  dummy, and homeownership status  dummy. The number of the covariates in $X$ is 18.} 
The figure also shows the $95\%$ confidence intervals that assume $\pi_0=\pi_1=0$, i.e., no misclassification. 
The confidence intervals that assume $\pi_0=\pi_1=0$ are exactly the same as the confidence intervals proposed in \citet[Section 3]{chernozhukov/hansen:2008}. 
All the confidence intervals are point-wise, that is, computed separately for each quantile index $\tau$.  

In this empirical exercise, the confidence intervals proposed in this paper are comparable in lengths to those that assume no misclassification. 
There are several features of this empirical exercise that make the confidence intervals tight (compared to those that assume no misclassification).  
First, in this empirical exercise, $E_n[D\mid Z=z_0]=0$ implies $\pi_0=0$, so under-identification for $q(\cdot,\tau)$ mainly comes from under-identification for the scalar parameter $\pi_1$.  It makes the degree of under-identification  smaller than when $\pi_0$ and $\pi_1$ are both under-identified. 
Second, the confidence intervals that assume $\pi_0=\pi_1=0$, are large enough to include the confidence interval with another value of $(p_0,p_1)$. 
Figure \ref{figEMP} (right) shows the confidence intervals that assume $(p_0,p_1)=(0,0)$ and that with $(p_0,p_1)=(0,0.31)$, where the points $(0,0)$ and $(0,0.31)$ are the two endpoints for $CI_1$.   
The confidence intervals that assume  $(p_0,p_1)=(0,0)$ include those that assume $(p_0,p_1)=(0,0.31)$ for almost all the values of $\tau$.

\subsection{Monte Carlo simulations}\label{sec:monte_calro}
Monte Carlo simulations are based on the following data generating process.
The instrumental  variable $Z$ takes $z_0$ with probability $0.5$ and $z_1$ with probability $0.5$. 
The error term $(\tilde{U},V)$ is a two-dimensional mean-zero normal random vector with variance $1$ and correlation coefficient $0.5$, independent of $Z$. 
The covariate $X$ is $X=(1,\tilde{X})$, where $\tilde{X}$ is a four-dimensional standard normal random vector with identity covariance matrix, independent of $(Z,\tilde{U},V)$. 
The latent treatment variable $D^\ast$ is determined by 
$$
D^\ast=1\{1\{Z=z_1\}+X'\beta+V>0\},
$$
and the outcome variable $Y$ is determined by 
$$
Y=\exp(\Phi(\tilde{U})-0.5)D^\ast+X'\beta+\tilde{U},
$$
where $\Phi$ is the standard normal cumulative distribution function and  $\beta=(-(2\times 1)^{-1},-(2\times 2)^{-1},\ldots,-(2\times 5)^{-1})'$. 
The binary measurement $D$ is determined by $Pr(D\ne D^\ast\mid D^\ast)=\pi_{D^\ast}$, where $(\pi_0,\pi_1)=(0,0),(0.1,0),(0,0.1),(0.2,0),(0.1,0.1),(0,0.2),(0.2,0.1),(0.1,0.2)$.
The true values for $\alpha_0$ is $\exp(\tau-0.5)$.

In this simulation exercise coverage frequencies are computed for two inference methods over $[\alpha_0-1,\alpha_0+1]$. One is the proposed inference method and the other is  the inference method that assumes no misclassification, that is,  $\pi_0=\pi_1=0$. 
All the results are based on $n=1,000$, $(\mathrm{size}_1,\mathrm{size}_2)=(1\%,4\%)$, and 5,000 simulations.\footnote{ 
The details about the confidence interval is as follows. $CI_1=[0,\bar{c}_0]\times [0,\bar{c}_1]$ where the value of $\bar{c}_0$ comes from the one-tailed $t$-test for $\pi_0\leq E[D\mid Z=0]$ with size $0.5\%$ and the value of $\bar{c}_1$ comes from the one-tailed $t$-test for $\pi_1\leq E[1-D\mid Z=1]$ with size $0.5\%$.
As in the empirical exercise, $\nu_0(y,x,z)$ is estimated by the probit regression of $D$ on all the interactions of $(1,X,Z)$ and the cubic polynomials of $Y$.
Using 1\% grid points $(p_0,p_1)\in\{0,0.01,0.02,\ldots,\bar{c}_0\}\times\{0,0.01,0.02,\ldots,\bar{c}_1\}$, $CI_{\alpha}(\mathrm{size}_1+\mathrm{size}_2)$ is $\bigcup_{p_1=0,0.01,0.02,\ldots,\bar{c}_0}\bigcup_{p_1=0,0.01,0.02,\ldots,\bar{c}_1}\left\{\alpha\in\mathcal{A}: T(\alpha;p_0,p_1)\leq cv\right\}$.}

\begin{figure}[ht]
\parbox{.4\textwidth}{
\includegraphics[height=.4\textwidth,keepaspectratio]{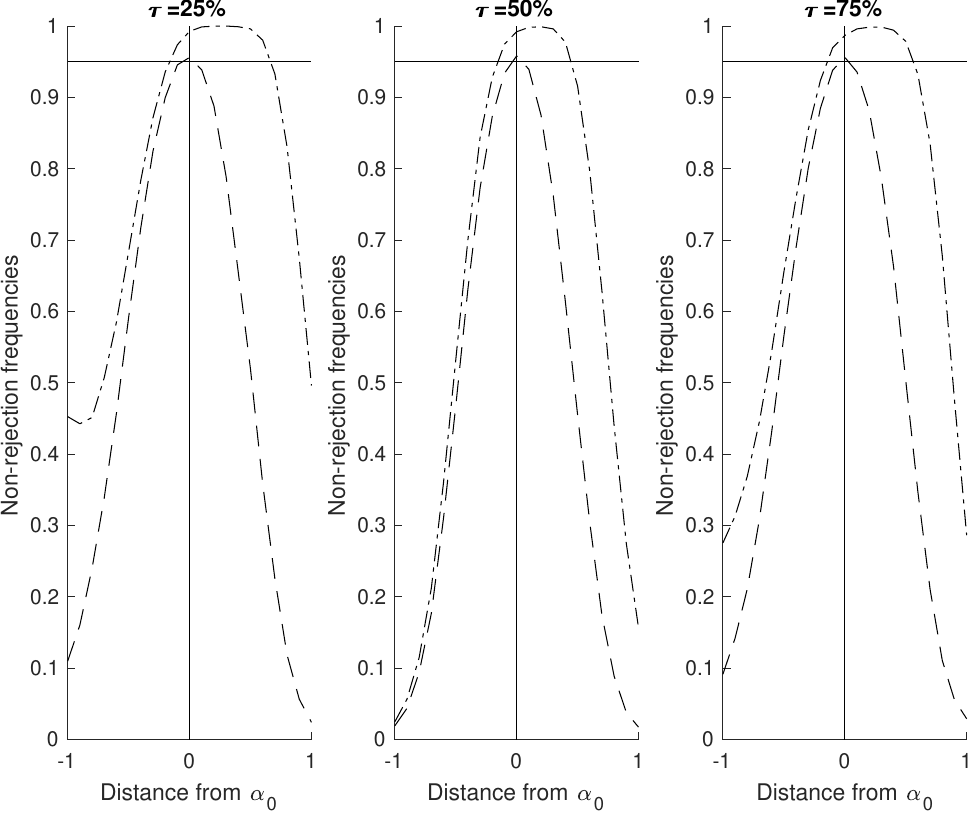}
\caption*{Figure \ref{fig_MC_diff1}.a: $\pi_0=\pi_1=0$.}
}
\hspace{.1\textwidth}
\parbox{.4\textwidth}{
\includegraphics[height=.4\textwidth,keepaspectratio]{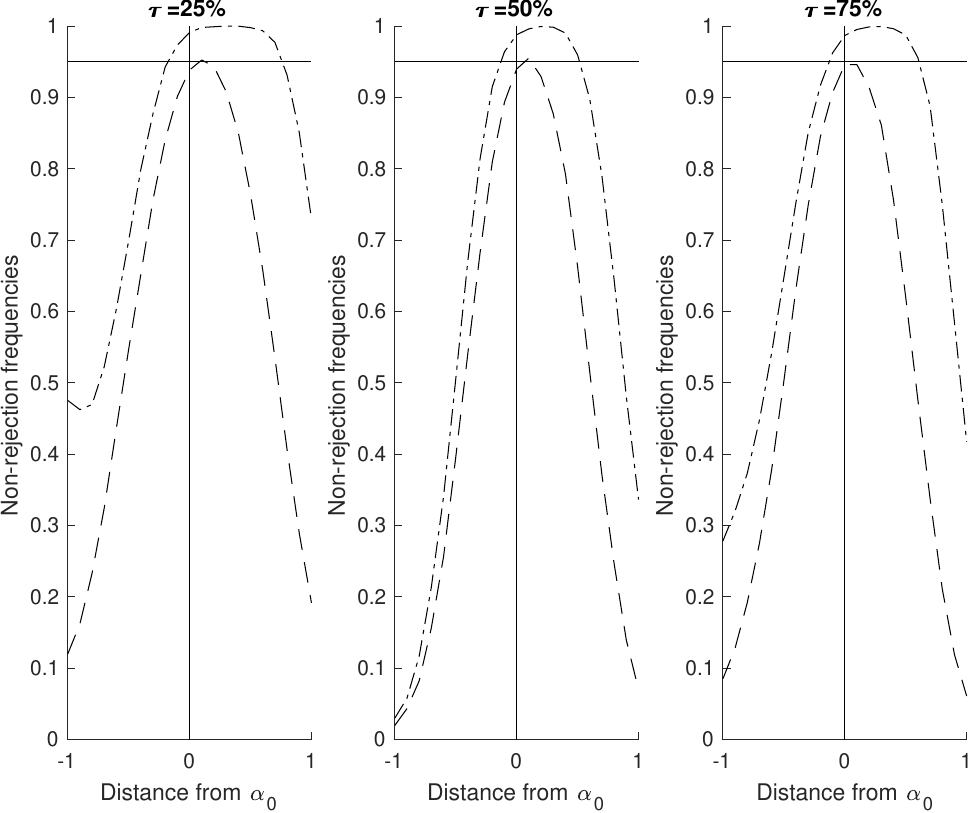}
\caption*{Figure \ref{fig_MC_diff1}.b: $\pi_0=0.1$ and $\pi_1=0$.}
}
\\
\bigskip
\bigskip
\\
\parbox{.4\textwidth}{
\includegraphics[height=.4\textwidth,keepaspectratio]{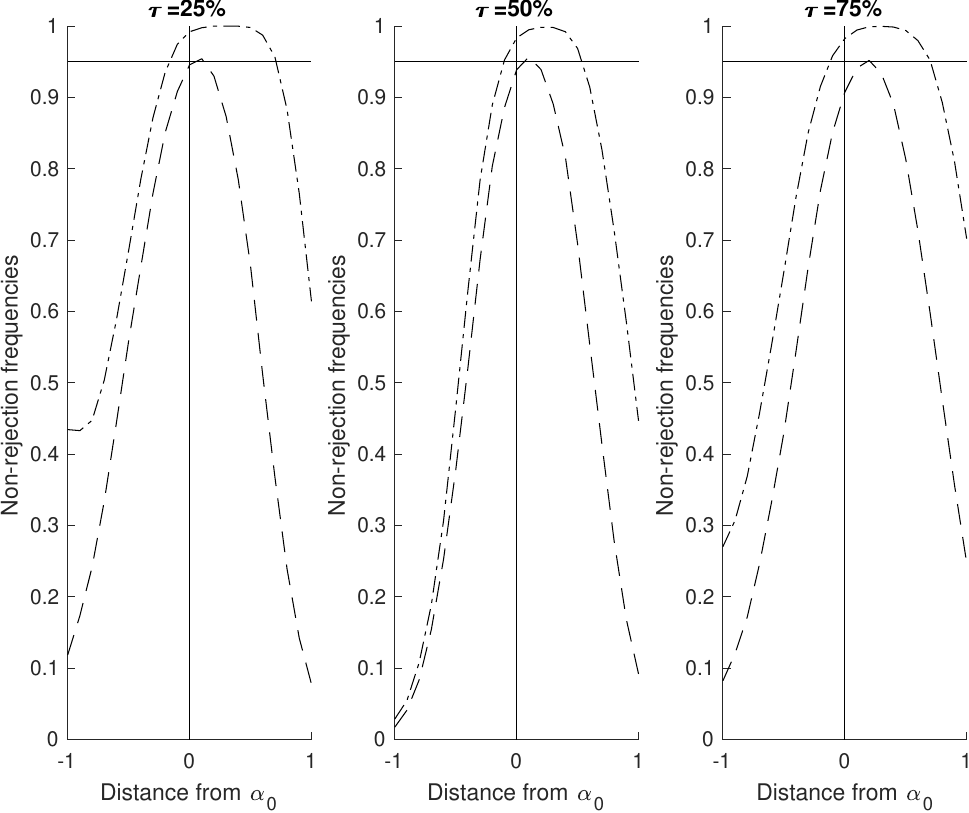}
\caption*{Figure \ref{fig_MC_diff1}.c: $\pi_0=0$ and $\pi_1=0.1$.}
}
\hspace{.1\textwidth}
\parbox{.4\textwidth}{
\includegraphics[height=.4\textwidth,keepaspectratio]{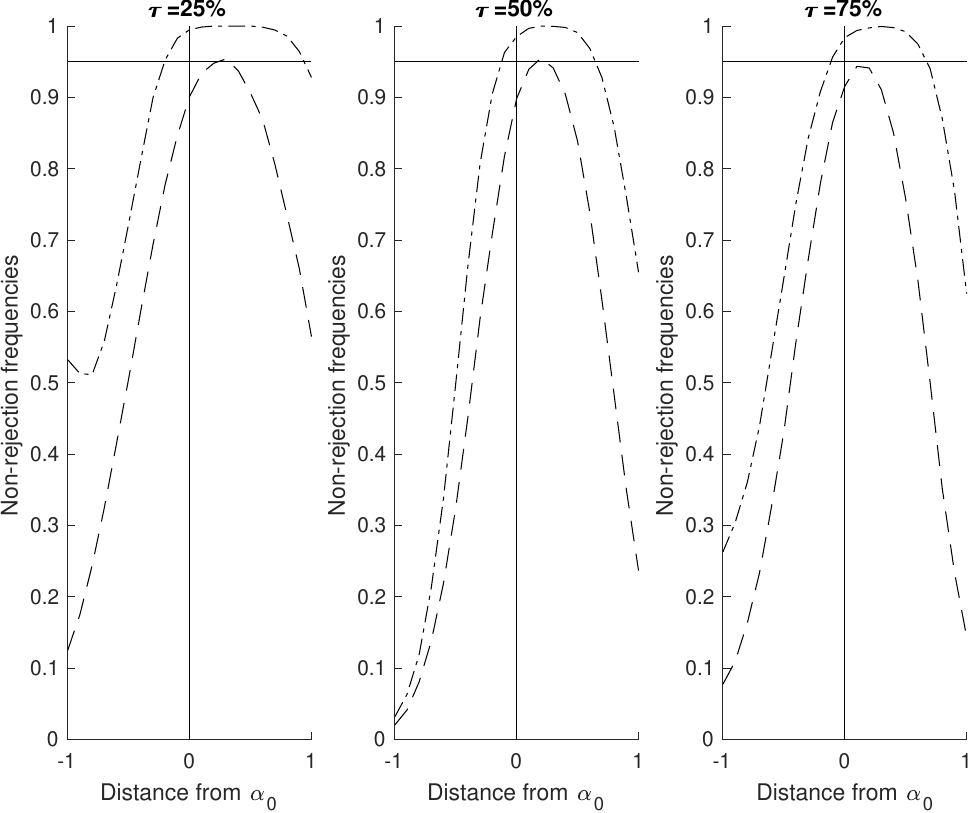}
\caption*{Figure \ref{fig_MC_diff1}.d: $\pi_0=0.2$ and $\pi_1=0$.}
}
\\
\bigskip
\bigskip
\\
\caption{Coverage frequencies. The dash-dot ($-.$) curve represents the proposed inference method, and the dashed ($--$) curve represents the inference method that assumes $\pi_0=\pi_1=0$.}
\label{fig_MC_diff1}
\end{figure}

Figure \ref{fig_MC_diff1}.a summarizes the simulation results when there is no misclassification, that is, $(\pi_0,\pi_1)=(0,0)$.
In this case, both the proposed method and the method that assumes $\pi_0=\pi_1=0$ have correct size, i.e., the coverage frequency at the true value of $\alpha$ is at least 95\%.  
The proposed inference method is less powerful than that with $\pi_0=\pi_1=0$, but this is the cost for achieving robustness to misclassification. 

Figures \ref{fig_MC_diff1}.b-\ref{fig_MC_diff1}.h summarize the simulation results when there is some misclassification. 
The method that assumes $\pi_0=\pi_1=0$ does not have correct size as $(\pi_0,\pi_1)$ becomes far from $(0,0)$, but the proposed method always has correct size.   
This is consistent with Theorem \ref{cov_theorem}, which shows that the proposed method has correct size even in the presence of misclassification. 

To summarize these simulation results, the proposed inference method covers the true parameter value at least with the pre-specified significance level in finite samples.  
A practitioner could obtain a narrower confidence interval by assuming no misclassification, but the confidence interval may not cover the true parameter with correct size when there is non-negligible misclassification. 

To investigate Monte Carlo simulations, Online Appendix provides a comparison between the proposed inference method and the infeasible method with knowing $(p_0,p_1)=(\pi_0,\pi_1)$.
The details are provided online in supplementary material associated with this article, available at Cambridge Journals Online (journals.cambridge.org/ect).

\section{Conclusion}
This paper extends the instrumental  variable quantile regression model \citep{chernozhukov/hansen:2005,chernozhukov/hansen:2013} for a binary regressor, to situations when this binary regressor is potentially misclassified. 
The first identification result is that under the rank similarity condition and the stochastic monotonicity condition, the reduced-form question effect, $Q_{Y\mid Z=z_1}(\tau)-Q_{Y\mid Z=z_0}(\tau)$, is biased towards zero compared to the structural quantile treatment effect $q(1,\tau)-q(0,\tau)$. 
The second identification result characterizes the sharp identified set for $q(d^\ast,\tau)$ under widely-used assumptions. 
An inference method for the structural quantile function is provided, and its finite sample performance is demonstrated in simulation studies and an empirical illustration.

\begin{figure}[ht]
\parbox{.4\textwidth}{
\centering
\includegraphics[height=.4\textwidth,keepaspectratio]{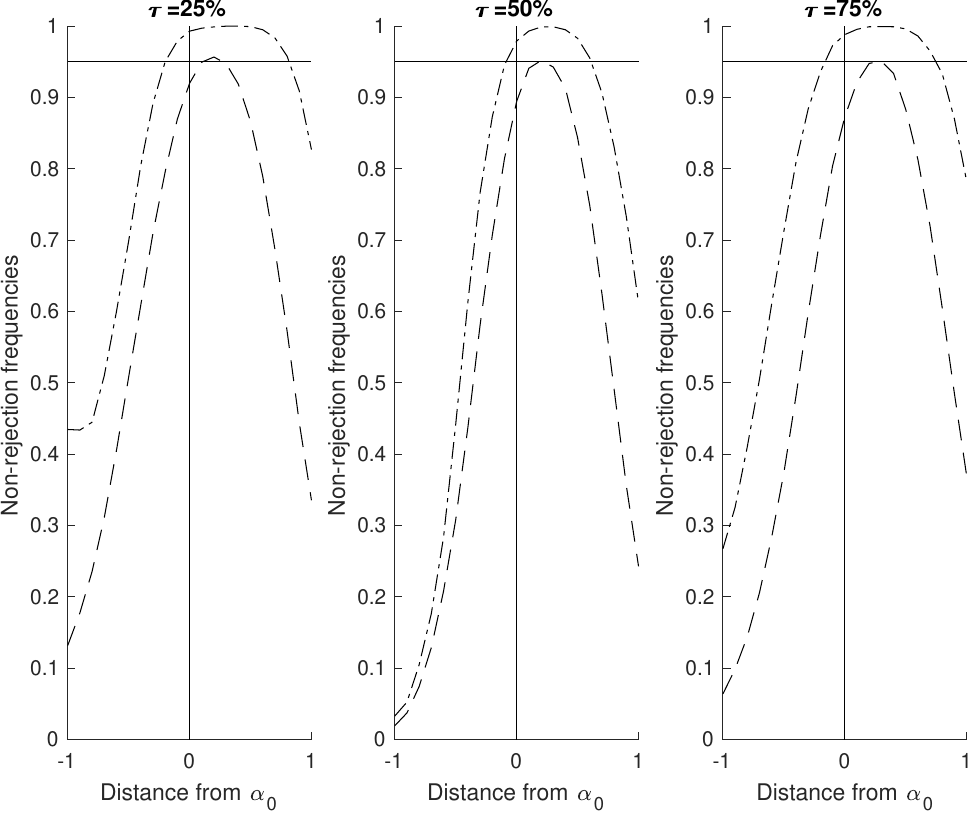}
\caption*{Figure \ref{fig_MC_diff1}.e: $\pi_0=\pi_1=0.1$.}
}
\hspace{.1\textwidth}
\parbox{.4\textwidth}{
\centering
\includegraphics[height=.4\textwidth,keepaspectratio]{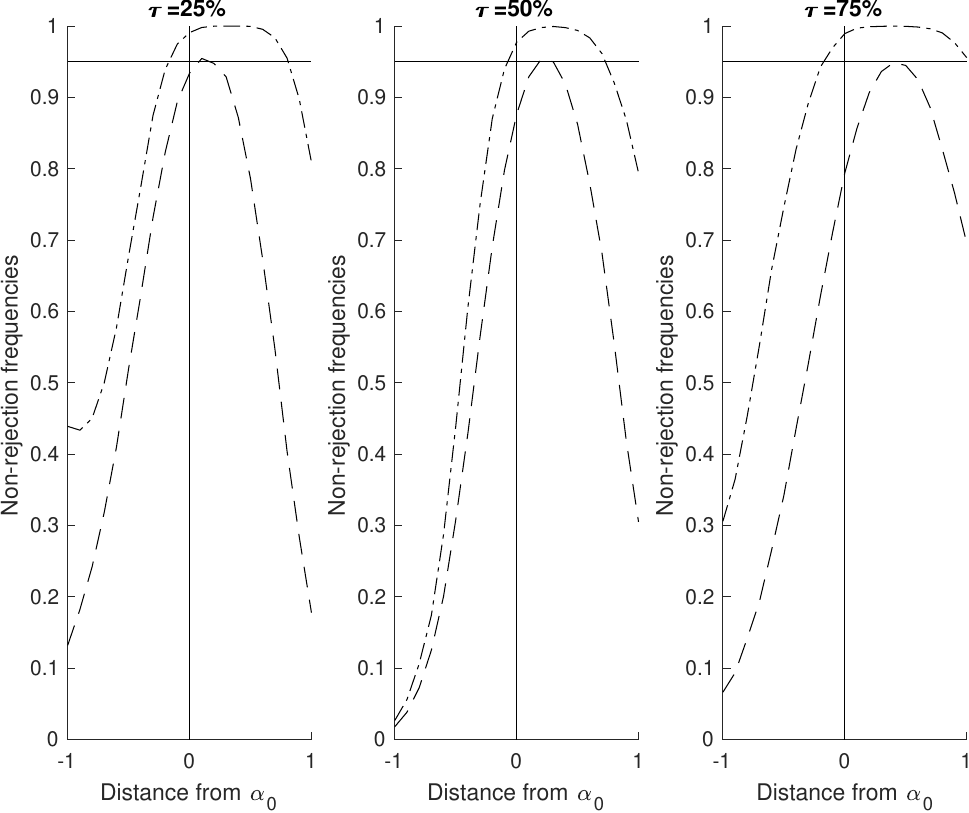}
\caption*{Figure \ref{fig_MC_diff1}.f: $\pi_0=0$ and $\pi_1=0.2$.}
}
\\
\bigskip
\bigskip
\\
\parbox{.4\textwidth}{
\centering
\includegraphics[height=.4\textwidth,keepaspectratio]{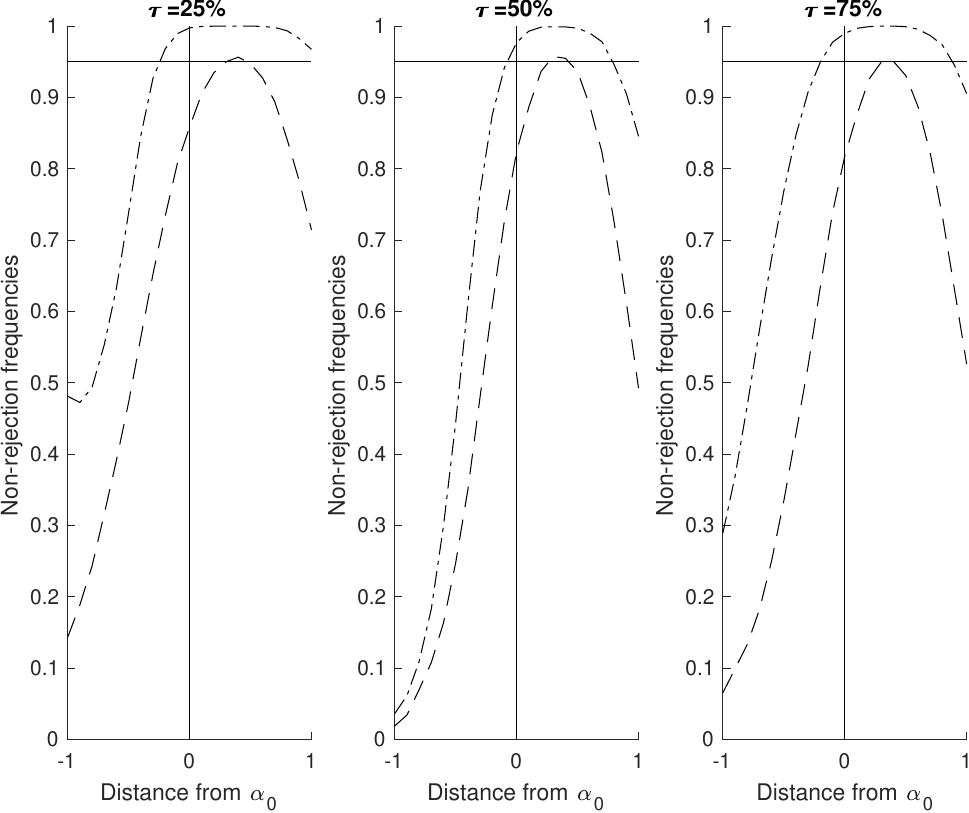}
\caption*{Figure \ref{fig_MC_diff1}.g: $\pi_0=0.2$ and $\pi_1=0.1$.}
}
\hspace{.1\textwidth}
\parbox{.4\textwidth}{
\centering
\includegraphics[height=.4\textwidth,keepaspectratio]{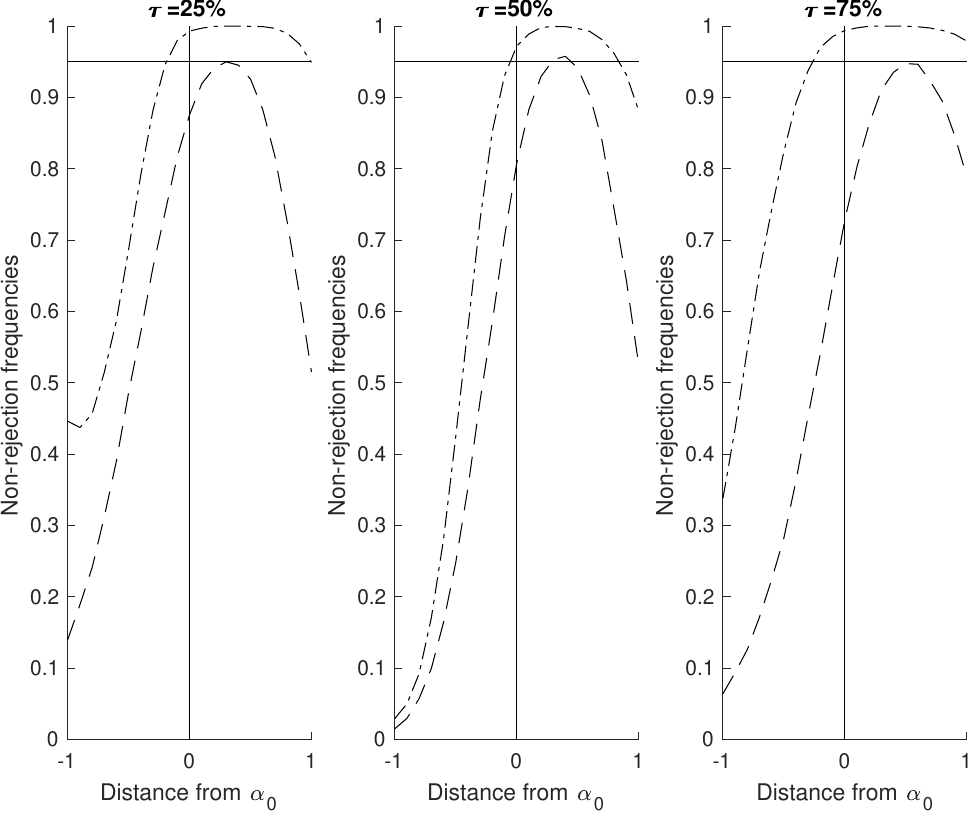}
\caption*{Figure \ref{fig_MC_diff1}.h: $\pi_0=0.1$ and $\pi_1=0.2$.}
}
\\
\bigskip
\bigskip
\\
\caption*{Figure \ref{fig_MC_diff1} (continued): Coverage frequencies. The dash-dot ($-.$) curve represents the proposed inference method, and the dashed ($--$) curve represents the inference method that assumes $\pi_0=\pi_1=0$.}
\end{figure}

\clearpage

\bibliography{/Users/tu10/Dropbox/Research/mybib}

\begin{thebibliography}{42}
\newcommand{\enquote}[1]{``#1''}
\expandafter\ifx\csname natexlab\endcsname\relax\def\natexlab#1{#1}\fi

\bibitem[\protect\citeauthoryear{Abadie, Angrist, and Imbens}{Abadie
  et~al.}{2002}]{abadie/angrist/imbens:2002}
\textsc{Abadie, A., J.~Angrist, and G.~Imbens} (2002): \enquote{Instrumental
  Variables Estimates of the Effect of Subsidized Training on the Quantiles of
  Trainee Earnings,} \emph{Econometrica}, 70, 91--117.

\bibitem[\protect\citeauthoryear{Angrist, Imbens, and Rubin}{Angrist
  et~al.}{1996}]{angrist/imbens/rubin:1996}
\textsc{Angrist, J.~D., G.~W. Imbens, and D.~B. Rubin} (1996):
  \enquote{Identification of Causal Effects using Instrumental Variables,}
  \emph{Journal of the American Statistical Association}, 91, 444--455.

\bibitem[\protect\citeauthoryear{Benjamin}{Benjamin}{2003}]{Benjamin:2003}
\textsc{Benjamin, D.~J.} (2003): \enquote{Does 401(k) Eligibility Increase
  Saving?: Evidence from Propensity Score Subclassification,} \emph{Journal of
  Public Economics}, 87, 1259 -- 1290.

\bibitem[\protect\citeauthoryear{Bitler, Hoynes, and Domina}{Bitler
  et~al.}{2016}]{bitler/hoynes/domina:2016}
\textsc{Bitler, M.~P., H.~W. Hoynes, and T.~Domina} (2016):
  \enquote{Experimental Evidence on Distributional Effects of Head Start,}
  Working paper.

\bibitem[\protect\citeauthoryear{Bound, Brown, and Mathiowetz}{Bound
  et~al.}{2001}]{bound/brown/mathiowetz:2001}
\textsc{Bound, J., C.~Brown, and N.~Mathiowetz} (2001): \enquote{Measurement
  Error in Survey Data,} in \emph{Handbook of Econometrics}, ed. by J.~Heckman
  and E.~Leamer, Elsevier, vol.~5, chap.~59, 3705--3843.

\bibitem[\protect\citeauthoryear{Buchinsky and Hahn}{Buchinsky and
  Hahn}{1998}]{buchinsky/hahn:1998}
\textsc{Buchinsky, M. and J.~Hahn} (1998): \enquote{An Alternative Estimator
  for the Censored Quantile Regression Model,} \emph{Econometrica}, 66, 653.

\bibitem[\protect\citeauthoryear{Calvi, Lewbel, and Tommasi}{Calvi
  et~al.}{2017}]{calvi/lewbel/tommasi:2017}
\textsc{Calvi, R., A.~Lewbel, and D.~Tommasi} (2017): \enquote{LATE With
  Mismeasured or Misspecified Treatment: An Application To Women's Empowerment
  in India,} Working paper.

\bibitem[\protect\citeauthoryear{Chernozhukov and Hansen}{Chernozhukov and
  Hansen}{2004}]{chernozhukov/hansen:2004}
\textsc{Chernozhukov, V. and C.~Hansen} (2004): \enquote{The Effects of 401(K)
  Participation on the Wealth Distribution: An Instrumental Quantile Regression
  Analysis,} \emph{Review of Economics and Statistics}, 86, 735--751.

\bibitem[\protect\citeauthoryear{Chernozhukov and Hansen}{Chernozhukov and
  Hansen}{2005}]{chernozhukov/hansen:2005}
---\hspace{-.1pt}---\hspace{-.1pt}--- (2005): \enquote{An IV Model of Quantile
  Treatment Effects,} \emph{Econometrica}, 73, 245--261.

\bibitem[\protect\citeauthoryear{Chernozhukov and Hansen}{Chernozhukov and
  Hansen}{2008}]{chernozhukov/hansen:2008}
---\hspace{-.1pt}---\hspace{-.1pt}--- (2008): \enquote{Instrumental Variable
  Quantile Regression: A Robust Inference Approach,} \emph{Journal of
  Econometrics}, 142, 379 -- 398.

\bibitem[\protect\citeauthoryear{Chernozhukov and Hansen}{Chernozhukov and
  Hansen}{2013}]{chernozhukov/hansen:2013}
---\hspace{-.1pt}---\hspace{-.1pt}--- (2013): \enquote{Quantile Models with
  Endogeneity,} \emph{Annual Review of Economics}, 5, 57--81.

\bibitem[\protect\citeauthoryear{Chesher}{Chesher}{1991}]{chesher:1991}
\textsc{Chesher, A.} (1991): \enquote{The Effect of Measurement Error,}
  \emph{Biometrika}, 78, 451--462.

\bibitem[\protect\citeauthoryear{Chesher}{Chesher}{2003}]{chesher:2003}
---\hspace{-.1pt}---\hspace{-.1pt}--- (2003): \enquote{Identification in
  Nonseparable Models,} \emph{Econometrica}, 71, 1405--1441.

\bibitem[\protect\citeauthoryear{Chesher}{Chesher}{2017}]{chesher:2017}
---\hspace{-.1pt}---\hspace{-.1pt}--- (2017): \enquote{Understanding the Effect
  of Measurement Error on Quantile Regressions,} \emph{Journal of
  Econometrics}, 200, 223 -- 237.

\bibitem[\protect\citeauthoryear{DiNardo and Lee}{DiNardo and
  Lee}{2011}]{dinardo/lee:2011}
\textsc{DiNardo, J. and D.~S. Lee} (2011): \enquote{Program Evaluation and
  Research Designs,} in \emph{Handbook of Labor Economics}, ed. by
  O.~Ashenfelter and D.~Card, Elsevier, vol. 4, Part A, chap.~5, 463--536.

\bibitem[\protect\citeauthoryear{DiTraglia and Garc\'{i}a-Jimeno}{DiTraglia and
  Garc\'{i}a-Jimeno}{2019}]{ditragliaand/garica-jimeno:2015}
\textsc{DiTraglia, F.~J. and C.~Garc\'{i}a-Jimeno} (2019): \enquote{On
  Mis-measured Binary Regressors: New Results and Some Comments on the
  Literature,} \emph{Journal of Econometrics}, 209, 376--390.

\bibitem[\protect\citeauthoryear{Doksum}{Doksum}{1974}]{doksum:1974}
\textsc{Doksum, K.} (1974): \enquote{Empirical Probability Plots and
  Statistical Inference for Nonlinear Models in the Two-Sample Case,}
  \emph{Annals of Statistics}, 2, 267--277.

\bibitem[\protect\citeauthoryear{Dong and Shen}{Dong and
  Shen}{2018}]{dong/shen:2019}
\textsc{Dong, Y. and S.~Shen} (2018): \enquote{Testing for Rank Invariance or
  Similarity in Program Evaluation,} \emph{Review of Economics and Statistics},
  100, 78--85.

\bibitem[\protect\citeauthoryear{Firpo, Galvao, and Song}{Firpo
  et~al.}{2017}]{firpo/galvao/song:2015}
\textsc{Firpo, S., A.~F. Galvao, and S.~Song} (2017): \enquote{Measurement
  Errors in Quantile Regression Models,} \emph{Journal of Econometrics}, 198,
  146--164.

\bibitem[\protect\citeauthoryear{Frandsen and Lefgren}{Frandsen and
  Lefgren}{2018}]{frandsen/lefgren:2018}
\textsc{Frandsen, B.~R. and L.~J. Lefgren} (2018): \enquote{Testing Rank
  Similarity,} \emph{Review of Economics and Statistics}, 100, 86--91.

\bibitem[\protect\citeauthoryear{Frazis and Loewenstein}{Frazis and
  Loewenstein}{2003}]{frazis/loewenstein:2003}
\textsc{Frazis, H. and M.~A. Loewenstein} (2003): \enquote{Estimating Linear
  Regressions with Mismeasured, Possibly Endogenous, Binary Explanatory
  Variables,} \emph{Journal of Econometrics}, 117, 151--178.

\bibitem[\protect\citeauthoryear{Galvao and Montes-Rojas}{Galvao and
  Montes-Rojas}{2009}]{galvao/montes-rojas:2009}
\textsc{Galvao, A.~F. and G.~Montes-Rojas} (2009): \enquote{Instrumental
  Variables Quantile Regression for Panel Data with Measurement Errors,}
  Working paper.

\bibitem[\protect\citeauthoryear{Gustman, Steinmeier, and Tabatabai}{Gustman
  et~al.}{2008}]{gustman/steinmeier/tabatabai:2007}
\textsc{Gustman, A.~L., T.~Steinmeier, and N.~Tabatabai} (2008): \enquote{Do
  Workers Know About Their Pension Plan Type? Comparing Workers' and Employers'
  Pension Information,} in \emph{Overcoming the Saving Slump; How to Increase
  the Effectiveness of Financial Education and Saving Programs}, ed. by
  A.~Lusardi, Chicago: University of Chicago Press, 47--81.

\bibitem[\protect\citeauthoryear{Hausman, Abrevaya, and Scott-Morton}{Hausman
  et~al.}{1998}]{hausman/abrevaya/schott-morton:1998}
\textsc{Hausman, J.~A., J.~Abrevaya, and F.~M. Scott-Morton} (1998):
  \enquote{{Misclassification of the Dependent Variable in a Discrete-Response
  Setting},} \emph{Journal of Econometrics}, 87, 239--269.

\bibitem[\protect\citeauthoryear{Heckman, Smith, and Clements}{Heckman
  et~al.}{1997}]{heckman/smith/clements:1997}
\textsc{Heckman, J.~J., J.~Smith, and N.~Clements} (1997): \enquote{Making the
  Most Out of Programme Evaluations and Social Experiments: Accounting For
  Heterogeneity in Programme Impacts,} \emph{The Review of Economic Studies},
  64, 487--535.

\bibitem[\protect\citeauthoryear{Henry, Kitamura, and Salani{\'e}}{Henry
  et~al.}{2014}]{henry/kitamura/salanie:2014}
\textsc{Henry, M., Y.~Kitamura, and B.~Salani{\'e}} (2014): \enquote{Partial
  Identification of Finite Mixtures in Econometric Models,} \emph{Quantitative
  Economics}, 5, 123--144.

\bibitem[\protect\citeauthoryear{Hu}{Hu}{2008}]{hu:2008}
\textsc{Hu, Y.} (2008): \enquote{Identification and Estimation of Nonlinear
  Models with Misclassification Error using Instrumental Variables: A General
  Solution,} \emph{Journal of Econometrics}, 144, 27--61.

\bibitem[\protect\citeauthoryear{Imbens and Angrist}{Imbens and
  Angrist}{1994}]{imbens/angrist:1994}
\textsc{Imbens, G.~W. and J.~D. Angrist} (1994): \enquote{Identification and
  Estimation of Local Average Treatment Effects,} \emph{Econometrica}, 62,
  467--75.

\bibitem[\protect\citeauthoryear{Kim and Park}{Kim and
  Park}{2018}]{kim/park:2016}
\textsc{Kim, J.~H. and B.~Park} (2018): \enquote{Testing Rank Similarity in the
  Local Average Treatment Effect Model,} Working paper.

\bibitem[\protect\citeauthoryear{Lewbel}{Lewbel}{2007}]{lewbel:2007}
\textsc{Lewbel, A.} (2007): \enquote{Estimation of Average Treatment Effects
  with Misclassification,} \emph{Econometrica}, 75, 537--551.

\bibitem[\protect\citeauthoryear{Mahajan}{Mahajan}{2006}]{mahajan:2006}
\textsc{Mahajan, A.} (2006): \enquote{Identification and Estimation of
  Regression Models with Misclassification,} \emph{Econometrica}, 74, 631--665.

\bibitem[\protect\citeauthoryear{Newey and McFadden}{Newey and
  McFadden}{1994}]{newey/mcfadden:1994}
\textsc{Newey, W.~K. and D.~McFadden} (1994): \enquote{Large Sample Estimation
  and Hypothesis Testing,} in \emph{Handbook of Econometrics}, ed. by R.~F.
  Engle and D.~L. McFadden, Elsevier, vol.~4, chap.~36, 2111--2245.

\bibitem[\protect\citeauthoryear{Nguimkeu, Denteh, and Tchernis}{Nguimkeu
  et~al.}{2019}]{Nguimkeu/Denteh/Tchernis:2017}
\textsc{Nguimkeu, P., A.~Denteh, and R.~Tchernis} (2019): \enquote{On the
  Estimation of Treatment Effects with Endogenous Misreporting,} \emph{Journal
  of Econometrics}, 208, 487--506.

\bibitem[\protect\citeauthoryear{Pollard}{Pollard}{1991}]{pollard:1991}
\textsc{Pollard, D.} (1991): \enquote{Asymptotics for Least Absolute Deviation
  Regression Estimators,} \emph{Econometric Theory}, 7, 186--199.

\bibitem[\protect\citeauthoryear{Schennach}{Schennach}{2008}]{schennach:2008}
\textsc{Schennach, S.~M.} (2008): \enquote{Quantile Regression with Mismeasured
  Covariates,} \emph{Econometric Theory}, 24, 1010--1043.

\bibitem[\protect\citeauthoryear{Small and Tan}{Small and
  Tan}{2007}]{small/tan:2007}
\textsc{Small, D.~S. and Z.~Tan} (2007): \enquote{A Stochastic Monotonicity
  Assumption for the Instrumental Variables Method,} Working paper.

\bibitem[\protect\citeauthoryear{Song}{Song}{2018}]{song:2016}
\textsc{Song, S.} (2018): \enquote{Nonseparable Triangular Models with Errors
  in Endogenous Variables,} Working paper.

\bibitem[\protect\citeauthoryear{Ura}{Ura}{2018}]{ura:2015}
\textsc{Ura, T.} (2018): \enquote{Heterogeneous Treatment Effects with
  Mismeasured Endogenous Treatment,} \emph{Quantitative Economics}, 9,
  1335--1370.

\bibitem[\protect\citeauthoryear{Wei and Carroll}{Wei and
  Carroll}{2009}]{wei/carroll:2009}
\textsc{Wei, Y. and R.~J. Carroll} (2009): \enquote{Quantile Regression With
  Measurement Error,} \emph{Journal of the American Statistical Association},
  104, 1129--1143.

\bibitem[\protect\citeauthoryear{W\"{u}thrich}{W\"{u}thrich}{2019}]{wuthrich:2016}
\textsc{W\"{u}thrich, K.} (2019): \enquote{A Comparison of Two Quantile Models
  with Endogeneity,} \emph{Journal of Business and Economic Statistics},
  forthcoming.

\bibitem[\protect\citeauthoryear{Yanagi}{Yanagi}{2019}]{yanagi:2017}
\textsc{Yanagi, T.} (2019): \enquote{Inference on Local Average Treatment
  Effects for Misclassified Treatment,} \emph{Econometric Reviews}, 38,
  938--959.

\bibitem[\protect\citeauthoryear{Yu}{Yu}{2017}]{yu:2017}
\textsc{Yu, P.} (2017): \enquote{Testing Conditional Rank Similarity With and
  Without Covariates,} Working paper.

\end{thebibliography}

\newpage
\appendix
\section*{Appendix} 
Appendix A discusses the identified set when the outcome variable is discrete, and Appendix B provides the proofs for the results in the main text.

\section*{Appendix A: Identified set with a discrete outcome variable}
This appendix demonstrates how to modify Theorem \ref{IS1} when the outcome variable is discrete.
Assumptions \ref{CHassumption1}-\ref{CHassumption2} are modified into the following two conditions. 

\begin{assumption}\label{CHassumption1_disc}
The mapping $u\mapsto q(d^\ast,u)$ is weakly increasing and left-continuous for every $u\in[0,1]$.
\end{assumption}
\begin{assumption}\label{CHassumption2_disc}
(i) $Pr(U_0\leq \tau\mid Z)\geq\tau$ and $Pr(U_1\leq \tau\mid Z)\geq\tau$.
(ii) $Pr(U_0\leq\tau\mid D^\ast,Z)=Pr(U_1\leq\tau\mid D^\ast,Z)$.
\end{assumption}

Under the above two assumptions, Theorem \ref{IS1} can be modified as follows: 
\begin{theorem}\label{IS1_disc}
Assume that all the elements in $\mathcal{Q}\times\mathcal{P}^\ast$ satisfy Assumptions \ref{missclass}, \ref{CHassumption1_disc}, and  \ref{CHassumption2_disc}.
(a) Given a distribution $P$ for the observed variables, if $(y_0,y_1)$ belongs to the sharp identified set for $q(\cdot,\tau)$, then 
\begin{equation}
P(Y\leq y_D\mid Z)-\tau\geq p_1(P(Y\leq y_0\mid Z)-\tau)+p_0(P(Y\leq y_1\mid Z)-\tau)\label{ModifTestImp_disc}
\end{equation}
for some $(p_0,p_1)$ with $p_0+p_1<1$
such that $0\leq p_0\leq P(D=1\mid Y,Z)\mbox{ a.s.}$ and $0\leq p_1\leq P(D=0\mid Y,Z)\mbox{ a.s.}$
(b) The converse is also true if $\mathcal{Q}\times\mathcal{P}^\ast$ includes all $(q,P^\ast)$'s satisfying Assumptions \ref{missclass}, \ref{CHassumption1_disc}, and  \ref{CHassumption2_disc}.
\end{theorem}
\begin{proof}
The proof for (a) is as follows. 
Define $(p_0,p_1)=(\pi_0,\pi_1)$.
By Assumption \ref{missclass}, 
\begin{equation}\label{mis_class_proof_eq11}
\left(\begin{array}{ccc}
P^\ast(D^\ast=0\mid Y,Z)\\
P^\ast(D^\ast=1\mid Y,Z)
\end{array}\right)
=
(1-p_0-p_1)^{-1}
\left(\begin{array}{ccc}
P(D=0\mid Y,Z)-p_1\\
P(D=1\mid Y,Z)-p_0\\
\end{array}\right).
\end{equation}
Using Eq. (\ref{mis_class_proof_eq11}),  
\begin{eqnarray*}
P(Y\leq y_{D}\mid Z)-p_1P(Y\leq y_0\mid Z)-p_0P(Y\leq y_1\mid Z)
=
(1-p_0-p_1)P(Y\leq y_{D^\ast}\mid Z).
\end{eqnarray*}
Since Assumption \ref{CHassumption1_disc} implies 
$P(Y\leq y_{D^\ast}\mid Z)=
P(q(D^\ast,U)\leq q(D^\ast,\tau)\mid Z)
\geq
P(U\leq\tau\mid Z)
=
P(U_0\leq\tau\mid Z)
\geq
\tau$,
it follows that  
$P(Y\leq y_{D}\mid Z)-p_1P(Y\leq y_0\mid Z)-p_0P(Y\leq y_1\mid Z)\geq (1-p_0-p_1)\tau$, and then Eq. (\ref{ModifTestImp_disc}) holds.
Moreover, by Eq. (\ref{mis_class_proof_eq11}), $P^\ast(D^\ast=d^\ast\mid Y,Z)\geq 0$ implies $P(D=0\mid Y,Z)\geq p_1$ and $P(D=1\mid Y,Z)\geq p_0$.

In the proof for (b), it is necessary to find $(\tilde{q},\tilde{P}^\ast)\in\mathcal{Q}\times\mathcal{P}^\ast$ such that $\tilde{q}(d^\ast,\tau)=y_{d^\ast}$ and that $P$ is the distribution for $(Y,D,Z)$ under $\tilde{P}^\ast$. 
For each $d^\ast=0,1$, there is a strictly increasing bijection $t_{d^\ast}:[0,1]\rightarrow[0,1]$ such that $t_{d^\ast}(\tau)=F_{Y\mid D=d^\ast}(y_{d^\ast})$.
For each $d^\ast=0,1$ and every $u\in[0,1]$, define $\tilde{q}(d^\ast,u)=Q_{Y\mid D=d^\ast}(t_{d^\ast}(u))$.
Define the distribution $\tilde{P}^\ast$ for $(D,Z,U_0,U_1,D^\ast)$ by 
\begin{eqnarray*}
\tilde{P}^\ast(D=1-d^\ast\mid Z,U_0,U_1,D^\ast)
&=&
p_{D^\ast}\\
\tilde{P}^\ast(Z\leq z,U_0\leq u_0,U_1\leq u_1,D^\ast=d^\ast)
&=&
\tilde{P}^\ast(Y\leq q(d^\ast,\min\{u_0,u_1\}),Z\leq z,D^\ast=d^\ast),
\end{eqnarray*}
where 
\begin{equation}\label{defin_dist_Y_proof_disc}
\left(\begin{array}{ccc}
\tilde{P}^\ast(Y\leq y,Z\leq z,D^\ast=0)\\
\tilde{P}^\ast(Y\leq y,Z\leq z,D^\ast=1)
\end{array}\right)
=
\left(\begin{array}{cc}
1-p_0&p_1\\
p_0&1-p_1
\end{array}\right)^{-1}
\left(\begin{array}{ccc}
P(Y\leq y,Z\leq z,D=0)\\
P(Y\leq y,Z\leq z,D=1)
\end{array}\right).
\end{equation}
By construction, the distribution for $(Y,D,Z)$ under $(\tilde{q},\tilde{P}^\ast)$ is $P$, and $\tilde{q}(d^\ast,\tau)=y_{d^\ast}$ for each $d^\ast=0,1$. 
To show $(\tilde{q},\tilde{P}^\ast)\in\mathcal{Q}\times\mathcal{P}^\ast$, it suffices to show that Assumptions \ref{CHassumption1_disc} and \ref{missclass} hold for $(\tilde{q},\tilde{P}^\ast)$. 
The rest of the proof is going to show Assumption \ref{CHassumption2} (i).\footnote{Assumption \ref{CHassumption1_disc} follows from the definition of $\tilde{q}$. 
Assumption \ref{CHassumption2} (ii) follows from the definition of $\tilde{P}^\ast$.
Assumption \ref{missclass}  follows from  $\tilde{P}^\ast(D\ne D^\ast\mid Z,U_0,U_1,D^\ast)=p_{D^\ast}$.}
By rearranging Eq. (\ref{ModifTestImp_disc}), 
\begin{eqnarray*}
\tau
&\leq&
\frac{1-p_1}{1-p_0-p_1}P(Y\leq y_0,D=0\mid Z)-\frac{p_1}{1-p_0-p_1}P(Y\leq y_0,D=1\mid Z)
\\&&-\frac{p_0}{1-p_0-p_1}P(Y\leq y_1,D=0\mid Z)+\frac{1-p_0}{1-p_0-p_1}P(Y\leq y_1,D=1\mid Z)
\end{eqnarray*}
Using Eq. (\ref{defin_dist_Y_proof_disc}), 
$\tau
\leq
\tilde{P}^\ast(Y\leq y_0,D^\ast=0\mid Z)+\tilde{P}^\ast(Y\leq y_1,D^\ast=1\mid Z)$.
The definition of $\tilde{q}$ and $\tilde{P}^\ast$ implies 
\begin{eqnarray*}
&&
\tilde{P}^\ast(U_0\leq \tau\mid Z)
=
\tilde{P}^\ast(Y\leq y_0,D^\ast=0\mid Z)+\tilde{P}^\ast(Y\leq y_1,D^\ast=1\mid Z)
\geq
\tau
\\
&&
\tilde{P}^\ast(U_1\leq \tau\mid Z)
=
\tilde{P}^\ast(Y\leq y_0,D^\ast=0\mid Z)+\tilde{P}^\ast(Y\leq y_1,D^\ast=1\mid Z)
\geq
\tau, 
\end{eqnarray*}
which is Assumption \ref{CHassumption2} (i). 
\end{proof}

\section*{Appendix B: Proofs of the results in the main text}

\subsection*{Proof of Theorem \ref{convcomb}}
\begin{lemma}\label{convcomb_lemma}
Under Assumptions \ref{CHassumption1} and \ref{CHassumption2}, $Q_{Y\mid Z}(\tau)$ is a convex combination of $q(1,\tau)$ and $q(0,\tau)$. 
\end{lemma}
\begin{proof}
By the monotonicity of $q(\cdot,d^\ast)$ and Lemma \ref{Thm1CH}, 
\begin{eqnarray*}
&&
\tau
=
P(Y\leq q(D^\ast,\tau)\mid Z)
\geq
P(Y\leq \min\{q(1,\tau),q(0,\tau)\}\mid Z)
=
F_{Y\mid Z}(\min\{q(1,\tau),q(0,\tau)\})
\\
&&
\tau
=
P(Y\leq q(D^\ast,\tau)\mid Z)
\leq
P(Y\leq \max\{q(1,\tau),q(0,\tau)\}\mid Z)
=
F_{Y\mid Z}(\max\{q(1,\tau),q(0,\tau)\}).
\end{eqnarray*}
Since $Q_{Y\mid Z}(\cdot)$ is monotonic, it follows that 
$\min\{q(1,\tau),q(0,\tau)\}
\leq
Q_{Y\mid Z}(\tau)
\leq
\max\{q(1,\tau),q(0,\tau)\}$.
\end{proof}

By Lemma \ref{convcomb_lemma}, the statement of theorem holds if $q(1,\tau)=q(0,\tau)$. 
The rest of the proof is going to focus on $q(1,\tau)>q(0,\tau)$. 
By Lemma \ref{convcomb_lemma}, there is some $\kappa\in[-1,1]$ such that $Q_{Y\mid Z=z_1}(\tau)-Q_{Y\mid Z=z_0}(\tau)=\kappa(q(1,\tau)-q(0,\tau))$.
Here it is sufficient to show $Q_{Y\mid Z=z_1}(\tau)\geq Q_{Y\mid Z=z_0}(\tau)$ for Theorem \ref{convcomb} (a) and $Q_{Y\mid Z=z_1}(\tau)>Q_{Y\mid Z=z_0}(\tau)$ for Theorem \ref{convcomb} (b).
Since $Q_{Y\mid Z}(\tau)$ is a convex combination of $q(1,\tau)$ and $q(0,\tau)$, it follows $q(1,\tau)\geq Q_{Y\mid Z}(\tau)\geq q(0,\tau)$. 
Since 
\begin{eqnarray*}
&&
P(Y\leq Q_{Y\mid Z}(\tau),D^\ast=0\mid Z)+P(Y\leq Q_{Y\mid Z}(\tau),D^\ast=1\mid Z)
\\
&&\quad =
F_{Y\mid Z}(Q_{Y\mid Z}(\tau))\\
&&\quad =
\tau\\
&&\quad =
P(q(D^\ast,U)\leq q(D^\ast,\tau)\mid Z)\\
&&\quad =
P(q(0,U)\leq q(0,\tau),D^\ast=0\mid Z)+P(q(1,U)\leq q(1,\tau),D^\ast=1\mid Z),
\end{eqnarray*}
it follows that $P(q(0,\tau)<q(0,U)\leq Q_{Y\mid Z}(\tau),D^\ast=0\mid Z)=P(Q_{Y\mid Z}(\tau)<q(1,U)\leq q(1,\tau),D^\ast=1\mid Z)$.
Using the monotonicity of $u\mapsto q(d^\ast,u)$, 
$P(\tau<U\leq q^{-1}(Q_{Y\mid Z}(\tau),0),D^\ast=0\mid Z)
=P(q^{-1}(Q_{Y\mid Z}(\tau),1)<U\leq\tau,D^\ast=1\mid Z)$.
Using the density function $f_{U_{d^\ast},D^\ast\mid Z}(y,d^\ast)$, the above equation can be rewritten as 
\begin{equation}\label{Diff_test_impl}
\int_{\tau}^{q^{-1}(Q_{Y\mid Z}(\tau),0)}f_{U_0,D^\ast\mid Z}(u,0)du=\int_{q^{-1}(Q_{Y\mid Z}(\tau),1)}^{\tau} f_{U_1,D^\ast\mid Z}(u,1)du.
\end{equation}

The first half of this proof is going to show Theorem \ref{convcomb} (a). 
Evaluate Eq. (\ref{Diff_test_impl}) at $Z=z_0$ and then 
$$
\int_{\tau}^{q^{-1}(Q_{Y\mid Z=z_0}(\tau),0)}f_{U_0,D^\ast\mid Z=z_0}(u,0)du=\int_{q^{-1}(Q_{Y\mid Z=z_0}(\tau),1)}^{\tau} f_{U_1,D^\ast\mid Z=z_0}(u,1)du.
$$
Since $f_{U_0,D^\ast\mid Z=z_1}(u,0)\leq f_{U_0,D^\ast\mid Z=z_0}(u,0)$ and $f_{U_1,D^\ast\mid Z=z_1}(u,1)\geq f_{U_1,D^\ast\mid Z=z_0}(u,1)$, 
$$
\int_{\tau}^{q^{-1}(Q_{Y\mid Z=z_0}(\tau),0)}f_{U_0,D^\ast\mid Z=z_1}(u,0)du\leq\int_{q^{-1}(Q_{Y\mid Z=z_0}(\tau),1)}^{\tau} f_{U_1,D^\ast\mid Z=z_1}(u,1)du.
$$
Subtracting the above equation from Eq. (\ref{Diff_test_impl}) at $Z=z_1$, and then 
$$
\int_{q^{-1}(Q_{Y\mid Z=z_1}(\tau),0)}^{q^{-1}(Q_{Y\mid Z=z_0}(\tau),0)}f_{U_0,D^\ast\mid Z=z_1}(u,0)du\leq\int_{q^{-1}(Q_{Y\mid Z=z_0}(\tau),1)}^{q^{-1}(Q_{Y\mid Z=z_1}(\tau),1)} f_{U_1,D^\ast\mid Z=z_1}(u,1)du.
$$
Using $Y=q(D^\ast,U)$, 
$$
\int_{Q_{Y\mid Z=z_1}(\tau)}^{Q_{Y\mid Z=z_0}(\tau)}f_{Y,D^\ast\mid Z=z_1}(y,0)dy\leq\int_{Q_{Y\mid Z=z_0}(\tau)}^{Q_{Y\mid Z=z_1}(\tau)} f_{Y,D^\ast\mid Z=z_1}(y,1)dy,
$$
and then $\int_{Q_{Y\mid Z=z_0}(\tau)}^{Q_{Y\mid Z=z_1}(\tau)}f_{Y,D^\ast\mid Z=z_1}(u)du\geq 0$,
which implies $\tau-F_{Y,D^\ast\mid Z=z_1}(Q_{Y\mid Z=z_0}(\tau))\geq 0$.

The second half of the proof is going to show Theorem \ref{convcomb} (b). 
To the contrary, suppose $Q_{Y\mid Z=z_0}(\tau)\leq Q_{Y\mid Z=z_1}(\tau)$. 
By Theorem \ref{convcomb} (a), $Q_{Y\mid Z=z_0}(\tau)=Q_{Y\mid Z=z_1}(\tau)$. 
Using $\Delta f_{U_{d^\ast},D^\ast\mid Z}(u,d^\ast)=f_{U_{d^\ast},D^\ast\mid Z=z_1}(u,d^\ast)-f_{U_{d^\ast},D^\ast\mid Z=z_0}(u,d^\ast)$, 
Eq. (\ref{Diff_test_impl}) implies  
$$
\int_{\tau}^{q^{-1}(Q_{Y\mid Z=z_0}(\tau),0)}\Delta f_{U_0,D^\ast\mid Z}(u,0)du=\int_{q^{-1}(Q_{Y\mid Z=z_0}(\tau),1)}^{\tau} \Delta f_{U_1,D^\ast\mid Z}(u,1)du.
$$
Since $f_{U_0,D^\ast\mid Z=z_1}(u,0)\leq f_{U_0,D^\ast\mid Z=z_0}(u,0)$ and $f_{U_1,D^\ast\mid Z=z_1}(u,1)\geq f_{U_1,D^\ast\mid Z=z_0}(u,1)$, 
$$
\int_{\tau}^{q^{-1}(Q_{Y\mid Z=z_0}(\tau),0)}|\Delta f_{U_0,D^\ast\mid Z}(u,0)|du=-\int_{q^{-1}(Q_{Y\mid Z=z_0}(\tau),1)}^{\tau} |\Delta f_{U_1,D^\ast\mid Z}(u,1)|du.
$$
Since $f_{U_0,D^\ast\mid Z=z_1}(u,0)<f_{U_0,D^\ast\mid Z=z_0}(u,0)$ and $f_{U_1,D^\ast\mid Z=z_1}(u,1)>f_{U_1,D^\ast\mid Z=z_0}(u,1)$ in a neighborhood of $\tau$, 
the above equation implies $q^{-1}(Q_{Y\mid Z=z_0}(\tau),0)=\tau=q^{-1}(Q_{Y\mid Z=z_0}(\tau),1)$, which contradicts $q(1,\tau)>q(0,\tau)$.

\subsection*{Proof of Theorem \ref{IS1}}
The proof for (a) is as follows. 
By Assumption \ref{missclass}, 
\begin{eqnarray*}
\left(\begin{array}{ccc}
P^\ast(D^\ast=0\mid Y,Z)\\
P^\ast(D^\ast=1\mid Y,Z)
\end{array}\right)
&=&
(1-\pi_0-\pi_1)^{-1}
\left(\begin{array}{ccc}
P(D=0\mid Y,Z)-\pi_1\\
P(D=1\mid Y,Z)-\pi_0\\
\end{array}\right).
\end{eqnarray*}
Eq. (\ref{originalTestableImpe}) in Lemma \ref{Thm1CH} becomes Eq. (\ref{ModifTestImp}). 
Moreover, $P^\ast(D^\ast=d^\ast\mid Y,Z)\geq 0$ implies $P(D=0\mid Y,Z)\geq\pi_1$ and $P(D=1\mid Y,Z)\geq\pi_0$.

In the proof for (b), it is necessary to find $(\tilde{q},\tilde{P}^\ast)\in\mathcal{Q}\times\mathcal{P}^\ast$ such that $\tilde{q}(d^\ast,\tau)=y_{d^\ast}$ and that $P$ is the distribution for $(Y,D,Z)$ under $\tilde{P}^\ast$. 
For each $d^\ast=0,1$, there is a strictly increasing bijection $t_{d^\ast}:[0,1]\rightarrow[0,1]$ such that $t_{d^\ast}(\tau)=F_{Y\mid D=d^\ast}(y_{d^\ast})$.
For each $d^\ast=0,1$ and every $u\in[0,1]$, define $\tilde{q}(d^\ast,u)=Q_{Y\mid D=d^\ast}(t_{d^\ast}(u))$.
Define the distribution $\tilde{P}^\ast$ for $(D,Z,U_0,U_1,D^\ast)$ by 
\begin{eqnarray*}
\tilde{P}^\ast(D\ne D^\ast\mid Z,U_0,U_1,D^\ast)
&=&
p_{D^\ast}\\
\tilde{P}^\ast(Z\leq z,U_0\leq u_0,U_1\leq u_1,D^\ast=d^\ast)
&=&
\tilde{P}^\ast(Y\leq q(d^\ast,\min\{u_0,u_1\}),Z\leq z,D^\ast=d^\ast),
\end{eqnarray*}
where 
\begin{equation}\label{defin_dist_Y_proof}
\left(\begin{array}{ccc}
\tilde{P}^\ast(Y\leq y,Z\leq z,D^\ast=0)\\
\tilde{P}^\ast(Y\leq y,Z\leq z,D^\ast=1)
\end{array}\right)
=
\left(\begin{array}{cc}
1-p_0&p_1\\
p_0&1-p_1
\end{array}\right)^{-1}
\left(\begin{array}{ccc}
P(Y\leq y,Z\leq z,D=0)\\
P(Y\leq y,Z\leq z,D=1)
\end{array}\right).
\end{equation}
By construction, the distribution for $(Y,D,Z)$ under $(\tilde{q},\tilde{P}^\ast)$ is $P$, and $\tilde{q}(d^\ast,\tau)=y_{d^\ast}$ for each $d^\ast=0,1$. 
To show $(\tilde{q},\tilde{P}^\ast)\in\mathcal{Q}\times\mathcal{P}^\ast$, it suffices to show that Assumptions \ref{CHassumption1}, \ref{CHassumption2}, and \ref{missclass} hold for $(\tilde{q},\tilde{P}^\ast)$. 
The rest of the proof is going to show Assumption \ref{CHassumption2} (i).
By rearranging Eq. (\ref{ModifTestImp}), 
\begin{eqnarray*}
\tau
&=&
\frac{1-p_1}{1-p_0-p_1}P(Y\leq y_0,D=0\mid Z)-\frac{p_1}{1-p_0-p_1}P(Y\leq y_0,D=1\mid Z)
\\&&-\frac{p_0}{1-p_0-p_1}P(Y\leq y_1,D=0\mid Z)+\frac{1-p_0}{1-p_0-p_1}P(Y\leq y_1,D=1\mid Z)
\end{eqnarray*}
Using Eq. (\ref{defin_dist_Y_proof}), 
$
\tau
=
\tilde{P}^\ast(Y\leq y_0,D^\ast=0\mid Z)+\tilde{P}^\ast(Y\leq y_1,D^\ast=1\mid Z)$.
The definition of $\tilde{q}$ and $\tilde{P}^\ast$ implies 
\begin{eqnarray*}
&&
\tilde{P}^\ast(U_0\leq \tau\mid Z)
=
\tilde{P}^\ast(Y\leq y_0,D^\ast=0\mid Z)+\tilde{P}^\ast(Y\leq y_1,D^\ast=1\mid Z)
=
\tau\\
&&
\tilde{P}^\ast(U_1\leq \tau\mid Z)
=
\tilde{P}^\ast(Y\leq y_0,D^\ast=0\mid Z)+\tilde{P}^\ast(Y\leq y_1,D^\ast=1\mid Z)
=
\tau.
\end{eqnarray*}

\subsection*{Proof of Corollary \ref{coro1}}
Use $(p_0,p_1)=(0,0)$. Then $(y_0,y_1)$ satisfies the conditions in Theorem \ref{IS1}. 

\subsection*{Proof of Corollary \ref{coro2}}
In this proof, assume $Q_{Y\mid Z=z_0}(\tau)\leq Q_{Y\mid Z=z_1}(\tau)$ without loss of generality. 
The ``only if'' part of this corollary is shown as follows.
By Theorem \ref{IS1}, Eq. (\ref{ModifTestImp}) holds for some $(p_0,p_1)$ with $0\leq p_0\leq P(D=1\mid Y,Z)$ a.s. and $0\leq p_1\leq P(D=0\mid Y,Z)$ a.s. 
Using $(y_0,y_1)=(Q_{Y\mid Z=z_0}(\tau),Q_{Y\mid Z=z_1}(\tau))$, 
Eq. (\ref{ModifTestImp}) becomes 
$$
P(y_0<Y\leq y_1,D=1\mid Z=z_0)=p_0P(y_0<Y\leq y_1\mid Z=z_0)
$$
$$
-P(y_0<Y\leq y_1,D=0\mid Z=z_1)=-p_1P(y_0<Y\leq y_1\mid Z=z_1).
$$
Since $p_0\leq P(D=1\mid Y,Z)$ a.s. and $p_1\leq P(D=0\mid Y,Z)$ a.s., it follows that 
$P(D=1\mid y_0<Y\leq y_1,Z=z_0)\leq P(D=1\mid Y,Z)$ a.s. and $P(D=0\mid y_0<Y\leq y_1,Z=z_1)\leq P(D=0\mid Y,Z)$ a.s.

The ``if'' part of this corollary is shown by choosing $p_0=P(D=1\mid  y_0<Y\leq y_1,Z=z_0)$ and $p_1=P(D=0\mid  y_0<Y\leq y_1,Z=z_1)$.

\subsection*{Proof of Theorem \ref{underident}}
Assume $\bar{d}^\ast=0$ for simplicity. 
Note that $q(d^\ast,u)=Q_{Y\mid D^\ast=d^\ast,Z}(u)$ from Condition (iii).
Take sufficiently small $\varepsilon>0$ and define 
\begin{eqnarray*}
\left(\tilde{q}(0,u),\tilde{q}(1,u)\right)&=&\left(q\left(0,u+\frac{\varepsilon}{1-\pi_0-\pi_1}(u-P^\ast(Y\leq q(1,u)\mid D^\ast=0))\right),q(1,u)\right)\\
({p}_0,{p}_1)&=&(\pi_0-\varepsilon,\pi_1).
\end{eqnarray*}
Condition (iv) guarantees $p_0\geq 0$. 
Consider $U_0=U_1$ and define $\tilde{P}^\ast$ by 
\begin{eqnarray*}
\tilde{P}^\ast(\tilde{q}(0,U_0)\leq y,D^\ast=0,Z\leq z)&=&\frac{1-\pi_0-\pi_1}{1-\pi_0-\pi_1+\varepsilon}P^\ast(Y\leq y,D^\ast=0,Z\leq z)\\
\tilde{P}^\ast(\tilde{q}(1,U_1)\leq y,D^\ast=1,Z\leq z)&=&P^\ast(Y\leq y,D^\ast=1,Z\leq z)\nonumber\\&&\quad+\frac{\varepsilon}{1-\pi_0-\pi_1+\varepsilon}P^\ast(Y\leq y,D^\ast=0,Z\leq z)\\
\tilde{P}^\ast(D=0\mid Z,U_1,D^\ast=1)
&=&
\pi_1\\
\tilde{P}^\ast(D=1\mid Z,U_0,D^\ast=0)
&=&
\pi_0-\varepsilon.
\end{eqnarray*}
By Condition (ii), $u\ne P^\ast(Y\leq q(1,u)\mid D^\ast=0)$ and
$\tilde{q}(0,\tau)
\ne 
q\left(0,u\right)$ as long as $\varepsilon$ is positive.
To establish the statement of this theorem, the rest of the proof is going to show that $(\tilde{q},\tilde{P}^\ast)\in\mathcal{Q}\times\mathcal{P}^\ast$ and that $(\tilde{q},\tilde{P}^\ast)$ is observationally equivalent to $(q,P^\ast)$.

First, $(\tilde{q},\tilde{P}^\ast)\in\mathcal{Q}\times\mathcal{P}^\ast$. 
The Lipschitz continuity in Condition (i) guarantees that $u\mapsto t(u)$ is a strictly increasing bijection of $[0,1]$ into $[0,1]$ for sufficiently small $\varepsilon$, where $t(u)=u+\frac{\varepsilon}{1-\pi_0-\pi_1}(u-P^\ast(Y\leq q(1,u)\mid D^\ast=0)$.
Therefore, Condition (v) implies $\tilde{q}\in\mathcal{Q}$ for sufficiently small $\varepsilon$.
To show $\tilde{P}^\ast\in\mathcal{P}^\ast$, it suffices to show Assumption \ref{CHassumption2} (i) because Assumption \ref{CHassumption2} (ii) holds for $U=U_0=U_1$.  
By the definition of $\tilde{q}$ and independence between $Z$ and $Y$ given $D^\ast$,  
\begin{eqnarray*}
P^\ast(Y\leq \tilde{q}(0,\tau),D^\ast=0\mid Z)
&=&
P^\ast(D^\ast=0\mid Z)P^\ast(Y\leq \tilde{q}(0,\tau)\mid D^\ast=0)\\
&=&
P^\ast(D^\ast=0\mid Z)
\left(
\tau+\frac{\varepsilon}{1-\pi_0-\pi_1}(\tau-P^\ast(Y\leq q(1,\tau)\mid D^\ast=0))
\right)\\
&=&
P^\ast(D^\ast=0\mid Z)
\left(
\tau+\frac{\varepsilon}{1-\pi_0-\pi_1}(\tau-P^\ast(Y\leq q(1,\tau)\mid D^\ast=0,Z))
\right),
\end{eqnarray*}
where the second equality uses $q(0,u)=Q_{Y\mid D^\ast=0}(u)$.
By the definition of $\tilde{P}^\ast$ and $\tilde{q}$, 
\begin{eqnarray*}
\tilde{P}^\ast(U\leq\tau\mid Z)
&=&
\frac{1-\pi_0-\pi_1}{1-\pi_0-\pi_1+\varepsilon}P^\ast(D^\ast=0\mid Z)
\left(
\tau+\frac{\varepsilon}{1-\pi_0-\pi_1}(\tau-P^\ast(Y\leq q(1,\tau)\mid D^\ast=0,Z))
\right)\\&&\quad+P^\ast(Y\leq q(1,\tau),D^\ast=1\mid Z)+\frac{\varepsilon}{1-\pi_0-\pi_1+\varepsilon}P^\ast(Y\leq q(1,\tau),D^\ast=0\mid Z)\\
&=&
\tau P^\ast(D^\ast=0\mid Z)+P^\ast(Y\leq q(1,\tau),D^\ast=1\mid Z)\\
&=&
\tau, 
\end{eqnarray*}
where the last equality uses $q(1,u)=Q_{Y\mid D^\ast=1,Z}(u)$.

Second, $(\tilde{q},\tilde{P}^\ast)$ is observationally equivalent to $(q,P^\ast)$. 
By the definition of $\tilde{P}^\ast$, 
\begin{eqnarray*}
\tilde{P}^\ast(Y\leq y,D^\ast=1,Z\leq z)
&=&
\tilde{P}^\ast(\tilde{q}(1,U)\leq y,D^\ast=1,Z\leq z)\\
&=&
P^\ast(Y\leq y,D^\ast=1,Z\leq z)+\frac{\varepsilon}{1-\pi_0-\pi_1+\varepsilon}P^\ast(Y\leq y,D^\ast=0,Z\leq z)\label{note_eq2}\\
\tilde{P}^\ast(Y\leq y,D^\ast=0,Z\leq z)
&=&
\tilde{P}^\ast(\tilde{q}(0,U)\leq y,D^\ast=0,Z\leq z)\\
&=&
\frac{1-\pi_0-\pi_1}{1-\pi_0-\pi_1+\varepsilon}P^\ast(Y\leq y,D^\ast=0,Z\leq z).\label{note_eq1}
\end{eqnarray*}
Therefore, 
\begin{eqnarray*}
\tilde{P}(Y\leq y,D=0,Z\leq z)
&=&
\tilde{P}^\ast(D=0\mid Z,U_0,U_1,D^\ast=0)\tilde{P}(Y\leq y,D^\ast=0,Z\leq z)\\&&\quad+\tilde{P}^\ast(D=0\mid Z,U_0,U_1,D^\ast=1)\tilde{P}(Y\leq y,D^\ast=1,Z\leq z)\\
&=&
(1-\pi_0+\varepsilon)\frac{1-\pi_0-\pi_1}{1-\pi_0-\pi_1+\varepsilon}P^\ast(Y\leq y,D^\ast=0,Z\leq z)\\&&
+\pi_1P^\ast(Y\leq y,D^\ast=1,Z\leq z)\nonumber\\&&+\pi_1
\frac{\varepsilon}{1-\pi_0-\pi_1+\varepsilon}P^\ast(Y\leq y,D^\ast=0,Z\leq z)\\
&=&
(1-\pi_0)P^\ast(Y\leq y,D^\ast=0,Z\leq z)\\&&
+\pi_1P^\ast(Y\leq y,D^\ast=1,Z\leq z)\\
&=&
P^\ast(Y\leq y,D=0,Z\leq z)
\end{eqnarray*}
and it can be similarly shown that $\tilde{P}(Y\leq y,D=1,Z\leq z)=P^\ast(Y\leq y,D=1,Z\leq z)$.

\subsection*{Proof of Lemma \ref{lemma_hu}}
Under Assumption \ref{point_identi_assn} (i), 
\begin{eqnarray*}
f_{(D,V)\mid Z}(d,v)
&=&
\left(
\begin{array}{cc}
f_{D\mid D^\ast=0,Z}(d)&f_{D\mid D^\ast=1,Z}(d)
\end{array}
\right)
\left(
\begin{array}{cc}
f_{D^\ast\mid Z}(0)&0\\
0&f_{D^\ast\mid Z}(1)
\end{array}
\right)
\left(
\begin{array}{cc}
f_{V\mid D^\ast=0}(v)\\
f_{V\mid D^\ast=1}(v)
\end{array}
\right).
\end{eqnarray*}
Under Assumption \ref{missclass}, 
\begin{eqnarray*}
&&\left(
\begin{array}{cc}
f_{(D,V)\mid Z}(0,v_0)&f_{(D,V)\mid Z}(0,v_1)\\
f_{(D,V)\mid Z}(1,v_0)&f_{(D,V)\mid Z}(1,v_1)\\
\end{array}
\right)\nonumber
\\
&&\quad\quad\quad=
\left(
\begin{array}{cc}
1-\pi_0&\pi_1\\
\pi_0&1-\pi_1
\end{array}
\right)
\left(
\begin{array}{cc}
f_{D^\ast\mid Z}(0)&0\\
0&f_{D^\ast\mid Z}(1)
\end{array}
\right)
\left(
\begin{array}{cc}
f_{V\mid D^\ast=0}(v_0)&f_{V\mid D^\ast=0}(v_1)\\
f_{V\mid D^\ast=1}(v_0)&f_{V\mid D^\ast=1}(v_1)\\
\end{array}
\right).
\end{eqnarray*}
Under Assumption \ref{point_identi_assn} (ii) and (iii), the above matrix for $Z=z_1$ is invertible, so  
\begin{eqnarray*}
&&
\begin{pmatrix}
f_{(D,V)\mid Z=z_0}(0,v_0)&f_{(D,V)\mid Z=z_0}(0,v_1)\\
f_{(D,V)\mid Z=z_0}(1,v_0)&f_{(D,V)\mid Z=z_0}(1,v_1)\\
\end{pmatrix}
\begin{pmatrix}
f_{(D,V)\mid Z=z_1}(0,v_0)&f_{(D,V)\mid Z=z_1}(0,v_1)\\
f_{(D,V)\mid Z=z_1}(1,v_0)&f_{(D,V)\mid Z=z_1}(1,v_1)\\
\end{pmatrix}
^{-1}
\\&&=
\begin{pmatrix}
1-\pi_0&\pi_1\\
\pi_0&1-\pi_1
\end{pmatrix}
\begin{pmatrix}
f_{D^\ast\mid Z=z_0}(0)/f_{D^\ast\mid Z=z_1}(0)&0\\
0&f_{D^\ast\mid Z=z_0}(1)/f_{D^\ast\mid Z=z_1}(1)
\end{pmatrix}
\begin{pmatrix}
1-\pi_0&\pi_1\\
\pi_0&1-\pi_1
\end{pmatrix}^{-1}
\end{eqnarray*}
Under Assumption \ref{point_identi_assn} (iv), the eigenvalue decomposition of the above matrix is uniquely determined, so $(\pi_0,\pi_1)$ is identified.  
Since
$f_{(Y,D)\mid Z}(y,d)
=
\sum_{d^\ast=0,1}f_{D\mid Y=y,D^\ast=d^\ast,Z}(d)f_{(Y,D^\ast)\mid Z}(y,d^\ast)$,
Assumption \ref{missclass} implies 
\begin{eqnarray*}
f_{(Y,D)\mid Z}(y,0)
&=&
\pi_0f_{(Y,D^\ast)\mid Z}(y,1)
+(1-\pi_0)f_{(Y,D^\ast)\mid Z}(y,0)\\
f_{(Y,D)\mid Z}(y,1)
&=&
(1-\pi_1)f_{(Y,D^\ast)\mid Z}(y,1)
+\pi_1
f_{(Y,D^\ast)\mid Z}(y,0),
\end{eqnarray*}
so that $f_{(Y,D^\ast)\mid Z}$ is point identified. 

\subsection*{Proof of Lemma \ref{lemma_CH_31}}
It follows from Theorem 2 of \cite{chernozhukov/hansen:2013}.

\subsection*{Proof of Theorem \ref{theorem_hu_CH_31}}
It follows from Lemmas \ref{lemma_hu} and \ref{lemma_CH_31}.

\subsection*{Proof of Lemma \ref{lemma:gamma_0_minimize}}
Note that 
\begin{eqnarray*}
&&
(1-\pi_0-\pi_1)E[\rho_{\tau}(Y-\alpha_0D^\ast-W'\theta)]
\\
&&\quad=
(1-\pi_0-\pi_1)E[\rho_\tau(Y-W'\theta)(1-D^\ast)]+(1-\pi_0-\pi_1)E[\rho_\tau(Y-\alpha_0-W'\theta)D^\ast]\\
&&\quad=
(1-\pi_1)E[\rho_\tau(Y-W'\theta)(1-D)]-\pi_1E[\rho_\tau(Y-W'\theta)D]\\&&\quad-\pi_0E[\rho_\tau(Y-\alpha_0-W'\theta)(1-D)]+(1-\pi_0)E[\rho_\tau(Y-\alpha_0-W'\theta)D]\\
&&\quad=
E[\rho_\tau(Y-W'\theta)(1-\pi_1-D)]]+E[\rho_\tau(Y-\alpha_0-W'\theta)(D-\pi_0)]\\
&&\quad=
E[\rho_\tau(Y-W'\theta)(1-\pi_1-E[D\mid Y,X,Z])]]+E[\rho_\tau(Y-\alpha_0-W'\theta)(E[D\mid Y,X,Z]-\pi_0)],
\end{eqnarray*}
because 
$$
\left(
\begin{array}{cc}
1-\pi_0&\pi_1\\
\pi_0&1-\pi_1
\end{array}
\right)^{-1}
=
\frac{1}{1-\pi_0-\pi_1}
\left(
\begin{array}{cc}
1-\pi_1&-\pi_1\\
-\pi_0&1-\pi_0
\end{array}
\right).
$$
Since 
$Pr(Y-\alpha_0D^\ast\leq X'\beta_0+0\cdot Z\mid X,Z)
=
Pr(q(D^\ast,X,U)\leq q(D^\ast,X,\tau)\mid X,Z)
=
\tau$, 
\citet[][p.383]{chernozhukov/hansen:2008} derives $0\in\argmin_{\gamma}\left(\min_{\beta}E[\rho_{\tau}(Y-\alpha_0D^\ast-W'\theta)]\right)$.
Therefore, 
$$
0\in\argmin_{\gamma}\left(\min_{\beta}E[\rho_\tau(Y-W'\theta)(1-\pi_1-\nu_0(Y,X,Z))]+E[\rho_\tau(Y-\alpha_0-W'\theta)(\nu_0(Y,X,Z)-\pi_0)]\right).
$$
Note that $\nu_0(Y,X,Z)-\pi_0=(\nu_0(Y,X,Z)-\pi_0)_+$ and $1-\pi_1-\nu_0(Y,X,Z)=(1-\pi_1-\nu_0(Y,X,Z))_+$, because 
\begin{eqnarray*}
&&
E[D\mid Y,X,Z]
=
\pi_0+(1-\pi_0-\pi_1)E[D^\ast\mid Y,X,Z]
\geq
\pi_0
\\
&&
1-E[D\mid Y,X,Z]
=
\pi_1+(1-\pi_0-\pi_1)E[1-D^\ast\mid Y,X,Z]
\geq
\pi_1.
\end{eqnarray*}

\subsection*{Proof of Theorem \ref{cov_theorem}}
By Lemma \ref{betagamma_conv} and \ref{omega_conv} below, 
$P(T(\alpha_0;\pi_0,\pi_1)\leq cv)\rightarrow 1-\mathrm{size}_2$ as $n\rightarrow\infty$. Then the theorem follows from 
\begin{eqnarray*}
P(\alpha_0\in CI_{\alpha}(\mathrm{size}_1+\mathrm{size}_2))
&\geq&
P(\{\alpha_0\in CI_{\alpha}(\mathrm{size}_1+\mathrm{size}_2)\}\cap\{(\pi_0,\pi_1)\in CI_1\})\\
&\geq&
P(\{T(\alpha_0;\pi_0,\pi_1)\leq cv\} \cap\{(\pi_0,\pi_1)\in CI_1\})\\
&\geq&
P(T(\alpha_0;\pi_0,\pi_1)\leq cv)-P((\pi_0,\pi_1)\notin CI_1)
\end{eqnarray*}
and then $\liminf_{n\rightarrow\infty}P(\alpha_0\in CI_{\alpha}(\mathrm{size}_1+\mathrm{size}_2))
\geq
1-(\mathrm{size}_1+\mathrm{size}_2)$.

The following proof assumes $1-\pi_1-\hat{\nu}(Y,X,Z)\geq 0$ and $\hat{\nu}(Y,X,Z)-\pi_0\geq 0$ without loss of generality, because $E[D^\ast\mid Y,X,Z]$ is bounded away from zero and one, and then  
\begin{eqnarray*}
&&
1-\pi_1-\hat{\nu}(Y,X,Z)
=
(1-\pi_0-\pi_1)E[1-D^\ast\mid Y,X,Z]-(\hat{\nu}(Y,X,Z)-{\nu}(Y,X,Z))
>
0
\\
&&
\hat{\nu}(Y,X,Z)-\pi_0
=
(1-\pi_0-\pi_1)E[D^\ast\mid Y,X,Z]+(\hat{\nu}(Y,X,Z)-{\nu}(Y,X,Z))
>
0
\end{eqnarray*}
with probability approaching one. 

\begin{lemma}\label{lemma_Qfun_con}
$\sup_{\theta\in\Theta}|\bar{Q}_n(\theta)-\bar{Q}_0(\theta)|=o_p(1)$, where $\bar{Q}_n(\theta)=Q_n(\theta;\alpha_0,\pi_0,\pi_1)$ and $\bar{Q}_0(\theta)=Q_0(\theta;\alpha_0,\pi_0,\pi_1)$.
\end{lemma}
\begin{proof}
The proof of this lemma is to check the conditions in \citet[][Lemma 2.9]{newey/mcfadden:1994}.  
Since $\Theta$ is compact and $\bar{Q}_0$ is continuous, it suffices to show $\bar{Q}_n(\theta)=\bar{Q}_0(\theta)+o_p(1)$ for every $\theta\in\Theta$ and $|\bar{Q}_n(\tilde\theta)-\bar{Q}_n(\theta)|\leq 2E_n\left[\|W\|\right]\cdot\|\tilde\theta-\theta\|$ for every $\tilde\theta,\theta\in\Theta$. This proof uses 
$$
\bar{Q}_n^\ast(\theta)=E_n[\rho_\tau(Y-W'\theta)(1-\pi_1-\nu_0(Y,X,Z))_+]+E_n[\rho_\tau(Y-\alpha_0-W'\theta)(\nu_0(Y,X,Z)-\pi_0)_+].
$$

The pointwise convergence of $\bar{Q}_n(\theta)$ to $\bar{Q}_0(\theta)$ is shown by demonstrating $\bar{Q}_n(\theta)-\bar{Q}_n^\ast(\theta)=o_p(1)$ and $\bar{Q}_n^\ast(\theta)-\bar{Q}_0(\theta)=o_p(1)$.
Since 
\begin{eqnarray*}
|\bar{Q}_n(\theta)-\bar{Q}_n^\ast(\theta)|
&\leq&
E_n[\rho_\tau(Y-W'\theta)|\hat{\nu}(Y,X,Z)-\nu_0(Y,X,Z)|]\\&&\qquad+E_n[\rho_\tau(Y-\alpha_0-W'\theta)|\hat{\nu}(Y,X,Z)-\nu_0(Y,X,Z)|]\\
&\leq&
\sup_{(y,x,z)}|\hat{\nu}(y,x,z)-\nu_0(y,x,z)|
\left(
E_n[|\rho_\tau(Y-W'\theta)|+|\rho_\tau(Y-\alpha_0-W'\theta)|]
\right),
\end{eqnarray*}
Assumption \ref{assn:condition_nu} (ii) implies $\bar{Q}_n(\theta)-\bar{Q}_n^\ast(\theta)=o_p(1)$. 
Moreover, $\bar{Q}_n^\ast(\theta)-\bar{Q}_0(\theta)=\bar{Q}_n^\ast(\theta)-E[\bar{Q}_n^\ast(\theta)]=o_p(1)$ is shown by checking the second moment of $\bar{Q}_n^\ast(\theta)$: $E[\bar{Q}_n^\ast(\theta)^2]=O(1/n)$ follows from  
\begin{eqnarray*}
&&
E\left[\left(\rho_\tau(Y-W'\theta)(1-\pi_1-D)+\rho_\tau(Y-\alpha_0-W'\theta)(D-\pi_0)\right)^2\right]^{1/2}
\\
&&\quad\leq
E\left[\left(Y-W'\theta\right)^2\right]^{1/2}
+
E\left[\left(Y-\alpha_0-W'\theta\right)^2\right]^{1/2}\\
&&\quad\leq
2E\left[Y^2\right]^{1/2}
+
\alpha_0
+
2E\left[\|W\|^2\right]^{1/2}\left\|\theta\right\|\\
&&\quad<
\infty. 
\end{eqnarray*}

The in-probability Lipschitz condition is shown as follows. 
Since 
$|\rho_\tau(Y-W'\theta)-\rho_\tau(Y-W'\tilde\theta)|\leq|W'(\tilde\theta-\theta)|\leq\|W\|\cdot \left\|\tilde\theta-\theta\right\|$ and 
$|\rho_\tau(Y-\alpha_0-W'\theta)-\rho_\tau(Y-\alpha_0-W'\tilde\theta)|\leq\|W\|\cdot \left\|\tilde\theta-\theta\right\|$,
it follows that 
$|\bar{Q}_n(\tilde\theta)-\bar{Q}_n(\theta)|
\leq
E_n\left[\|W\|\cdot \left\|\tilde\theta-\theta\right\| \right]+E_n\left[\|W\|\cdot \left\|\tilde\theta-\theta\right\| \right]
\leq
2E_n\left[\|W\|\right]\cdot  \left\|\tilde\theta-\theta\right\|$.
\end{proof}

\begin{lemma}\label{theta_conv_lema}
$\hat\theta_0-\theta_0=o_p(1)$, where $\hat\theta_0=\hat\theta(\alpha_0;\pi_0,\pi_1)$.
\end{lemma}
\begin{proof}
The proof of this lemma is to check the conditions in \citet[][Theorem 2.1]{newey/mcfadden:1994} to establish the consistency.  
Since $\Theta$ is compact, $\bar{Q}_0$ is continuous, and Lemma \ref{lemma_Qfun_con} establishes the uniform convergence of $\bar{Q}_n(\theta)$, it suffices to show that $\bar{Q}_0$ is uniquely minimized at $\theta_0$. 
As in the proof of Lemma \ref{lemma:gamma_0_minimize}, 
\begin{eqnarray*}
\bar{Q}_0(\theta)
&=&
E\left[\rho_\tau(Y-W'\theta)(1-\pi_1-D)+\rho_\tau(Y-\alpha_0-W'\theta)(D-\pi_0)\right]\\
&=&
(1-\pi_0-\pi_1)E\left[\rho_\tau(Y-\alpha_0D^\ast-W'\theta)\right]
\end{eqnarray*}
Since 
\begin{eqnarray*}
&&
\frac{\partial}{\partial\theta}\bar{Q}_0(\theta)
=
(1-\pi_0-\pi_1)\frac{\partial}{\partial\theta}E\left[\rho_\tau(Y-\alpha_0D^\ast-W'\theta)\right]
=
(1-\pi_0-\pi_1)E\left[(F_{Y-\alpha_0D^\ast\mid X,Z}(W'\theta)-\tau)W\right]\\
&&
\frac{\partial^2}{\partial\theta\partial\theta'}\bar{Q}_0(\theta)
=
(1-\pi_0-\pi_1)E\left[f_{Y-\alpha_0D^\ast\mid X,Z}(W'\theta)WW'\right],
\end{eqnarray*}
it follows that $\frac{\partial}{\partial\theta}\bar{Q}_0(\theta_0)=0$ and $\frac{\partial^2}{\partial\theta\partial\theta'}\bar{Q}_0(\theta)$ is  positive semidefinite everywhere and positive definite at $\theta_0$. 
Therefore, $\bar{Q}_0$ is uniquely minimized at $\theta_0$. 
\end{proof}

\begin{lemma}\label{lemma:newey_lemma5}
$E_n\left[\xi_0(\hat{\nu})W\right]=E_n[s_0]+o_p(n^{-1/2})$, where $s_0=\xi_0(\nu_0)W+E\left[W\Xi_0(\delta_0)\right]\psi_{\delta}$.
\end{lemma}
\begin{proof}
By Assumption \ref{assn:condition_nu} (iii)-(iv), 
$E_n\left[\xi_0(\hat{\nu})W\right]
=
E_n\left[\xi_0(\nu_0)W\right]
+E\left[W\Xi_0(\delta_0)\right]E_n[\psi_{\delta}]
+o_p(n^{-1/2})$.
\end{proof}

\begin{lemma}\label{lemma:conv_r_lemma}
Define 
\begin{eqnarray*}
\lambda_0(\boldsymbol{\tau})
&=&
f_{Y\mid D,Z,X}(W'\theta_0-n^{-1/2}W'\boldsymbol{\tau})(1-\pi_1-D)+f_{Y\mid D,Z,X}(\alpha_0+W'\theta_0-n^{-1/2}W'\boldsymbol{\tau})(D-\pi_0)\\
g(\boldsymbol{\tau},\nu)
&=&
(\rho_\tau(Y-W'\theta_0-n^{-1/2}W'\boldsymbol{\tau})-\rho_\tau(Y-W'\theta_0))(1-\pi_1-\nu(Y,X,Z))_+
\\&&+
(\rho_\tau(Y-\alpha_0-W'\theta_0-n^{-1/2}W'\boldsymbol{\tau})
-\rho_\tau(Y-\alpha_0-W'\theta_0))
(\nu(Y,X,Z)-\pi_0)_+
\\
r_n(\boldsymbol{\tau})
&=&
nE_n[g(\boldsymbol{\tau},\hat{\nu})]
-\frac{1}{2}\boldsymbol{\tau}'E[\lambda_0(0)WW']\boldsymbol{\tau}
+\sqrt{n}
\boldsymbol{\tau}'
E_n\left[\xi_0(\hat{\nu})W\right].
\end{eqnarray*}
Then $\sup_{\boldsymbol{\tau}}|r_n(\boldsymbol{\tau})|=o_p(1)$.
\end{lemma}
\begin{proof}
If $r_n(\boldsymbol{\tau})=o_p(1)$ pointwise in $\boldsymbol{\tau}$, it is possible to the convexity lemma in \cite{pollard:1991} to $\boldsymbol{\tau}\mapsto nE_n[g(\boldsymbol{\tau},\hat{\nu})]+\sqrt{n}\boldsymbol{\tau}'E_n\left[\xi_0(\hat{\nu})W\right]$, and then $\sup_{\boldsymbol{\tau}}|r_n(\boldsymbol{\tau})|=o_p(1)$.
Therefore, it suffices to show that $r_n(\boldsymbol{\tau})=o_p(1)$ pointwise in $\boldsymbol{\tau}$. 
Note that 
\begin{eqnarray*}
E[\lambda_0(0)WW']
&=&
E\left[f_{Y\mid D,Z,X}(W'\theta_0)(1-\pi_1-\nu_0(Y,X,Z))_+WW'\right]
\\&&+
E\left[f_{Y\mid D,Z,X}(\alpha_0+W'\theta_0)(\nu_0(Y,X,Z)-\pi_0)_+WW'\right],
\end{eqnarray*}
where the equality uses the law of iterated expectation, $(1-\pi_1-\nu_0(Y,X,Z))_+=1-\pi_1-\nu_0(Y,X,Z)$ and $(\nu_0(Y,X,Z)-\pi_0)_+=\nu_0(Y,X,Z)-\pi_0$. 
Therefore,  
$
\left.\frac{\partial}{\partial \boldsymbol{\tau}}E[ng(\boldsymbol{\tau},\nu_0)]\right|_{\boldsymbol{\tau}=0}
=
-n^{1/2}
E\left[\xi_0(\nu_0)W\right]
\ =
0
$
and 
$\frac{\partial^2}{\partial \boldsymbol{\tau}\partial \boldsymbol{\tau}'}E[ng(\tau,\nu_0)]
=
E[\lambda_0(\boldsymbol{\tau})WW']
=
E[\lambda_0(0)WW']+o(1)$, so that 
$$
E[ng(\boldsymbol{\tau},\nu_0)]=\frac{1}{2}\boldsymbol{\tau}'E[\lambda_0(0)WW']\boldsymbol{\tau}+o(1).
$$
Therefore,  
\begin{eqnarray*}
r_n(\boldsymbol{\tau})
&=&
nE_n[g(\boldsymbol{\tau},\hat{\nu})]
-nE[g(\boldsymbol{\tau},\nu_0)]
+\sqrt{n}
\boldsymbol{\tau}'
E_n\left[\xi_0(\hat{\nu})W\right]+o(1).
\\
&=&
E_n\left[ng(\boldsymbol{\tau},\hat{\nu})+\sqrt{n}\boldsymbol{\tau}'\xi_0(\hat{\nu})W\right]
-E_n\left[ng(\boldsymbol{\tau},\nu_0)+\sqrt{n}\boldsymbol{\tau}'\xi_0(\nu_0)W\right]
\\&&
+E_n\left[ng(\boldsymbol{\tau},\nu_0)+\sqrt{n}\boldsymbol{\tau}'\xi_0(\nu_0)W\right]
-E[ng(\boldsymbol{\tau},\nu_0)]
+o(1)
\\
&=&
E_n\left[n(g(\boldsymbol{\tau},\hat{\nu})-g(\boldsymbol{\tau},\nu_0))+\sqrt{n}\boldsymbol{\tau}'(\xi_0(\hat{\nu})-\xi_0(\nu_0))W\right]
\\&&
+(E_n-E)[ng(\boldsymbol{\tau},\nu_0)+\sqrt{n}\boldsymbol{\tau}'\xi_0(\nu_0)W]
+o(1),
\end{eqnarray*}
where the last equality follows from $E[\boldsymbol{\tau}'\xi_0(\nu_0)W]=0$.

First, $E_n\left[n(g(\boldsymbol{\tau},\hat{\nu})-g(\boldsymbol{\tau},\nu_0))+\sqrt{n}\boldsymbol{\tau}'(\xi_0(\hat{\nu})-\xi_0(\nu_0))W\right]$ converges to zero in $L^1$. 
By the definitions of $g(\boldsymbol{\tau},{\nu})$ and $\xi_0(\nu)$, 
\begin{eqnarray*}
&&
\hspace*{-1cm}
n(g(\boldsymbol{\tau},\hat{\nu})-g(\boldsymbol{\tau},\nu_0))+\sqrt{n}\boldsymbol{\tau}'(\xi_0(\hat{\nu})-\xi_0(\nu_0))W
\\
&=&
n\times\mathrm{term}_1\times((1-\pi_1-\hat{\nu}(Y,X,Z))_+-(1-\pi_1-\nu_0(Y,X,Z)))
\\&&+
n\times\mathrm{term}_2\times((\hat{\nu}(Y,X,Z)-\pi_0)_+-(\nu_0(Y,X,Z)-\pi_0)),
\end{eqnarray*}
where 
$\mathrm{term}_1=\rho_\tau(Y-W'(\theta_0-n^{-1/2}\boldsymbol{\tau}))-\rho_\tau(Y-W'\theta_0)+(\tau-1\{Y-W'\theta_0\leq 0\})n^{-1/2}W'\boldsymbol{\tau}$ and $\mathrm{term}_2=\rho_\tau(Y-\alpha_0-W'(\theta_0-n^{-1/2}\boldsymbol{\tau}))-\rho_\tau(Y-\alpha_0-W'\theta_0)+(\tau-1\{Y-\alpha_0-W'\theta_0\leq 0\})n^{-1/2}W'\boldsymbol{\tau}$.
By the definition of $\rho_\tau$, the two terms, $\mathrm{term}_1$ and $\mathrm{term}_2$, can be bounded as follows: 
\begin{eqnarray}
\left|\mathrm{term}_1\right|
&&
\leq
1\{|Y-W'\theta_0|\leq |n^{-1/2}W'\boldsymbol{\tau}|\}
\left(|Y-W'\theta_0|+|n^{-1/2}W'\boldsymbol{\tau}|\right)
\nonumber\\
&&
\leq
1\{|Y-W'\theta_0|\leq |n^{-1/2}W'\boldsymbol{\tau}|\}
\times 2|n^{-1/2}W'\boldsymbol{\tau}|
\label{eq:term1_bound}\\
\left|\mathrm{term}_2\right|
&&
\leq
1\{|Y-\alpha_0-W'\theta_0|\leq |n^{-1/2}W'\boldsymbol{\tau}|\}
\left(|Y-\alpha_0-W'\theta_0|+|n^{-1/2}W'\boldsymbol{\tau}|\right)
\nonumber\\
&&
\leq
1\{|Y-\alpha_0-W'\theta_0|\leq |n^{-1/2}W'\boldsymbol{\tau}|\}
\times 2|n^{-1/2}W'\boldsymbol{\tau}|.
\label{eq:term2_bound}
\end{eqnarray}
As long as $1-\pi_1-\hat{\nu}(Y,X,Z)\geq 0$ and $\hat{\nu}(Y,X,Z)-\pi_0\geq 0$,  
\begin{eqnarray}
|(1-\pi_1-\hat{\nu}(Y,X,Z))_+-(1-\pi_1-\nu_0(Y,X,Z))|
&=&
|(1-\pi_1-\hat{\nu}(Y,X,Z))-(1-\pi_1-\nu_0(Y,X,Z))|
\nonumber\\
&\leq&
\sup_{(y,x,z)}|\hat{\nu}(y,x,z)-\nu_0(y,x,z)|
\label{eq:coeff_bound1}
\\
|(\hat{\nu}(Y,X,Z)-\pi_0)_+-(\nu_0(Y,X,Z)-\pi_0)|
&=&
|(\hat{\nu}(Y,X,Z)-\pi_0)-(\nu_0(Y,X,Z)-\pi_0)|
\nonumber\\
&\leq&
\sup_{(y,x,z)}|\hat{\nu}(y,x,z)-\nu_0(y,x,z)|.
\label{eq:coeff_bound2}
\end{eqnarray}
Now it is possible to bound $E\left[\left\|E_n\left[n(g(\boldsymbol{\tau},\hat{\nu})-g(\boldsymbol{\tau},\nu_0))+\sqrt{n}\boldsymbol{\tau}'(\xi_0(\hat{\nu})-\xi_0(\nu_0))W\right]\right\|\right]$ as follows: 
\begin{eqnarray*}
&&
\hspace*{-1cm}
E\left[\left\|E_n\left[n(g(\boldsymbol{\tau},\hat{\nu})-g(\boldsymbol{\tau},\nu_0))+\sqrt{n}\boldsymbol{\tau}'(\xi_0(\hat{\nu})-\xi_0(\nu_0))W\right]\right\|\right]
\\
&\leq&
E\left[\left|n(g(\boldsymbol{\tau},\hat{\nu})-g(\boldsymbol{\tau},\nu_0))+\sqrt{n}\boldsymbol{\tau}'(\xi_0(\hat{\nu})-\xi_0(\nu_0))\right|\left\|W\right\|\right]
\\
&\leq&
2n
E[1\{|Y-W'\theta_0|\leq |n^{-1/2}W'\boldsymbol{\tau}|\}|n^{-1/2}W'\boldsymbol{\tau}|\left\|W\right\|]\sup_{(y,x,z)}|\hat{\nu}(y,x,z)-\nu_0(y,x,z)|
\\&&+2n
E[1\{|Y-\alpha_0-W'\theta_0|\leq |n^{-1/2}W'\boldsymbol{\tau}|\}|n^{-1/2}W'\boldsymbol{\tau}|\left\|W\right\|]\sup_{(y,x,z)}|\hat{\nu}(y,x,z)-\nu_0(y,x,z)|
\end{eqnarray*}
Since $|f_{Y\mid X,Z}|\leq C$,  it follows that 
\begin{eqnarray*}
E[1\{|Y-W'\theta_0|\leq |n^{-1/2}W'\boldsymbol{\tau}|\}|n^{-1/2}W'\boldsymbol{\tau}|\left\|W\right\|]
\leq
2Cn^{-1}E[|W'\boldsymbol{\tau}|^2\left\|W\right\|]
\\
E[1\{|Y-\alpha_0-W'\theta_0|\leq |n^{-1/2}W'\boldsymbol{\tau}|\}|n^{-1/2}W'\boldsymbol{\tau}|\left\|W\right\|]
\leq
2Cn^{-1}E[|W'\boldsymbol{\tau}|^2\left\|W\right\|],
\end{eqnarray*}
so that 
\begin{eqnarray*}
&&
E\left[\left|n(g(\boldsymbol{\tau},\hat{\nu})-g(\boldsymbol{\tau},\nu_0))+\sqrt{n}\boldsymbol{\tau}'(\xi_0(\hat{\nu})-\xi_0(\nu_0))\right|\left\|W\right\|\right]
\\
&&\qquad\leq
8CE[|W'\boldsymbol{\tau}|^2\left\|W\right\|]\sup_{(y,x,z)}|\hat{\nu}(y,x,z)-\nu_0(y,x,z)|
\\
&&\qquad=
o(1).
\end{eqnarray*}

Next, $(E_n-E)[ng(\boldsymbol{\tau},\nu_0)+\sqrt{n}\boldsymbol{\tau}'\xi_0(\nu_0)W]$ converges to zero in $L^2$.
Using the bounds in Eq. (\ref{eq:term1_bound})-(\ref{eq:coeff_bound2}), 
\begin{eqnarray*}
&&
\hspace*{-1cm}
E[(ng(\boldsymbol{\tau},\nu_0)+\sqrt{n}\boldsymbol{\tau}'\xi_0(\nu_0)W)^2]^{1/2}\\
&\leq&
nE[\left|\mathrm{term}_1\right|^2\times((1-\pi_1-\hat{\nu}(Y,X,Z))_+-(1-\pi_1-\nu_0(Y,X,Z)))^2]^{1/2}
\\&&+
nE[\left|\mathrm{term}_2\right|^2\times((\hat{\nu}(Y,X,Z)-\pi_0)_+-(\nu_0(Y,X,Z)-\pi_0))^2]^{1/2}
\\
&\leq&
4nE[1\{|Y-W'\theta_0|\leq |n^{-1/2}W'\boldsymbol{\tau}|\}
\times |n^{-1/2}W'\boldsymbol{\tau}|
^2]^{1/2}\sup_{(y,x,z)}|\hat{\nu}(y,x,z)-\nu_0(y,x,z)|.
\\&&+
4nE[1\{|Y-\alpha_0-W'\theta_0|\leq |n^{-1/2}W'\boldsymbol{\tau}|\}
\times |n^{-1/2}W'\boldsymbol{\tau}|
^2]^{1/2}\sup_{(y,x,z)}|\hat{\nu}(y,x,z)-\nu_0(y,x,z)|
\\
&\leq&
8n^{1/2}E[|W'\boldsymbol{\tau}|^2]^{1/2}\sup_{(y,x,z)}|\hat{\nu}(y,x,z)-\nu_0(y,x,z)|\\
&=&
o(n^{1/2}),
\end{eqnarray*}
and then $E[((E_n-E)[ng(\boldsymbol{\tau},\nu_0)+\sqrt{n}\boldsymbol{\tau}'\xi_0(\nu_0)W])^2]=
n^{-1}E[(ng(\boldsymbol{\tau},\nu_0)+\sqrt{n}\boldsymbol{\tau}'\xi_0(\nu_0)W)^2]=o(1)$.
\end{proof}

\begin{lemma}\label{betagamma_conv}
$\sqrt{n}(\hat\theta_0-\theta_0)=\sqrt{n}E[\lambda_0(0)WW']^{-1}E_n[s_0]+o_p(1)$ and therefore $\sqrt{n}(\hat\theta_0-\theta_0)
\rightarrow_d
N\left(0,\Omega_0\right)$ as $n\rightarrow\infty$,
where 
$
\Omega_0=
E[\lambda_0(0)WW']^{-1}E[s_0s_0']E[\lambda_0(0)WW']^{-1}.$
\end{lemma}
\begin{proof}
Define 
$\eta_n=\sqrt{n}E[\lambda_0(0)WW']^{-1}E_n[s_0]$.
It suffices to show $\sqrt{n}(\hat\theta_0-\theta_0)=\eta_n+o_p(1)$.
By Lemma \ref{lemma:newey_lemma5}, 
$
\eta_n=\sqrt{n}E[\lambda_0(0)WW']^{-1}
E_n\left[\xi_0(\hat{\nu})W\right]+o_p(1).$
Using the definition of $r_n(\cdot)$,   
\begin{eqnarray*}
-\frac{1}{2}\eta_n'E[\lambda_0(0)WW']\eta_n+r_n(\eta_n)
&=&
nE_n[g(\eta_n,\hat{\nu})]
\\
&\geq&
nE_n[g(\sqrt{n}(\hat\theta_0-\theta_0),\hat{\nu})]
\\
&=&
\frac{1}{2}(\sqrt{n}(\hat\theta_0-\theta_0)-\eta_n)'E[\lambda_0(0)WW'](\sqrt{n}(\hat\theta_0-\theta_0)-\eta_n)
\\&&-\frac{1}{2}\eta_n'E[\lambda_0(0)WW']\eta_n
+r_n(\sqrt{n}(\hat\theta_0-\theta_0))\\
&\geq&
\frac{1}{2}\|\sqrt{n}(\hat\theta_0-\theta_0)-\eta_n\|^2\mathrm{eig}_{\min}\left(E[\lambda_0(0)WW']\right)
\\&&-\frac{1}{2}\eta_n'E[\lambda_0(0)WW']\eta_n
+r_n(\sqrt{n}(\hat\theta_0-\theta_0)),
\end{eqnarray*}
where the first inequality uses $\sqrt{n}(\hat\theta_0-\theta_0)=\argmin_{\boldsymbol{\tau}}E_n[g(\boldsymbol{\tau},\hat{\nu})]$ and $\mathrm{eig}_{\min}\left(E[\lambda_0(0)WW']\right)$ is the minimum eigenvalue of $E[\lambda_0(0)WW']$.
Therefore, 
$r_n(\eta_n)-r_n(\sqrt{n}(\hat\theta_0-\theta_0))
\geq
\frac{1}{2}\|\sqrt{n}(\hat\theta_0-\theta_0)-\eta_n\|^2\mathrm{eig}_{\min}\left(E[\lambda_0(0)WW']\right)$, 
so that Lemma \ref{lemma:conv_r_lemma} implies $\sqrt{n}(\hat\theta_0-\theta_0)-\eta_n=o_p(1)$.
\end{proof}

\begin{lemma}\label{lemma:prob_diminising}
$Pr(|Y-W'\theta_0|\leq \|(\hat\theta_0-\theta_0)\|\|W\|\})=o(1)$ and 
$Pr(|Y-\alpha_0-W'\theta_0|\leq \|(\hat\theta_0-\theta_0)\|\|W\|\})=o(1)$.
\end{lemma}
\begin{proof}
By Lemma \ref{betagamma_conv} and Assumption \ref{assn_inverty} (iii), the first part of this lemma follows from 
$Pr(|Y-W'\theta_0|\leq \|(\hat\theta_0-\theta_0)\|\|W\|\})
\leq
Pr(\|(\hat\theta_0-\theta_0)\|\leq n^{-1/2}\log(n))
+
Pr(|Y-W'\theta_0|\leq n^{-1/2}\log(n)\|W\|\})
=
o(1)$.
The second part can be shown similarly.
\end{proof}

\begin{lemma}\label{omega_s_conv}
$E_n[\hat{s}(\alpha_0;\pi_0,\pi_1)\hat{s}(\alpha_0;\pi_0,\pi_1)']=E[s_0s_0']+o_p(1)$. 
\end{lemma}
\begin{proof}
The weak law of large numbers implies $E_n[s_0s_0']=E[s_0s_0']+o_p(1)$, 
and then it suffices to show $E_n[\hat{s}(\alpha_0;\pi_0,\pi_1)\hat{s}(\alpha_0;\pi_0,\pi_1)'-s_0s_0']=o_p(1)$.
Since 
\begin{eqnarray*}
&&
\hspace*{-1cm}
E_n[(\hat{s}(\alpha_0;\pi_0,\pi_1)-s_0)(\hat{s}(\alpha_0;\pi_0,\pi_1)-s_0)']
\\
&=&
E_n[(\widehat{\xi}(\alpha_0;\pi_0,\pi_1)-\xi_0(\nu_0))W
W'(\widehat{\xi}(\alpha_0;\pi_0,\pi_1)-\xi_0(\nu_0))']
\\&&+
E_n[(\widehat{\xi}(\alpha_0;\pi_0,\pi_1)-\xi_0(\nu_0))W
\widehat{\psi_{\delta}}']E_n\left[W(\widehat{\Xi}(\alpha_0;\pi_0,\pi_1)-\Xi_0(\delta_0))\right]'
\\&&+
E_n[(\widehat{\xi}(\alpha_0;\pi_0,\pi_1)-\xi_0(\nu_0))W
\widehat{\psi_{\delta}}'](E_n-E)\left[W\Xi_0(\delta_0)\right]'
\\&&+
E_n[(\widehat{\xi}(\alpha_0;\pi_0,\pi_1)-\xi_0(\nu_0))W
(\widehat{\psi_{\delta}}-\psi_{\delta})']E\left[W\Xi_0(\delta_0)\right]'
\\&&+
E_n\left[W(\widehat{\Xi}(\alpha_0;\pi_0,\pi_1)-\Xi_0(\delta_0))\right]E_n[\widehat{\psi_{\delta}}
W'(\widehat{\xi}(\alpha_0;\pi_0,\pi_1)-\xi_0(\nu_0))']
\\&&+
E_n\left[W(\widehat{\Xi}(\alpha_0;\pi_0,\pi_1)-\Xi_0(\delta_0))\right]E_n[\widehat{\psi_{\delta}}
\widehat{\psi_{\delta}}']E_n\left[W(\widehat{\Xi}(\alpha_0;\pi_0,\pi_1)-\Xi_0(\delta_0))\right]'
\\&&+
E_n\left[W(\widehat{\Xi}(\alpha_0;\pi_0,\pi_1)-\Xi_0(\delta_0))\right]E_n[\widehat{\psi_{\delta}}
\widehat{\psi_{\delta}}'](E_n-E)\left[W\Xi_0(\delta_0)\right]'
\\&&+
E_n\left[W(\widehat{\Xi}(\alpha_0;\pi_0,\pi_1)-\Xi_0(\delta_0))\right]E_n[\widehat{\psi_{\delta}}
(\widehat{\psi_{\delta}}-\psi_{\delta})']E\left[W\Xi_0(\delta_0)\right]'
\\&&+
(E_n-E)\left[W\Xi_0(\delta_0)\right]E_n[\widehat{\psi_{\delta}}
W'(\widehat{\xi}(\alpha_0;\pi_0,\pi_1)-\xi_0(\nu_0))']
\\&&+
(E_n-E)\left[W\Xi_0(\delta_0)\right]E_n[\widehat{\psi_{\delta}}
\widehat{\psi_{\delta}}']E_n\left[W(\widehat{\Xi}(\alpha_0;\pi_0,\pi_1)-\Xi_0(\delta_0))\right]'
\\&&+
(E_n-E)\left[W\Xi_0(\delta_0)\right]E_n[\widehat{\psi_{\delta}}
\widehat{\psi_{\delta}}'](E_n-E)\left[W\Xi_0(\delta_0)\right]'
\\&&+
(E_n-E)\left[W\Xi_0(\delta_0)\right]E_n[\widehat{\psi_{\delta}}
(\widehat{\psi_{\delta}}-\psi_{\delta})']E\left[W\Xi_0(\delta_0)\right]'
\\&&+
E\left[W\Xi_0(\delta_0)\right]E_n[(\widehat{\psi_{\delta}}-\psi_{\delta})
W'(\widehat{\xi}(\alpha_0;\pi_0,\pi_1)-\xi_0(\nu_0))']
\\&&+
E\left[W\Xi_0(\delta_0)\right]E_n[(\widehat{\psi_{\delta}}-\psi_{\delta})
\widehat{\psi_{\delta}}']E_n\left[W(\widehat{\Xi}(\alpha_0;\pi_0,\pi_1)-\Xi_0(\delta_0))\right]'
\\&&+
E\left[W\Xi_0(\delta_0)\right]E_n[(\widehat{\psi_{\delta}}-\psi_{\delta})
\widehat{\psi_{\delta}}'](E_n-E)\left[W\Xi_0(\delta_0)\right]'
\\&&+
E\left[W\Xi_0(\delta_0)\right]E_n[(\widehat{\psi_{\delta}}-\psi_{\delta})
(\widehat{\psi_{\delta}}-\psi_{\delta})']E\left[W\Xi_0(\delta_0)\right]'
\end{eqnarray*}
and 
\begin{eqnarray*}
&&
E_n[(\hat{s}(\alpha_0;\pi_0,\pi_1)-s_0)s_0']
\\
&&\quad=
E_n[(\widehat{\xi}(\alpha_0;\pi_0,\pi_1)-\xi_0(\nu_0))W
W'\xi_0(\nu_0)']
+E_n[(\widehat{\xi}(\alpha_0;\pi_0,\pi_1)-\xi_0(\nu_0))W\psi_{\delta}']E\left[W\Xi_0(\delta_0)\right]'
\\&&\qquad+
E_n\left[W(\widehat{\Xi}(\alpha_0;\pi_0,\pi_1)-\Xi_0(\delta_0))\right]E_n[\widehat{\psi_{\delta}}
W'\xi_0(\nu_0)']
\\&&\qquad+
E_n\left[W(\widehat{\Xi}(\alpha_0;\pi_0,\pi_1)-\Xi_0(\delta_0))\right]E_n[\widehat{\psi_{\delta}}
\psi_{\delta}']E\left[W\Xi_0(\delta_0)\right]'
\\&&\qquad+
(E_n-E)\left[W\Xi_0(\delta_0)\right]E_n[\widehat{\psi_{\delta}}
W'\xi_0(\nu_0)']
+(E_n-E)\left[W\Xi_0(\delta_0)\right]E_n[\widehat{\psi_{\delta}}
\psi_{\delta}']E\left[W\Xi_0(\delta_0)\right]'
\\&&\qquad+E\left[W\Xi_0(\delta_0)\right]E_n[(\widehat{\psi_{\delta}}-\psi_{\delta})
W'\xi_0(\nu_0)']
+E\left[W\Xi_0(\delta_0)\right]E_n[(\widehat{\psi_{\delta}}-\psi_{\delta}) 
\psi_{\delta}']E\left[W\Xi_0(\delta_0)\right]',
\end{eqnarray*}
it suffices to show that $(E_n-E)\left[W\Xi_0(\delta_0)\right]=o_p(1)$,
that
$E_n[\widehat{\psi_{\delta}}W'\xi_0(\nu_0)']$,
$E_n[\widehat{\psi_{\delta}}\psi_{\delta}']$, 
$E_n[\widehat{\psi_{\delta}}W'\xi_0(\nu_0)']$, 
and $E_n[\widehat{\psi_{\delta}}\psi_{\delta}']$ are $O_p(1)$,
 and that the sample averages of the following variables are $o_p(1)$: 
(1) $(\widehat{\xi}(\alpha_0;\pi_0,\pi_1)-\xi_0(\nu_0))W\psi_{\delta}'$, 
(2) $(\widehat{\xi}(\alpha_0;\pi_0,\pi_1)-\xi_0(\nu_0))WW'\xi_0(\nu_0)'$,
(3) $(\widehat{\xi}(\alpha_0;\pi_0,\pi_1)-\xi_0(\nu_0))WW'(\widehat{\xi}(\alpha_0;\pi_0,\pi_1)-\xi_0(\nu_0))'$,
(4) $(\widehat{\xi}(\alpha_0;\pi_0,\pi_1)-\xi_0(\nu_0))W\widehat{\psi_{\delta}}'$,
(5) $(\widehat{\xi}(\alpha_0;\pi_0,\pi_1)-\xi_0(\nu_0))W(\widehat{\psi_{\delta}}-\psi_{\delta})'$,
(6) $W(\widehat{\Xi}(\alpha_0;\pi_0,\pi_1)-\Xi_0(\delta_0))$,
(7) $\widehat{\psi_{\delta}}W'(\widehat{\xi}(\alpha_0;\pi_0,\pi_1)-\xi_0(\nu_0))'$,
(8) $\widehat{\psi_{\delta}}(\widehat{\psi_{\delta}}-\psi_{\delta})'$,
(9) $(\widehat{\psi_{\delta}}-\psi_{\delta})W'\xi_0(\nu_0)'$,
(10) $(\widehat{\psi_{\delta}}-\psi_{\delta})W'(\widehat{\xi}(\alpha_0;\pi_0,\pi_1)-\xi_0(\nu_0))'$, and 
(11) $(\widehat{\psi_{\delta}}-\psi_{\delta})(\widehat{\psi_{\delta}}-\psi_{\delta})'$.
The convergence $(E_n-E)\left[W\Xi_0(\delta_0)\right]=o_p(1)$ comes from the weak law of large numbers, and the other parts come from a combination of Assumption \ref{assn:condition_nu} (iv)-(vii) and the following equalities: 
\begin{eqnarray}
&&
E[\|\xi_0(\nu_0)W\|^2]
<\infty
\label{eq:minimin_goal1}\\
&&
E[\|(\widehat{\xi}(\alpha_0;\pi_0,\pi_1)-\xi_0(\nu_0))W\|^2]
=
o(1)
\label{eq:minimin_goal2}\\
&&
E\left[W(\widehat{\Xi}(\alpha_0;\pi_0,\pi_1)-\Xi_0(\delta_0))\right]
=
o_p(1).
\label{eq:minimin_goal3}
\end{eqnarray}
Note that Eq. (\ref{eq:minimin_goal1}) follows from $|\xi_0(\nu_0)|\leq 2$ and $E[\|W\|^2]<\infty$. 

First, Eq. (\ref{eq:minimin_goal2}) is shown as follows. 
Since 
\begin{eqnarray*}
|\widehat{\xi}(\alpha_0;\pi_0,\pi_1)-\xi_0(\nu_0)|
&\leq&
|(1\{Y-W'\theta_0\leq 0\}-1\{Y-W'\hat\theta_0\leq 0\})|\times |1-\pi_1-\hat\nu(Y,X,Z)|
\\&&+|\tau-1\{Y-W'\theta_0\leq 0\}|\times |\nu_0(Y,X,Z)-\hat\nu(Y,X,Z)|
\\&&+|1\{Y-\alpha_0-W'\theta_0\leq 0\}-1\{Y-\alpha_0-W'\hat\theta_0\leq 0\}|\times |\hat\nu(Y,X,Z)-\pi_0|
\\&&+|\tau-1\{Y-\alpha_0-W'\theta_0\leq 0\}|\times |\hat\nu(Y,X,Z)-\nu_0(Y,X,Z)|
\\
&\leq&
1\{|Y-W'\theta_0|\leq \|(\hat\theta_0-\theta_0)\|\|W\|\}
\\&&+1\{|Y-\alpha_0-W'\theta_0|
\leq \|(\hat\theta_0-\theta_0)\|\|W\|\}
\\&&+2|\hat\nu(Y,X,Z)-\nu_0(Y,X,Z)|, 
\end{eqnarray*}
it follows that 
\begin{eqnarray*}
E[\|(\widehat{\xi}(\alpha_0;\pi_0,\pi_1)-\xi_0(\nu_0))W\|^2]^{1/2}
&\leq&
Pr(|Y-W'\theta_0|\leq \|(\hat\theta_0-\theta_0)\|\|W\|\})^{1/4}E[\|\|W\|^4]^{1/4}
\\&&+
Pr(|Y-\alpha_0-W'\theta_0|\leq \|(\hat\theta_0-\theta_0)\|\|W\|)^{1/4}E[\|\|W\|^4]^{1/4}
\\&&+2E[|\hat\nu(Y,X,Z)-\nu_0(Y,X,Z)|^4]^{1/4}E[\|\|W\|^4]^{1/4}.
\end{eqnarray*}
By Assumption \ref{assn:condition_nu} (ii) and Lemma \ref{lemma:prob_diminising}, Eq. (\ref{eq:minimin_goal2}) holds. 

Next, Eq. (\ref{eq:minimin_goal3}) is shown as follows. 
Since 
\begin{eqnarray*}
\widehat{\Xi}(\alpha_0;\pi_0,\pi_1)-\Xi_0(\delta_0)
&=&
1\{Y-W'\hat\theta_0\leq 0\}\left(\widehat{\frac{\partial}{\delta'}\nu_{\delta}}(Y,X,Z)-{\frac{\partial}{\delta'}\nu_{\delta}}(Y,X,Z)\right)
\\&&
+\left(1\{Y-W'\hat\theta_0\leq 0\}-1\{Y-W'\theta_0\leq 0\}\right)\frac{\partial}{\delta'}\nu_{\delta}(Y,X,Z)
\\&&
-1\{Y-\alpha-W'\hat\theta_0\leq 0\}\left(\widehat{\frac{\partial}{\delta'}\nu_{\delta}}(Y,X,Z)-{\frac{\partial}{\delta'}\nu_{\delta}}(Y,X,Z)\right)
\\&&
-\left(1\{Y-\alpha-W'\hat\theta_0\leq 0\}-1\{Y-\alpha-W'\theta_0\leq 0\}\right){\frac{\partial}{\delta'}\nu_{\delta}}(Y,X,Z),
\end{eqnarray*}
it follows that 
\begin{eqnarray*}
&&
\hspace*{-1cm}
E\left[\left\|W(\widehat{\Xi}(\alpha_0;\pi_0,\pi_1)-\Xi_0(\delta_0))\right\|\right]
\\
&&\leq
2E\left[\left\|W\left(\widehat{\frac{\partial}{\delta'}\nu_{\delta}}(Y,X,Z)-{\frac{\partial}{\delta'}\nu_{\delta}}(Y,X,Z)\right)\right\|\right]
\\&&
\qquad+Pr(|Y-W'\theta_0|\leq \|(\hat\theta_0-\theta_0)\|\|W\|)^{1/2}
E\left[\left\|W\frac{\partial}{\delta'}\nu_{\delta}(Y,X,Z)\right\|^2\right]^{1/2}
\\&&
\qquad+Pr(|Y-\alpha_0-W'\theta_0|\leq \|(\hat\theta_0-\theta_0)\|\|W\|)^{1/2}
E\left[\left\|W\frac{\partial}{\delta'}\nu_{\delta}(Y,X,Z)\right\|^2\right]^{1/2}.
\end{eqnarray*}
By Lemma \ref{lemma:prob_diminising} and Assumption \ref{assn:condition_nu} (vi)-(vii), Eq. (\ref{eq:minimin_goal3}) holds. 
\end{proof}

\begin{lemma}\label{omega_lambda_conv}
$E_n[\hat{\lambda}(\alpha_0;\pi_0,\pi_1)WW']=E[\lambda_0(0)WW']+o_p(1)$.
\end{lemma}
\begin{proof}
Since 
\begin{eqnarray*}
E[\|\lambda_0(0)WW'\|^2]
&\leq&
E[\|f_{Y\mid D,Z,X}(W'\theta_0)(1-\pi_1-D)+f_{Y\mid D,Z,X}(\alpha_0+W'\theta_0)(D-\pi_0)
\|^2\cdot\|W\|^4]\\
&\leq&
E[(\|f_{Y\mid D,Z,X}(W'\theta_0)\|+\|f_{Y\mid D,Z,X}(\alpha_0+W'\theta_0)
\|)^2\cdot\|W\|^4]\\
&\leq&
4C^2E[\|W\|^4]\\
&<&
\infty,
\end{eqnarray*}
the weak law of large numbers implies $E_n[\lambda_0(0)WW']=E[\lambda_0(0)WW']+o_p(1)$.
It suffices to show that $E_n[\hat{\lambda}(\alpha_0;\pi_0,\pi_1)WW']=E_n[\lambda_0(0)WW']+o_p(1)$.
Using the mean value expansion,  
\begin{eqnarray*}
\hat{\lambda}(\alpha_0;\pi_0,\pi_1)-\lambda_0(0)
&=&
\left(K_h^{(1)}(\tilde{V})(1-\pi_1-D)+K_h^{(1)}(\tilde{V}-\alpha_0)(D-\pi_0)\right)W'
(\hat\theta_0-\theta_0)
\\&&+
\left(K_h(Y-W'\theta_0)-f_{Y\mid D,Z,X}(W'\theta_0)\right)(1-\pi_1-D)\\&&+\left(K_h(Y-\alpha_0-W'\theta_0)-f_{Y\mid D,Z,X}(\alpha_0+W'\theta_0)\right)(D-\pi_0),
\end{eqnarray*}
where $K_h(t)=K(t/h)/h$, and $\tilde{V}$ is a value between $Y-W'\hat\theta_0$  and $Y-W'\theta_0$.
Using the above mean value expansion, 
\begin{eqnarray*}
&&
\|E_n[\hat{\lambda}(\alpha_0;\pi_0,\pi_1)WW']-E_n[\lambda_0(0)WW']\|
\\
&&\quad=
\|E_n[(\hat{\lambda}(\alpha_ 0;\pi_0,\pi_1)-\lambda_0(0))WW']\|\\
&&\quad\leq
\left\|E_n\left[\left(K_h^{(1)}(\tilde{V})(1-\pi_1-D)+K_h^{(1)}(\tilde{V}-\alpha_0)(D-\pi_0)\right)W'(\hat\theta_0-\theta_0)WW'\right]\right\|
\\&&\quad\quad+
\|E_n[\left(K_h(Y-W'\theta_0)-f_{Y\mid D,Z,X}(W'\theta_0)\right)(1-\pi_1-D)WW']\|
\\&&\quad\quad+
\|E_n[\left(K_h(Y-\alpha_0-W'\theta_0)-f_{Y\mid D,Z,X}(\alpha_0+W'\theta_0)\right)(D-\pi_0)WW']\|\\
&&\quad\leq
2\sup_{v}|K_h^{(1)}(v)|\cdot\left\|(\hat\theta_0-\theta_0)\right\|\cdot E_n[\|W\|^3]
\\&&\quad\quad+
\|E_n[\left(K_h(Y-W'\theta_0)-f_{Y\mid D,Z,X}(W'\theta_0)\right)(1-\pi_1-D)WW']\|
\\&&\quad\quad+
\|E_n[\left(K_h(Y-\alpha_0-W'\theta_0)-f_{Y\mid D,Z,X}(\alpha_0+W'\theta_0)\right)(D-\pi_0)WW']\|\\
&&\quad\leq
O_p(n^{-1/2})+
\|E_n[\left(K_h(Y-W'\theta_0)-f_{Y\mid D,Z,X}(W'\theta_0)\right)(1-\pi_1-D)WW']\|
\\&&\quad\quad+
\|E_n[\left(K_h(Y-\alpha_0-W'\theta_0)-f_{Y\mid D,Z,X}(\alpha_0+W'\theta_0)\right)(D-\pi_0)WW']\|.
\end{eqnarray*}
Since each entry in $WW'$ has a finite variance, it suffices to show that 
\begin{eqnarray}
&&E_n[\left(K_h(Y-W'\theta_0)-f_{Y\mid D,Z,X}(W'\theta_0)\right)(1-\pi_1-D)\omega]=o_p(1)\label{eq_ss_ss1}\\
&&E_n[\left(K_h(Y-\alpha_0-W'\theta_0)-f_{Y\mid D,Z,X}(\alpha_0+W'\theta_0)\right)(D-\pi_0)\omega]=o_p(1)\label{eq_ss_ss2}
\end{eqnarray}
for a random variable $\omega$ such that $\omega$ is a function of $(D,Z,X)$ and $E[\omega^2]<\infty$. 
The rest of the proof is going to focus on (\ref{eq_ss_ss1}) because the proof for (\ref{eq_ss_ss2}) is similar. 
The mean of the left-hand side of (\ref{eq_ss_ss1}) is $O(h)$, because 
\begin{eqnarray*}
|E[\left(K_h(Y-W'\theta_0)-f_{Y\mid D,Z,X}(W'\theta_0)\right)\omega]|
&\leq&
E[\int |K(v)|\left|f_{Y\mid X,Z,D}(W'\theta_0+vh)-f_{Y\mid D,Z,X}(W'\theta_0)\right|dv|\omega|]\\
&\leq&
ChE[|\omega|]\int |K(v)v|dv\\
&=&
O(h).
\end{eqnarray*}
The variance of the left-hand side of (\ref{eq_ss_ss1}) is $O((nh)^{-1})$, because 
\begin{eqnarray*}
&&
E[(\left(K_h(Y-W'\theta_0)-f_{Y\mid D,Z,X}(W'\theta_0)\right)\omega)^2]
\\
&&\qquad\leq
E[\int K_h(y-W'\theta_0)^2f_{Y\mid X,Z,D}(y)dy\omega^2]
+E[f_{Y\mid D,Z,X}(W'\theta_0)^2\omega^2]
\\&&\qquad\qquad-2E[\int K_h(y-W'\theta_0)f_{Y\mid X,Z,D}(y)dyf_{Y\mid D,Z,X}(W'\theta_0)\omega]\\
&&\qquad\leq
Ch^{-1}\int K(v)^2dvE[\omega^2]+C^2E[\omega^2]+2C^2E[|\omega|]\\
&&\qquad=
O(h^{-1}).
\end{eqnarray*}
\end{proof}

\begin{lemma}\label{omega_conv}
$\hat\Omega(\alpha_0;\pi_0,\pi_1)=\Omega_0+o_p(1)$. 
\end{lemma}
\begin{proof}
Since $E[\lambda_0(0)WW']=E\left[f_{Y-\alpha_0D^\ast\mid Z,X}(W'\theta_0)WW'\right]$ is invertible, 
the statement of this lemma follows from Lemmas \ref{omega_s_conv} and \ref{omega_lambda_conv}. 
\end{proof}

\newpage
\setcounter{page}{1}
\section*{Supplemental Online Appendix for ``Instrumental Variable Quantile Regression with Misclassification'' by Takuya Ura}
\section*{Additional simulation results}
This supplemental online appendix provides additional simulation results in the simulation designs in Section \ref{sec:monte_calro}. 
The proposed inference method uses the Bonferroni size correction for the parameters $(\pi_0,\pi_1)$. 
To evaluate the conservatism of the Bonferroni correction, the proposed inference method is compared with the infeasible method with knowing $(p_0,p_1)=(\pi_0,\pi_1)$.
The infeasible inference method is based on the true parameter $(\pi_0,\pi_1)$, and uses $T(\alpha;\pi_0,\pi_1)$ and $cv$ in Section \ref{inference_section} with $\mathrm{size}_2=5\%$.

For all the simulation designs, the proposed method rejects the alternatives less often than the infeasible method. 
The power comparisons are different between the alternatives larger and smaller than $\alpha_0$.
On one hand, the power of the proposed method for the alternative smaller than $\alpha_0$ is comparable to that of the infeasible method. 
On the other hand, the powers are significantly different for the alternative larger $\alpha_0$.  
I conjecture that the value of $(p_0,p_1)$ might not be relevant to reject the alternative smaller than $\alpha_0$, because, as in Section \ref{sec:reducdee_fourm}, it is possible to get a lower bound for $\alpha_0$ without using the variable $D$.

\newpage

\begin{figure}[ht]
\parbox{.4\textwidth}{
\centering
\includegraphics[height=.4\textwidth,keepaspectratio]{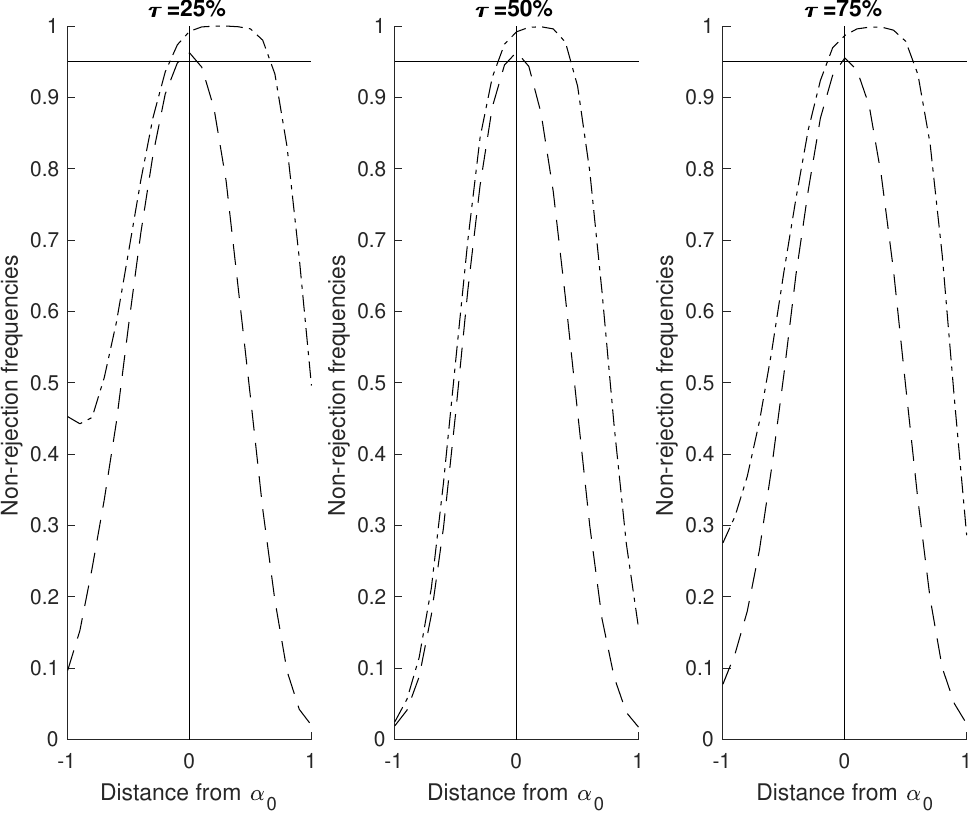}
\caption*{Figure \ref{fig_appendix1}.a:  $\pi_0=\pi_1=0$.}
}
\hspace{.1\textwidth}
\parbox{.4\textwidth}{
\centering
\includegraphics[height=.4\textwidth,keepaspectratio]{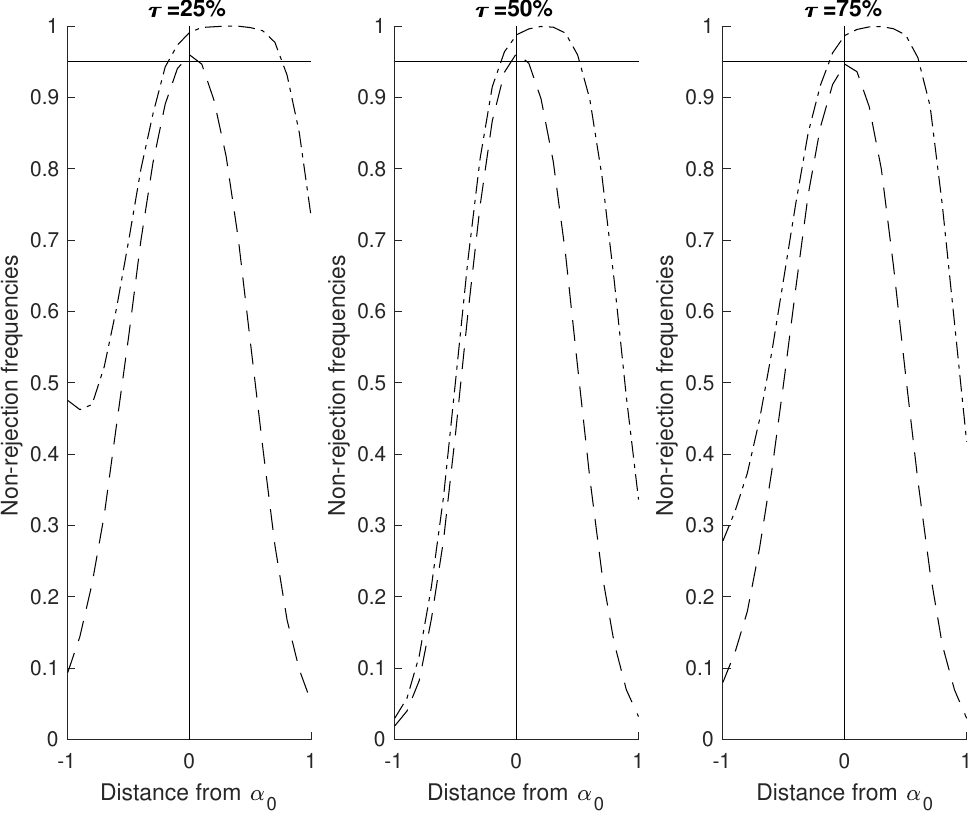}
\caption*{Figure \ref{fig_appendix1}.b:  $\pi_0=0.1$ and $\pi_1=0$.}
}
\\
\bigskip
\bigskip
\\
\parbox{.4\textwidth}{
\centering
\includegraphics[height=.4\textwidth,keepaspectratio]{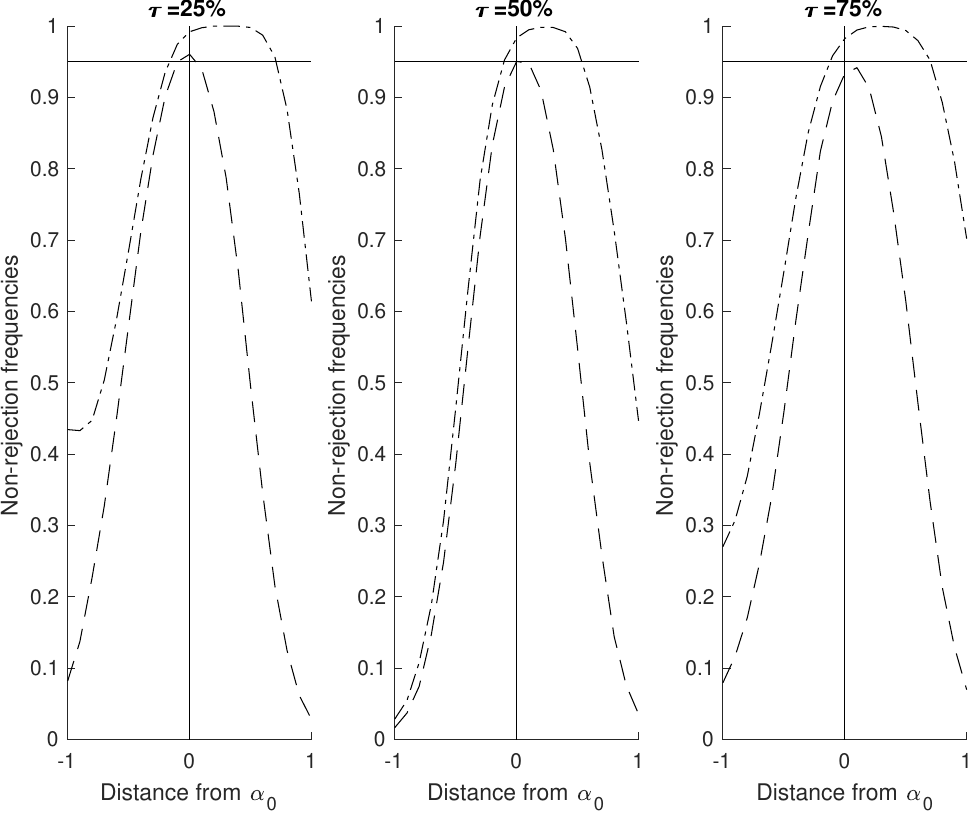}
\caption*{Figure \ref{fig_appendix1}.c: $\pi_0=0$ and $\pi_1=0.1$.}
}
\hspace{.1\textwidth}
\parbox{.4\textwidth}{
\centering
\includegraphics[height=.4\textwidth,keepaspectratio]{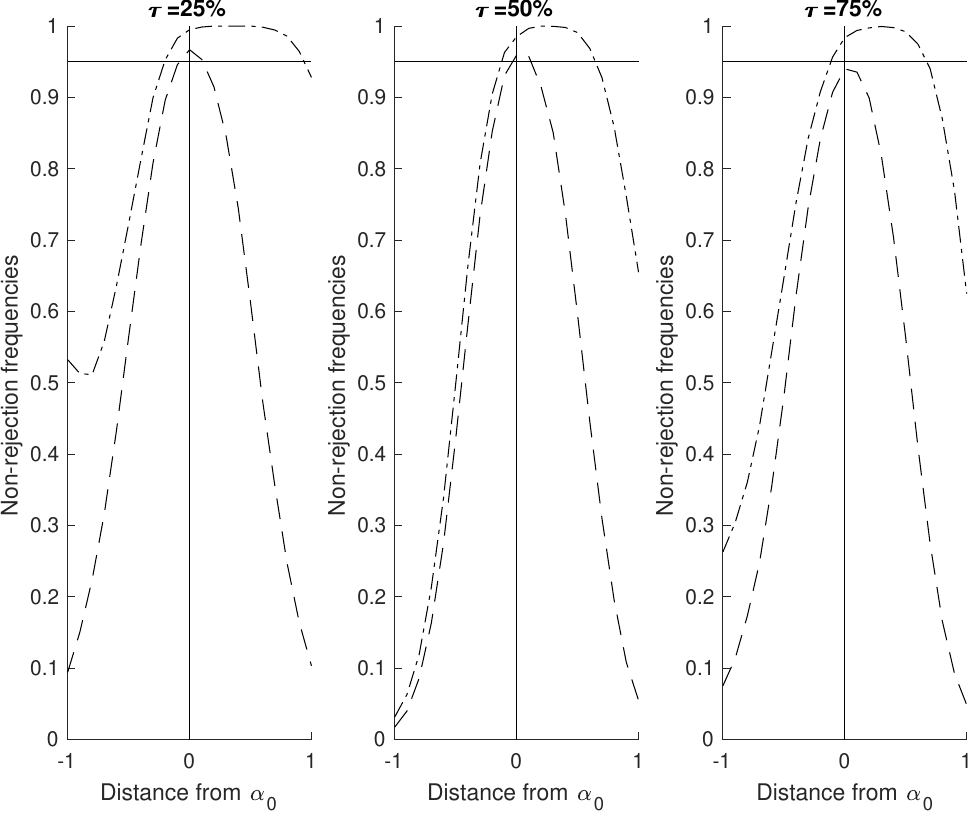}
\caption*{Figure \ref{fig_appendix1}.d: $\pi_0=0.2$ and $\pi_1=0$.}
}
\\
\bigskip
\bigskip
\\
\caption{Coverage frequencies. The dash-dot ($-.$) curve represents the proposed inference method, and the dashed ($--$) curve represents the infeasible inference method with knowing $(p_0,p_1)=(\pi_0,\pi_1)$.}
\label{fig_appendix1}
\end{figure}

\begin{figure}[ht]
\parbox{.4\textwidth}{
\centering
\includegraphics[height=.4\textwidth,keepaspectratio]{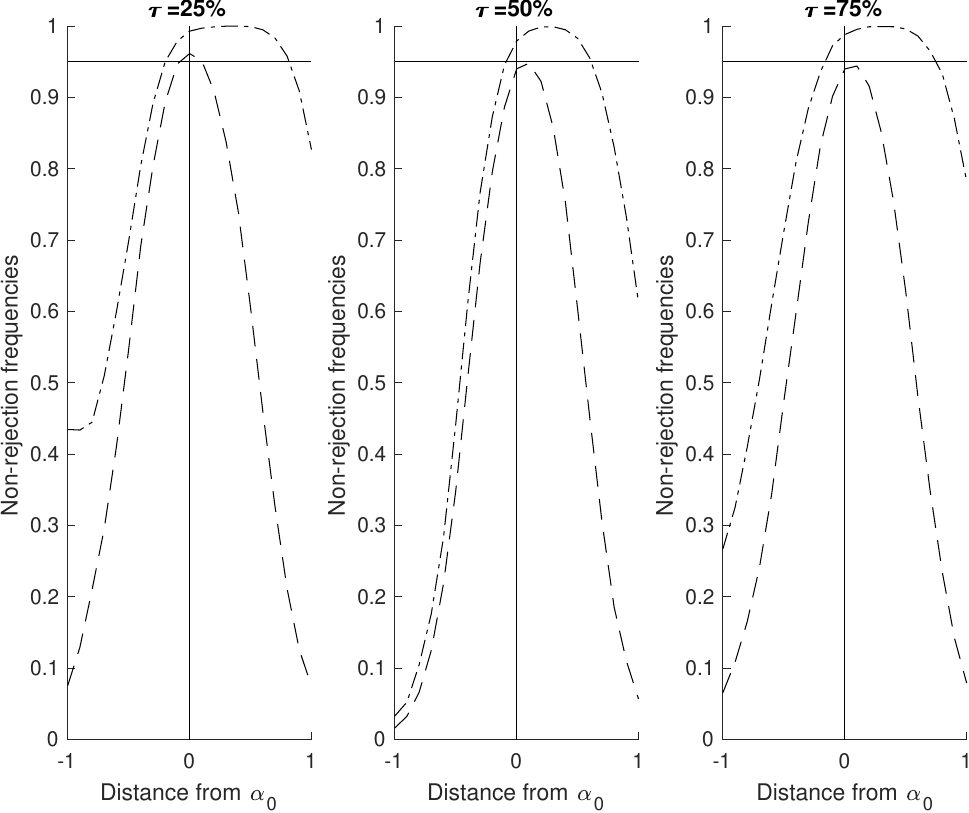}
\caption*{Figure \ref{fig_appendix1}.e:  $\pi_0=\pi_1=0.1$.}
}
\hspace{.1\textwidth}
\parbox{.4\textwidth}{
\centering
\includegraphics[height=.4\textwidth,keepaspectratio]{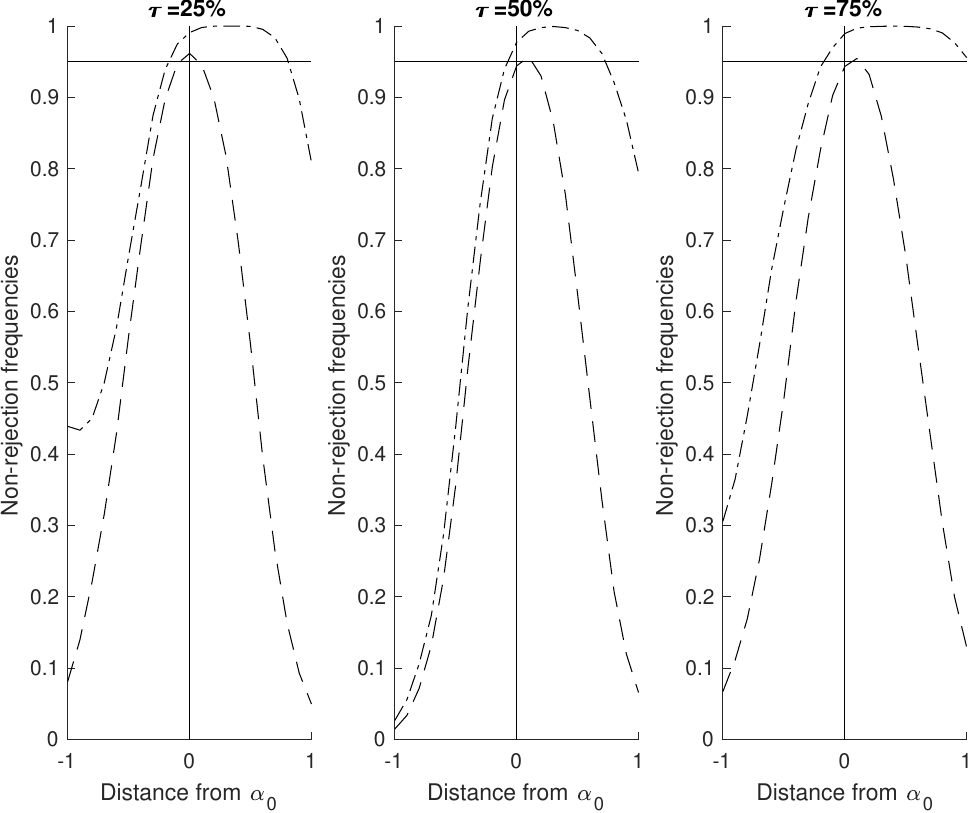}
\caption*{Figure \ref{fig_appendix1}.f:  $\pi_0=0$ and $\pi_1=0.2$.}
}
\\
\bigskip
\bigskip
\\
\parbox{.4\textwidth}{
\centering
\includegraphics[height=.4\textwidth,keepaspectratio]{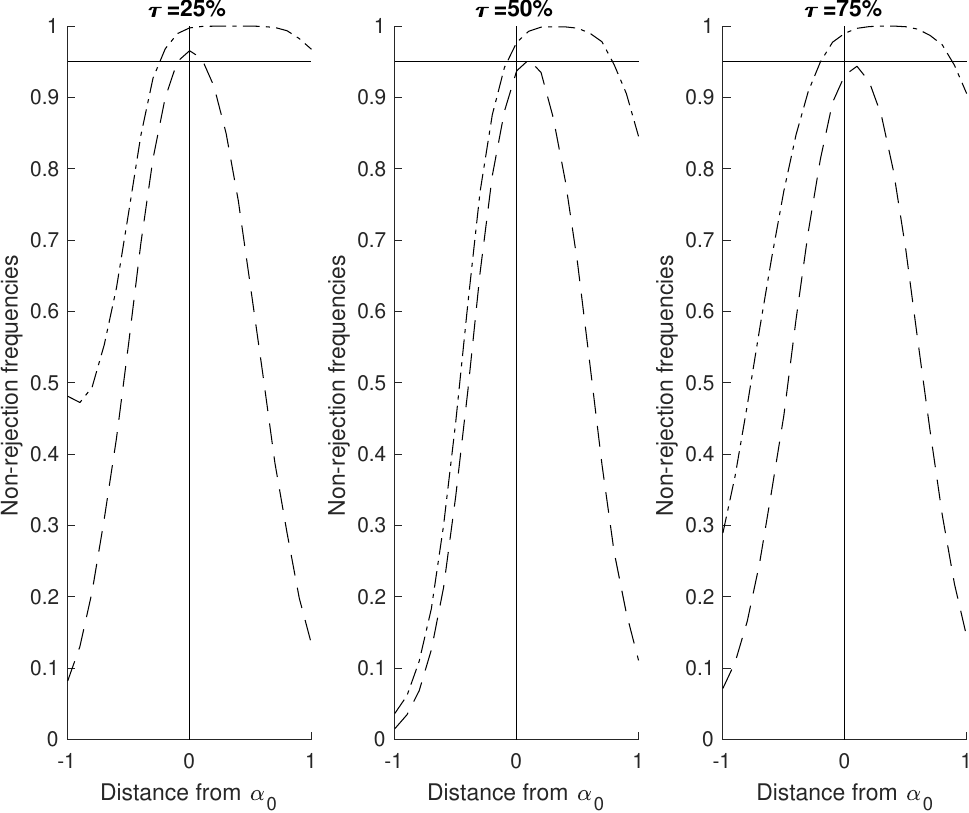}
\caption*{Figure \ref{fig_appendix1}.g: $\pi_0=0.2$ and $\pi_1=0.1$.}
}
\hspace{.1\textwidth}
\parbox{.4\textwidth}{
\centering
\includegraphics[height=.4\textwidth,keepaspectratio]{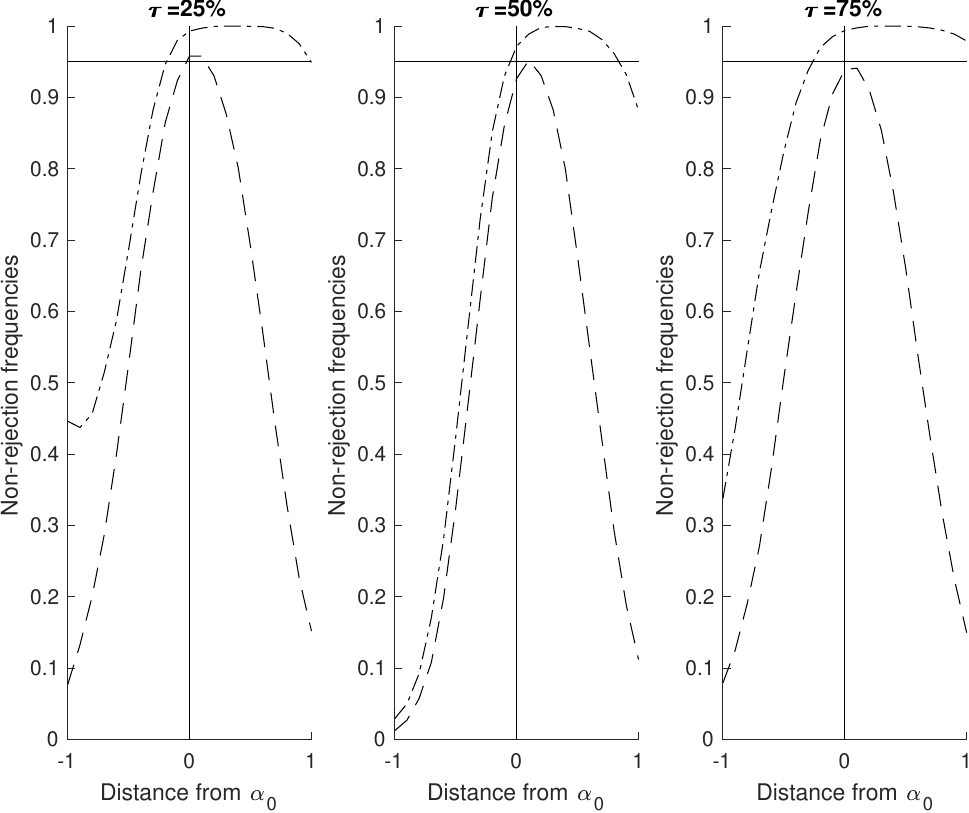}
\caption*{Figure \ref{fig_appendix1}.h: $\pi_0=0.1$ and $\pi_1=0.2$.}
}
\\
\bigskip
\bigskip
\\
\caption*{Figure \ref{fig_appendix1} (continued): Coverage frequencies. The dash-dot ($-.$) curve represents the proposed inference method, and the dashed ($--$) curve represents the infeasible inference method with knowing $(p_0,p_1)=(\pi_0,\pi_1)$.}
\end{figure}

\end{document}